\documentclass[12pt,a4paper]{article}
\usepackage{amsfonts,amssymb,graphicx,fullpage}
\usepackage[tight]{subfigure}
\usepackage{rotating}
\usepackage{sidecap}
\pdfoutput=1

\usepackage{ifpdf}
\ifpdf

\else
  
\fi
\usepackage{a4wide}
\usepackage{longtable}
\usepackage{graphicx}
\usepackage{subfigure}
\usepackage[centertags]{amsmath}
\usepackage{multirow}
\usepackage{jheppub} 
 \usepackage{color}

\newcommand{\unity}{{\footnotesize\mbox{1\!\!I}}}

\def\muc{\multicolumn}

\def\Z{\mathbb{Z}}
\def\R{\mathbb{R}}
\def\unity{1\!\!{\rm I}}
\def\ov{\overline}

\def\Sym{\mathbf{Sym}}
\def\Anti{\mathbf{Anti}}
\def\Adj{\mathbf{Adj}}

\def\ov{\overline}
\def\1{{\bf 1}}
\def\2{{\bf 2}}
\def\3{{\bf 3}}
\def\4{{\bf 4}}
\def\6{{\bf 6}}
\def\OR{\Omega\mathcal{R}}

\def\pp{\uparrow\uparrow}

\def\targ#1#2{\genfrac{[}{]}{0pt}{}{#1}{#2}}

\def\targ2#1#2{\genfrac{}{}{0pt}{}{#1}{#2}}

\definecolor{blus}{rgb}{0.1,0.1,0.8}
\definecolor{GreenYellow}{cmyk}{0.15,0,0.69,0}
\definecolor{Yellow}{cmyk}{0,0,1,0}
\definecolor{Goldenrod}{cmyk}{0,0.10,0.84,0}
\definecolor{Dandelion}{cmyk}{0,0.29,0.84,0}
\definecolor{Apricot}{cmyk}{0,0.32,0.52,0}
\definecolor{Peach}{cmyk}{0,0.50,0.70,0}
\definecolor{Melon}{cmyk}{0,0.46,0.50,0}
\definecolor{YellowOrange}{cmyk}{0,0.42,1,0}
\definecolor{Orange}{cmyk}{0,0.61,0.87,0}
\definecolor{BurntOrange}{cmyk}{0,0.51,1,0}
\definecolor{Bittersweet}{cmyk}{0,0.75,1,0.24}
\definecolor{RedOrange}{cmyk}{0,0.77,0.87,0}
\definecolor{Mahogany}{cmyk}{0,0.85,0.87,0.35}
\definecolor{Maroon}{cmyk}{0,0.87,0.68,0.32}
\definecolor{BrickRed}{cmyk}{0,0.89,0.94,0.28}
\definecolor{Red}{cmyk}{0,1,1,0}
\definecolor{OrangeRed}{cmyk}{0,1,0.50,0}
\definecolor{RubineRed}{cmyk}{0,1,0.13,0}
\definecolor{WildStrawberry}{cmyk}{0,0.96,0.39,0}
\definecolor{Salmon}{cmyk}{0,0.53,0.38,0}
\definecolor{CarnationPink}{cmyk}{0,0.63,0,0}
\definecolor{Magenta}{cmyk}{0,1,0,0}
\definecolor{VioletRed}{cmyk}{0,0.81,0,0}
\definecolor{Rhodamine}{cmyk}{0,0.82,0,0}
\definecolor{Mulberry}{cmyk}{0.34,0.90,0,0.02}
\definecolor{RedViolet}{cmyk}{0.07,0.90,0,0.34}
\definecolor{Fuchsia}{cmyk}{0.47,0.91,0,0.08}
\definecolor{Lavender}{cmyk}{0,0.48,0,0}
\definecolor{Thistle}{cmyk}{0.12,0.59,0,0}
\definecolor{Orchid}{cmyk}{0.32,0.64,0,0}
\definecolor{DarkOrchid}{cmyk}{0.40,0.80,0.20,0}
\definecolor{Purple}{cmyk}{0.45,0.86,0,0}
\definecolor{Plum}{cmyk}{0.50,1,0,0}
\definecolor{Violet}{cmyk}{0.79,0.88,0,0}
\definecolor{RoyalPurple}{cmyk}{0.75,0.90,0,0}
\definecolor{BlueViolet}{cmyk}{0.86,0.91,0,0.04}
\definecolor{Periwinkle}{cmyk}{0.57,0.55,0,0}
\definecolor{CadetBlue}{cmyk}{0.62,0.57,0.23,0}
\definecolor{CornflowerBlue}{cmyk}{0.65,0.13,0,0}
\definecolor{MidnightBlue}{cmyk}{0.98,0.13,0,0.43}
\definecolor{NavyBlue}{cmyk}{0.94,0.54,0,0}
\definecolor{RoyalBlue}{cmyk}{1,0.50,0,0}
\definecolor{Blue}{cmyk}{1,1,0,0}
\definecolor{Cerulean}{cmyk}{0.94,0.11,0,0}
\definecolor{Cyan}{cmyk}{1,0,0,0}
\definecolor{ProcessBlue}{cmyk}{0.96,0,0,0}
\definecolor{SkyBlue}{cmyk}{0.62,0,0.12,0}
\definecolor{Turquoise}{cmyk}{0.85,0,0.20,0}
\definecolor{TealBlue}{cmyk}{0.86,0,0.34,0.02}
\definecolor{Aquamarine}{cmyk}{0.82,0,0.30,0}
\definecolor{BlueGreen}{cmyk}{0.85,0,0.33,0}
\definecolor{Emerald}{cmyk}{1,0,0.50,0}
\definecolor{JungleGreen}{cmyk}{0.99,0,0.52,0}
\definecolor{SeaGreen}{cmyk}{0.69,0,0.50,0}
\definecolor{Green}{cmyk}{1,0,1,0}
\definecolor{ForestGreen}{cmyk}{0.91,0,0.88,0.12}
\definecolor{PineGreen}{cmyk}{0.92,0,0.59,0.25}
\definecolor{LimeGreen}{cmyk}{0.50,0,1,0}
\definecolor{YellowGreen}{cmyk}{0.44,0,0.74,0}
\definecolor{SpringGreen}{cmyk}{0.26,0,0.76,0}
\definecolor{OliveGreen}{cmyk}{0.64,0,0.95,0.40}
\definecolor{RawSienna}{cmyk}{0,0.72,1,0.45}
\definecolor{Sepia}{cmyk}{0,0.83,1,0.70}
\definecolor{Brown}{cmyk}{0,0.81,1,0.60}
\definecolor{Tan}{cmyk}{0.14,0.42,0.56,0}
\definecolor{Gray}{cmyk}{0,0,0,0.50}
\definecolor{Black}{cmyk}{0,0,0,1}
\definecolor{White}{cmyk}{0,0,0,0}

\newcommand{\bCaptionfonts}{\small}
\makeatletter
\long\def\@makecaption#1#2{%
  \vskip\abovecaptionskip
  \sbox\@tempboxa{{\bCaptionfonts #1: #2}}%
  \ifdim \wd\@tempboxa >\hsize
    {\bCaptionfonts #1: #2\par}
  \else
    \hbox to\hsize{\hfil\box\@tempboxa\hfil}%
  \fi
  \vskip\belowcaptionskip}
\makeatother

\makeatletter
\let\ORIGINALlatex@openbib@code=\@openbib@code
\renewcommand{\@openbib@code}{\ORIGINALlatex@openbib@code\setlength{\itemsep}{1ex plus.5ex minus.5ex}\setlength{\parsep}{0pt}}
\makeatother

\renewcommand{\arraystretch}{1.3}

\allowdisplaybreaks[4]

\title{To Tilt or Not To Tilt: \\ Discrete Gauge Symmetries in Global  Intersecting D-Brane Models}
\author[a]{Gabriele Honecker,}
\author[a]{Wieland Staessens,}

\affiliation[a]{PRISMA Cluster of Excellence \&
Institut f\"ur Physik  (WA THEP), Johannes-Gutenberg-Universit\"at, D-55099 Mainz, Germany}

\emailAdd{Gabriele.Honecker@uni-mainz.de}
\emailAdd{Wieland.Staessens@uni-mainz.de}

\abstract{Discrete gauge symmetries in global intersecting D-brane models constrain the exact form of the perturbative as well as non-perturbative superpotential. We derive the complete set of conditions on the existence of discrete $\Z_n$ gauge symmetries on toroidal orbifolds, $T^6/\Z_N$ and $T^6/\Z_2 \times \Z_{2M}$, with fractional or rigid D6-branes on tilted tori, for which global models of particle physics are known. Several examples of global left-right symmetric and Pati-Salam models are presented. Some discrete `stringy' $\Z_n$ symmetries are trivial from the field theory point of view, while others have not been identified before.}

\begin{document}
\hspace{12cm}{MZ-TH/13-019}

\maketitle
\flushbottom

\section{Introduction}\label{S:intro}
Symmetries form an important gateway to a deeper understanding of the physics in our universe. Local symmetries are intimately connected with particle interactions, while global symmetries in field theory are often
interpreted as accidental. However, discrete global symmetries play a major role in Beyond the Standard Model (BSM) physics. For example, different $\Z_n$ symmetries 
such as  R-parity, baryon triality or proton hexality  have been proposed to prevent too fast proton decay in the Minimal Supersymmetric extension of the Standard Model (MSSM) of particle physics~\cite{Ibanez:1991pr,Dreiner:2005rd}.

 In string theory, a global $U(1)$ symmetry is a perturbative remnant of a local gauge symmetry. The corresponding gauge boson acquires a mass at the string scale by a St\"uckelberg coupling to some axion,
whose complexification is the closed string modulus that  enters the supersymmetry condition on the corresponding D-brane in the language of compactifications of Type II string theory. Non-perturbative effects such as D-brane instantons further
 break the global {\it continuous} symmetry~\cite{Blumenhagen:2009qh} as expected in any  consistent quantum gravity theory~\cite{Abbott:1989jw,Coleman:1989zu,Kallosh:1995hi,Banks:2010zn,Banks:1988yz,Hellerman:2010fv}.
However, as recently realised in~\cite{BerasaluceGonzalez:2011wy} in general some {\it discrete} subgroup of the massive gauge symmetry survives. 
In~\cite{BerasaluceGonzalez:2011wy}, discrete symmetries in compactifications with D6-branes on the (orientifolded) six-torus and $\Z_2\times \Z_2$ orbifold were investigated for particular (non)-supersymmetric ``protomodels" (models for which the gauge and the visible chiral matter is already fixed yet not all torus wrapping numbers are completely determined). Whether a discrete $\Z_n$ symmetry exists, is closely linked to the values of the undetermined wrapping numbers, but many of the MSSM discrete symmetries turn out to be viable in these models as well. 
In~\cite{Ibanez:2012wg},  Gepner models with particle physics spectra were scanned for discrete remnants of massive gauge symmetries with the result that only  $\Z_2$-symmetries (R-parity) and some $\Z_3$-symmetries,
but no baryon triality, were found. 
In~\cite{Anastasopoulos:2012zu}, local models based on MSSM-like quivers were studied and an attempt to constrain the existence of global family-independent $\Z_n$ symmetries was made.
However, global D-brane models crucially differ from the most simple local gauge quivers in various ways. Global models generically contain several massive $U(1)$ gauge symmetries and additional matter, which
includes vector-like exotics with respect to the Standard Model gauge group as well as matter in hidden sectors. While for $U(1) \subset U(3)=SU(3)_{QCD} \times U(1)$ a discrete $\Z_3$ baryon number symmetry is naively 
expected~\cite{Ibanez:2001nd}, for Pati-Salam models, a $\Z_4 \subset U(1) \subset U(4)$ seems natural. Also the mixing among different massive $U(1)$ factors, such as e.g. the diagonal factor of $U(2)_L \times U(2)_R$, is seen here for the first time. Last but not least, our search provides family-dependent as well as family-independent $\Z_n$ symmetries.

This article is organised as follows: in section~\ref{S:DGS}, the basic considerations on the existence of a discrete $\Z_n$ symmetry are briefly reviewed. The constraints on the untilted six-torus are here for the first time 
generalised to all orbifold backgrounds with known global models containing the MSSM or some GUT. In subsection~\ref{Ss:DGSSUSYFT} the classification of $\Z_n$ symmetries in field theory is briefly reviewed. Section~\ref{S:GICBM}
contains a systematic search for discrete $\Z_n$ symmetries in all known MSSM and Pati-Salam models on fractional or rigid D6-branes in the $T^6/\Z_6$ and $T^6/\Z_6'$ and $T^6/\Z_2 \times \Z_6'$ orientifolds.
The discussion of our results and our conclusions are given in section~\ref{S:Conclusions}.

\section{Discrete Gauge Symmetries}\label{S:DGS}

In section~\ref{Ss:NecSufConDbranes} we briefly review the known constraints~\cite{BerasaluceGonzalez:2011wy} on discrete $\Z_n$ symmetries in intersecting D6-brane models on Type IIA orientifolds
and then proceed in section~\ref{Ss:DiscreteOrbifolds} to extend the known formulas from the six-torus to all toroidal orbifolds, for which global fractional or rigid D6-brane models with particle physics spectra are known~\cite{Honecker:2004kb,Gmeiner:2007zz,Gmeiner:2008xq,Gmeiner:2009fb,Honecker:2012qr}.  In section~\ref{Ss:DGSSUSYFT}, we briefly review the field theoretical classification of discrete $\Z_n$ symmetries in the MSSM~\cite{Ibanez:1991pr,Dreiner:2005rd}. 
This section establishes all new techniques required in section~\ref{S:GICBM} for the study of global D6-brane models with MSSM-like and Pati-Salam spectra, where we also discuss the difference of `stringy' discrete $\Z_n$ symmetries
as compared to the purely field theoretical considerations.

\subsection{Conditions on the existence of discrete $\Z_n$ symmetries in intersecting D6-brane worlds}\label{Ss:NecSufConDbranes}

In Type IIA orientifolds, the worldsheet parity $\Omega$ needs to be combined with an anti-holomorphic involution ${\cal R}$ on the Calabi-Yau threefold ${\cal CY}_3$~\cite{Ibanez:2012zz,Blumenhagen:2006ci}. 
If ${\cal CY}_3$ is given by an toroidal orbifold or a hypersurface~\cite{Palti:2009bt} in 
some weighted projective space, the involution ${\cal R}$ is simply taken to be complex conjugation per two-torus, ${\cal R}: z^i \to \ov{z}^i$ for $i \in \{1,2,3\}$. Supersymmetric D6-branes in Type IIA string theory on ${\cal CY}_3/\OR$ 
wrap special Lagrangian three-cycles. The $b_3=2 \, h_{21}+2$ dimensional lattice of three-cycles can in the most simple case of the untilted six-torus
be decomposed into a symplectic basis of  $\OR$-even cycles $\Pi^{\text{even}}_i$ and  $\OR$-odd cycles $\Pi^{\text{odd}}_j$ with\footnote{In toroidal orbifold compactifications with at least one tilted 
two-torus, the statement has to be modified as follows: the $\OR$-even cycles $\Pi^{\text{even}}_i$ and  $\OR$-odd cycles $\Pi^{\text{odd}}_j$  form in general an integral lattice within the full lattice of three-cycles, but their intersection form is not unimodular. The wrapping numbers $A^i_a, B^i_a$ take values in $\mathbb{Q}$ instead of $\Z$, and the corresponding conditions in equation~(\ref{Eq:massless-U1}) 
and~(\ref{Eq:discrete-Zn}) used before in~\cite{BerasaluceGonzalez:2011wy,Anastasopoulos:2012zu} require an appropriate rescaling. 
Spelling out the conditions in terms of intersection numbers in equation~(\ref{Eq:Zn-condition})  avoids this problem as exemplified for various orbifold backgrounds in section~\ref{Ss:DiscreteOrbifolds}.
}
\begin{equation}\label{Eq:even-odd-intersection}
\Pi^{\text{even}}_i \circ \Pi^{\text{odd}}_j =  \delta_{ij}
\qquad \text{for} \qquad
i,j \in \{0, \ldots, h_{21}\}
.
\end{equation}
A generic three-cycle $\Pi_a$ wrapped by the D6-brane $a$ along ${\cal CY}_3/\OR$ can likewise be decomposed in terms of $\OR$-even and $\OR$-odd components,
\begin{equation}\label{Eq:3-cycle_even+odd}
\Pi_a= \sum_{i=0}^{h_{21}} \left(A^i_a \, \Pi^{\text{even}}_i + B^i_a \, \Pi^{\text{odd}}_i  \right)
,
\qquad 
\Pi_{a'}= \sum_{i=0}^{h_{21}} \left(A^i_a \, \Pi^{\text{even}}_i - B^i_a \, \Pi^{\text{odd}}_i  \right)
,
\end{equation} 
where $\Pi_{a'}$ denotes the $\OR$-image three-cycle of $\Pi_a$, and $A^i_a, B^i_a \in \Z$ denote the wrapping numbers in the symplectic unimodular basis.
 In practice, usually a different basis of three-cycles from the above one is used, and the coefficients $A^i_a$, $B^i_a$ in the decomposition~(\ref{Eq:3-cycle_even+odd}) are read off from  
$\frac{\Pi_a +\Pi_{a'} }{2}= \sum_{i=0}^{h_{21}} A^i_a \, \Pi^{\text{even}}_i$ and $\frac{\Pi_a - \Pi_{a'}}{2}= \sum_{i=0}^{h_{21}} B^i_a \, \Pi^{\text{odd}}_i$, respectively.

The symplectic basis  of three-cycles is important for D6-brane model building in several respects:
\begin{enumerate}
\item
Stacks of $N_a$ identical D6-branes $a$ wrapping generic three-cycles $\Pi_a$ as in equation~(\ref{Eq:3-cycle_even+odd}) support $U(N_a)$ gauge groups.
If the D6-branes wrap only an orientifold even three-cycle, $ \Pi_a=\Pi_{a'}= \Pi^{\text{even}}_i$ for some $i\in \{0, \ldots, h_{21}\}$, additional string excitations with antisymmetric or symmetric Chan-Paton factors
become massless, and the dimension of the gauge group is enhanced to $SO(2N_a)$ or $USp(2N_a)$. The type of gauge enhancement depends on the cycle $\Pi^{\text{even}}_i$ 
and can to date only be determined on a case-by-case basis as detailed in section~\ref{Ss:DiscreteOrbifolds} for various toroidal orbifold backgrounds.
\item
According to~\cite{Uranga:2000xp,Uranga:2002vk,MarchesanoBuznego:2003hp}, the K-theory constraint on globally consistent D6-brane models can be expressed in terms of intersection numbers
with all those `probe' D6-branes wrapped on three-cycles supporting $USp(2)_i$ gauge factors, 
\begin{equation}\label{Eq:K-theory}
\Pi^{\text{even}}_i  \circ \sum_a N_a \Pi_a \stackrel{!}{=} 0 \text{  mod } 2
\qquad
\forall i \text{ with }
U(1)_i \hookrightarrow USp(2)_i.
\end{equation}
The same three-cycles  $ \Pi^{\text{even}}_i$ can be used for model building with $SU(2)_L \simeq USp(2)_i$ gauge factors instead of $SU(2)_L \subset U(2)$.
\item
The four-dimensional closed string spectrum contains $h_{21}+1$ axionic scalars $\phi_i$ and their dual two-forms $B^i_{(2)}$
which stem from the dimensional reduction of the ten-dimensional RR-form $C_{(3)}$ and its dual $C_{(5)}$ along the symplectic basis of three-cycles, 
\begin{equation}\label{Eq:Def-Axions}
\phi_i \equiv \frac{1}{\ell_s^3} \int_{ \Pi^{\text{even}}_i} C_{(3)}
,
\qquad
B^i_{(2)} \equiv \frac{1}{\ell_s^5} \int_{\Pi^{\text{odd}}_i} C_{(5)}
\qquad \text{with} \qquad
 \ell_s \equiv 2\pi \sqrt{\alpha'}
.
\end{equation}
The dimensional reduction of the Chern-Simons actions along all D6-branes $a$ in a given global D-brane model, 
\begin{equation}\label{Eq:GS-couplings}
\sum_a \sum_{i=0}^{h_{21}} A^i_a \; \int_{\R^{1,3}} \phi_i \; \text{tr} \left(F_a \wedge F_a \right)
,
\qquad
\sum_a N_a  \sum_{i=0}^{h_{21}}  B^i_a \;  \int_{\R^{1,3}}  B^i_{(2)} \wedge F_{U(1)_a}
,
\end{equation}
with wrapping numbers $A^i_a, B^i_a \in \Z$ defined in equation~(\ref{Eq:3-cycle_even+odd})
provides the Green-Schwarz couplings to cancel all mixed one-loop gauge anomalies of the form $U(1)_a-U(1)_b^2$ and $U(1)_a - SU(N_b)^2$.
The St\"uckelberg terms on the right hand side of equation~(\ref{Eq:GS-couplings}) provide masses for the corresponding Abelian gauge bosons.
\item
A linear combination $U(1)_X=\sum_a q_a U(1)_a$ remains anomaly-free and massless if the corresponding 
St\"uckelberg coupling in equation~(\ref{Eq:GS-couplings}) vanishes, or equivalently if the following constraint is satisfied in homology,
\begin{equation}\label{Eq:massless-U1}
 \sum_a q_a \, N_a \frac{\Pi_a - \Pi_{a'}}{2}=0
 \quad \Leftrightarrow \quad 
 \sum_a q_a \, N_a B^i_a =0 \quad \forall i
\qquad 
\text{with} \quad q_a \in \mathbb{Q}.
\end{equation}
The factor $N_a$ arises from the trace normalizations of the diagonal $U(1)_a \subset U(N_a)$ factors~\cite{Ghilencea:2002da,Gmeiner:2008xq}.
\end{enumerate}

As argued in~\cite{BerasaluceGonzalez:2011wy}, the condition~(\ref{Eq:massless-U1}) can be generalised for discrete $\Z_n$ subgroups of massive Abelian gauge symmetries,
\begin{equation}\label{Eq:discrete-Zn}
 \sum_a k_a \, N_a \frac{\Pi_a - \Pi_{a'}}{2}=0 \text{ mod } n
\quad \Leftrightarrow \quad 
\sum_a k_a \, N_a B^i_a =0  \text{ mod } n  \quad \forall i 
\qquad 
\text{with} \quad k_a \in \Z
,
\end{equation}
where `$0 \text{ mod } n$' in homology means that the right hand side is $n$ times a linear combination of the basic $\OR$-odd three-cycles $\Pi^{\text{odd}}_i$. 
Due to the $\Z_n$ symmetry, the coefficients $k_a$ in the linear combinations can be chosen to lie in the interval $0 \leqslant k_a < n$ with $\gcd(k_a,k_b \ldots n)=1$.

In practice, on toroidal orbifolds it is more convenient to work with a basis of three-cycles which differs from the symplectic one with $\OR$-even and $\OR$-odd elements. It is then useful
to rewrite the constraints on the existence of massless Abelian gauge symmetries~(\ref{Eq:massless-U1}) or discrete $\Z_n$ symmetries~(\ref{Eq:discrete-Zn})  analogously to the K-theory constraint~(\ref{Eq:K-theory})
as~\cite{BerasaluceGonzalez:2011wy}
\begin{equation}\label{Eq:Zn-condition}
\begin{array}{ll}
 \Pi^{\text{even}}_i \circ \sum_a q_a \, N_a \, \Pi_a =0 & \quad \text{ for a massless } U(1)_X=\sum_a q_a U(1)_a
 ,
\\
 \Pi^{\text{even}}_i \circ \sum_a k_a \, N_a \, \Pi_a =0 \text{ mod } n \quad \forall i
 & \quad \text{ for a discrete } \Z_n \subset \sum_a k_a U(1)_a
 .
 \end{array}
\end{equation}
On orbifolds, where all gauge enhancements are of the type $U(1)_i \hookrightarrow USp(2)_i$ and none $U(1)_i \hookrightarrow SO(2)_i$, the K-theory constraint in equation~(\ref{Eq:K-theory}) is identical to the second line in equation~(\ref{Eq:Zn-condition}) with $k_a \equiv 1$ and $n=2$ and  therefore implies the existence of a `diagonal' discrete $\Z_2$ symmetry. This `diagonal' $\Z_2$ symmetry occurs for instance in all globally consistent D6-brane models on $T^6/(\Z_2 \times \Z_2 \times \OR)$  and $T^6/(\Z_2 \times \Z_4 \times \OR)$ orbifolds without discrete torsion~\cite{Forste:2000hx,Cvetic:2001tj,Cvetic:2001nr,Honecker:2003vq,Honecker:2004np}. 

On orbifolds where some gauge enhancement of the type $U(1)_i \hookrightarrow SO(2)_i$ occurs, the `diagonal' discrete $\Z_2$ symmetry does not necessarily exist. 
In particular, the orbifolds $T^6/\OR$, $T^6/(\Z_3 \times \OR)$, $T^6/(\Z_{2N} \times \OR)$ or $T^6/(\Z_2 \times \Z_{2M} \times \OR)$ with discrete torsion allow for both types of gauge enhancements~\cite{Honecker:2011sm,Honecker:2012qr}  - in some cases depending on the number of tilted two-tori. 
A more detailed discussion on the existence of the `diagonal' $\Z_2$ symmetry is given for each example on $T^6/(\Z_6 \times \OR)$, $T^6/(\Z_6' \times \OR)$ and $T^6/(\Z_2 \times \Z_6' \times \OR)$
with discrete torsion in sections~\ref{Sss:Z6} to~\ref{Sss:Z2Z6p}.

The formulation~(\ref{Eq:Zn-condition}) has the advantage that it is unnecessary to determine the basis $\Pi^{\text{odd}}_i$ of $\OR$-odd three-cycles explicitly, or the intersection form $\Pi^{\text{even}}_i \circ \Pi^{\text{odd}}_j= c_i \; \delta_{ij} $ with $c_i \in \Z$ is not required to be unimodular, as shown in the next section~\ref{Ss:DiscreteOrbifolds}.
 
 \subsubsection{$\Z_n$ action on closed string axions and D2-brane instantons}\label{Sss:Zn-axions+instantons}
 
The existence of the discrete $\Z_n$ symmetry in the low-energy effective field theory can be verified as follows: the axion and dual two-form of equation~(\ref{Eq:Def-Axions}) are 
for any integer valued intersection form, $c_i \in \Z$, related by 
\begin{equation}
d B_{(2)}^{i} =  c_i\, \star_4 d \phi_i.
\end{equation}
Dualization of the $B^i_{(2)}\wedge F_{U(1)_a}$ term in the four dimensional Chern Simons action~(\ref{Eq:GS-couplings}) thus implies that the axion is shifted as follows under a massive $U(1)$ gauge transformation,
\begin{equation}\label{eq:Discrete_GaugeTrafo}
A_\mu \rightarrow A_\mu + \partial_\mu \lambda, \hspace{0.2in}  \phi_i \rightarrow \phi_i + \left(\sum_a N_a k_a B_a^i c_i \right) \lambda.
\end{equation} 
Due to  the axionic shift symmetry $\phi_i \rightarrow \phi_i +1$, the discrete $\Z_n$ gauge symmetry is manifest, and the r.h.s. of equation~(\ref{Eq:discrete-Zn})
is naturally generalised to \mbox{$\sum_a N_a k_a B_a^i c_i = 0 \text{ mod } n \; \forall i$} for non-unimodular lattices with some $c_i \neq \pm 1$.

The discrete $\Z_n$ symmetries are left unbroken by Euclidean D2-brane instantons wrapped on three-cycles, as first observed in~\cite{BerasaluceGonzalez:2011wy}. This characteristic can 
be easily shown to hold for non-unimodular lattices as well. For $\OR$-invariant Euclidean D2-branes supporting $O(1)$ instantons, the four-dimensional effective action contains the factor $e^{-{\cal S}_{D2,O}}$ with the classical D2-instanton action
\begin{equation}\label{Eq:InstantonAction}
{\cal S}_{D2,O} = - \frac{\text{Vol} (D2_O)}{g_s} +  2 \pi i\, \phi_{O} \hspace{0.2in} \text{with} \hspace{0.1in} \phi_{O} = \frac{1}{\ell_s^3} \int_{\Pi_{D2,O}} \hspace{-2mm} C_{(3)} =  \sum_{i=0}^{h_{21}}  A_{D2,O}^i \phi_{(i)}.
\end{equation}
The axionic shift~(\ref{eq:Discrete_GaugeTrafo}) of the $\phi_i$'s under a massive $U(1)=\sum_a k_a U(1)_a$ gauge transformation induces a shift in the classical $O(1)$ instanton action, 
\begin{equation} 
{\cal S}_{D2,O} \rightarrow {\cal S}_{D2,O} + 2 \pi i\,  \sum_{i=0}^{h_{21}}  A_{D2,O}^i  \left(\sum_a N_a k_a B_a^i c_i \right) \, \lambda,
\end{equation}
which can be recast in terms of intersection numbers,
\begin{equation}\label{Eq:TopConInstDiscSymm}
\sum_{i=0}^{h_{21}}  A_{D2,O}^i  \left(\sum_a N_a k_a B_a^i c_i \right) =\Pi_{D2,O} \circ \sum_a k_a \, N_a \, \Pi_a = 0 \text{ mod } n,
\end{equation}
 since $\Pi_{D2,O}=\sum_{i=0}^{h_{21}} A_{D2,O}^i \, \Pi^{\text{even}}_i$ constitutes a special Lagrangian three-cycle in the unimodular lattice.
In the case of a $U(1)$ instanton, the $\phi$-term in~(\ref{Eq:InstantonAction}) receives an additional contribution from the $\OR$-image three-cycle, 
$\phi_U=\frac{1}{\ell_s^3} \int_{\Pi_{D2,U} +\Pi_{D2,U}^{\prime} } \hspace{-2mm} C_{(3)} $, and using $\Pi_{D2,U}^{\prime} \circ \Pi_a =- \Pi_{D2,U} \circ \Pi_a^{\prime}$ 
equation~(\ref{Eq:TopConInstDiscSymm}) is replaced by 
\begin{equation}
\Pi_{D2,U} \circ 2 \sum_a k_a \, N_a \, \frac{\Pi_a-\Pi_a^{\prime}}{2}=0 \text{ mod } 2n.
\end{equation}
The $U(1)$ instantons thus respect an extended discrete symmetry $\Z_{2n}$. However, the factor of two reappears in the counting of zero modes, with the consequence that the 
 effective action does not receive single $U(1)$ instanton contributions~\cite{Blumenhagen:2009qh}. Analogous arguments apply to $USp(2)$ instantons.

 \subsubsection{Discrete $\Z_n$ symmetries versus centers of non-Abelian gauge groups}\label{Sss:Zn-vs-center}

As remarked in~\cite{BerasaluceGonzalez:2011wy}, the gauge group on $N$ identical D-branes is given by the isomorphism $U(N) \simeq (SU(N)\times U(1))/{\rm ker}(\phi)$, where $\phi$ is the covering homomorphism
\begin{equation}
\begin{aligned}
\phi : SU(N) \times U(1) & \rightarrow U(N)\\
(e^{i\, \alpha^i T_i}, e^{i \theta}) & \mapsto e^{i \theta}  \cdot  \, e^{i\, \alpha^i T_i} ,
\end{aligned}
\end{equation}
with Lie algebra generators $T_i$ and group parameters $\theta,\alpha^i$. The kernel of the homomorphism is given by ${\rm ker}(\phi) = \Z_N$, and the center of $U(N)$ is given by $U(1)$.
As a consequence, the discrete $\Z_n$ symmetry arising from a single $U(N)$ gauge group can be larger than the center $\Z_N$ of the non-Abelian subgroup $SU(N)$.
  In subsections~\ref{Ss:T6Z6pNoHidden} and~\ref{Ss:T6Z6pWithHidden} we provide examples of a discrete $\Z_6$ gauge symmetry arising from a $U(2)$ gauge factor.  

The automatic $\Z_N$ subgroup of a single $U(N)$ factor defined by the condition~(\ref{Eq:Zn-condition}) does not correspond to the center of the non-Abelian subgroup $SU(N)$. This can e.g. be seen using the 
isomorphism between $USp(2)$ and $SU(2)$: the $USp(2)$ group does not lead to any discrete $\Z_2$ subgroup according to the condition~(\ref{Eq:Zn-condition}), and the closed string axions do not receive any gauge shift~(\ref{eq:Discrete_GaugeTrafo}) since there is no St\"uckelberg coupling  ($B_a^i = 0$ $\forall i$). The $U(1) \subset U(2)$ group on the other hand receives a St\"uckelberg mass as per~(\ref{Eq:GS-couplings}), and the shift of the closed string axions~(\ref{eq:Discrete_GaugeTrafo}) makes the $\Z_2$ symmetry within $U(1)$ manifest.

Another indication that discrete gauge symmetries cannot be simply reduced to the center of the non-Abelian semi-simple subgroup can be found in the spontaneous breaking of a gauge group $U(N_a+N_b) \longrightarrow U(N_a)\times U(N_b)$ due to a displacement of non-rigid D6-brane stacks. Both stacks remain wrapped on three-cycles in the same homology class, and the St\"uckelberg couplings are identical ($c_i B_a^i = c_i B^i_b $ $\forall$ $i$). Thus, the conditions for discrete $\Z_n$ symmetries arising from $U(1)_a\times U(1)_b$  read:
\begin{equation}   
(k_{a} N_a + k_{b} N_b) c_i B_a^i = 0 \text{ mod } n \qquad \forall i,
\end{equation}
with the intersection numbers $c_i \in \Z$ and ignoring further D6-brane stacks. The linear combination $N_b\, U(1)_a - N_a\, U(1)_b$ corresponds to a massless $U(1)$ gauge symmetry, while the orthogonal massive direction allows for the presence of a discrete $\Z_{N_a+N_b}$ symmetry. The effective gauge group for this configuration is thus $SU(N_a) \times SU(N_b) \times U(1)_{\text{massless}} \times \Z_{N_a + N_b}$, where
the group $\Z_{N_a + N_b}$ is in general smaller than the center of the non-Abelian factors  $\Z_{N_a} \times \Z_{N_b}$. 
Similarly, under the spontaneous breaking of $U(N+1) \longrightarrow U(N) \times U(1) $, the effective gauge group is $SU(N)_a \times U(1)_{\text{massless}} \times \Z_{N+1}$ with $\Z_{N+1}$ larger than 
the center of $SU(N)$. Explicit examples of this situation appear in Pati-Salam models with Higgsing by some vacuum expectation value in the adjoint representation, such as in the global models in section~\ref{Ss:PS-models}.
  
These considerations indicate that Abelian discrete gauge symmetries purely originate from the $U(1)$ center of the $U(N)$ gauge group, and not from the center of the non-Abelian $SU(N)$ subgroup. 
However, in the open string sector the discrete $\Z_n$ symmetries defined by~(\ref{Eq:Zn-condition}) will often appear trivial since they forbid couplings that were already excluded by the non-Abelian charges. Then again, antisymmetric representations of $U(2)$ are singlets under the non-Abelian group $SU(2)$, but the Abelian charge constrains both perturbative and non-perturbative couplings, provided that the discrete $\Z_n$ symmetry does not coincide with the automatic $\Z_2 \subset U(2)$. Explicit examples of this effect are given by the $\Z_6$ symmetry in subsection~\ref{Ss:T6Z6pNoHidden} and by the generation-dependent $\Z_4$ symmetry in subsection~\ref{Ss:PST6Z2Z6pmodel1}.

\subsection{Discrete $\Z_n$ symmetries on fractional D6-branes in toroidal orbifold models}\label{Ss:DiscreteOrbifolds}

The decomposition into a symplectic basis of $\OR$-even and $\OR$-odd three-cycles with intersection form given by equation~(\ref{Eq:even-odd-intersection}) is straightforward for untilted tori, but does not lead to the full unimodular 
lattice if some of the tori are tilted. Instead, the three-cycles $ \Pi^{\text{even}}_i$ and $ \Pi^{\text{odd}}_j$ only form an integral sublattice. We first briefly exemplify this in the case of the well-known 
$T^6/(\Z_2 \times \Z_2 \times \OR)$ orientifold without discrete torsion and with some tilted tori, see e.g.~\cite{Cvetic:2001tj,Cvetic:2001nr,Gmeiner:2005vz}. We then proceed to derive the conditions on the existence of $\Z_n$ 
symmetries for the orbifolds $T^6/(\Z_6 \times \OR)$ and $T^6/(\Z_6' \times \OR)$ with fractional D6-branes and for $T^6/(\Z_2 \times \Z_6' \times \OR)$ with discrete torsion with rigid D6-branes.

\subsubsection{The orbifold $T^6/(\Z_2 \times \Z_2 \times  \OR)$ without discrete torsion on tilted tori}\label{Sss:Z2Z2}

The unimodular lattice of three-cycles is spanned by eight half-bulk cycles, which can be expressed in terms of the basic one-cycles $\pi_{2i-1}$ and $\pi_{2i}$ per two-torus as follows,
\begin{equation}
\begin{aligned}
\Pi_{-1} &\equiv \frac{1}{2} \times \left( 4 \,\pi_1 \otimes \pi_3 \otimes \pi_5 \right),
\qquad\qquad 
\Pi_{0} =\frac{1}{2} \times \left( 4 \, \pi_2 \otimes \pi_4 \otimes \pi_6 \right),
\\
\Pi_{2i-1} &=\frac{1}{2} \times \left(4 \, \pi_{2i-1} \otimes \pi_{2j} \otimes \pi_{2k} \right),
\qquad
\Pi_{2i} =\frac{1}{2} \times \left( 4 \, \pi_{2i} \otimes \pi_{2j-1} \otimes \pi_{2k-1} \right),
\end{aligned}
\end{equation}
with $\Pi_{-1} \circ \Pi_0 = \Pi_{2i-1} \circ \Pi_{2i} =-1$ and $i\neq j \neq k \neq i \in \{1,2,3\}$. For untilted tori ($b_1=b_2=b_3=0$), the cycles $\Pi_{-1} , \Pi_{2i-1}$ are $\OR$-even and the cycles
$\Pi_{0} , \Pi_{2i}$ are $\OR$-odd.  If, however, a two-torus $T^2_{(i)}$ is tilted ($b_i=\frac{1}{2}$),  the $\OR$-even direction is given by $\tilde{\pi}_{2i-1}=\frac{\pi_{2i-1} - b_i \, \pi_{2i}}{1-b_i}$,
and the integral lattice of $\OR$-even and $\OR$-odd three cycles is spanned by
\begin{equation}\label{Eq:def_even+odd_Z2Z2}
\begin{aligned}
\Pi_0^{\text{even}} &= \frac{ \Pi_{-1} - \sum_{i=1}^3 \left( b_i \,\Pi_{2i}   - b_jb_k \,\Pi_{2i-1}   \right) - b_1b_2b_3 \, \Pi_{0b} }{\prod_{l=1}^3 (1-b_l)} ,
\qquad\qquad\qquad\quad
\Pi_0^{\text{odd}} =\Pi_{0},
\\
\Pi_i^{\text{even}} &=\frac{\Pi_{2i-1} - b_i \, \Pi_{0}}{1-b_i},
\qquad\qquad\qquad\quad
\Pi_i^{\text{odd}} = \frac{\Pi_{2i} - b_j \, \Pi_{2k-1}  - b_k \, \Pi_{2j-1}   + b_jb_k \, \Pi_{0}}{(1-b_j)(1-b_k)},
\end{aligned}
\end{equation}
with intersection form
\begin{equation}\label{Eq:Z2Z2-inters_even+odd}
\Pi_I^{\text{even}} \circ \Pi_J^{\text{odd}}= - \frac{\delta_{IJ}}{\prod_{l=1}^3 (1-b_l)}
\qquad
\text{ with }
I,J \in \{ 0 \ldots 3\}
.
\end{equation}
With the notation of bulk wrapping numbers $(X^I_a, Y^I_a)_{I \in \{0 \ldots 3\}}$ in terms of one-cycle wrapping numbers $(n^i_a,m^i_a)_{i\in \{1 \ldots 3\}}$~\cite{Blumenhagen:2004xx},
\begin{equation}
X^0_a = n^1_a n^2_a n^3_a,
\qquad
 X^i_a = n^i_a m^j_a m^k_a,
 \qquad
Y^0_a =m^1_a m^2_a m^3_a,
\qquad 
Y^i_a = m^i_a n^j_a n^k_a,
\end{equation}
the wrapping numbers $(A^I_a, B^I_a)_{I \in \{0 \ldots 3\}}$ in the expansion~(\ref{Eq:3-cycle_even+odd}) of $\OR$-even and $\OR$-odd cycles 
are multiples of $\frac{1}{8},\frac{1}{4},\frac{1}{2},1$ for three, two, one or no tilted two-torus, respectively. They
take the following form,
\begin{equation}\label{Eq:AB-wrappings_Z2Z2}
\begin{array}{lll}
A^0_a =& \left[\prod_{l=1}^3 (1-b_l)\right] \,  X^0_a 
,
\qquad \qquad\qquad\quad
B^0_a =  Y^0_a + \sum_{i=1}^3 \left(b_i \, X^i_a  + b_jb_k \,Y^i_a   \right) + b_1b_2b_3 \,  X^0_a 
,
\\
A^i_a =& (1-b_i) \, \left[ X^i_a +  b_j \,Y^k_a +  b_k \,Y^j_a + b_jb_k \,X^0_a \right]
,
\qquad\quad
B^i_a =  (1-b_j)(1-b_k) \, \left[ Y^i_a + b_i \, X^0_a \right]
.
\end{array}
\end{equation}
Due to intersection numbers~(\ref{Eq:Z2Z2-inters_even+odd}) being multiples of eight, four, two or one for three, two, one or no tilted two-torus, the conditions~(\ref{Eq:Zn-condition}) on the existence of a discrete $\Z_n$ symmetry
boil down to
\begin{equation}\label{Eq:discrete_Z2Z2}
- \sum_a k_a N_a \;   \frac{B^I_a}{\prod_{l=1}^3 (1-b_l)} \stackrel{!}{=} 0 \text{ mod } n
 \qquad 
 \forall I \in \{0 \ldots 3\}.
\end{equation}
This generalises the discussion in appendix~A of~\cite{BerasaluceGonzalez:2011wy}, which was valid for one $b_i \neq 0$ and only the values $I \in \{0,i\}$.
Contrary to a claim in~\cite{Anastasopoulos:2012zu}, the combination of fractional wrapping numbers~(\ref{Eq:AB-wrappings_Z2Z2}) and integer intersection numbers~(\ref{Eq:Z2Z2-inters_even+odd})
 in the condition for a discrete $\Z_n$ symmetry~(\ref{Eq:discrete_Z2Z2}) always takes values  $\frac{B^I_a}{\prod_{l=1}^3 (1-b_l)} \in \Z$.  The analogous statement is true for all other orbifolds investigated,
 and it underlines the statement in~\cite{BerasaluceGonzalez:2011wy} that for a $\Z_n$ symmetry only values $0 \leqslant k_a < n$ are physically distinct.
 
On  $T^6/(\Z_2 \times \Z_2 \times \OR)$ without discrete torsion, all $\Pi_I^{\text{even}}$ support $USp(2N_I)$ gauge groups~\cite{Berkooz:1996dw,Forste:2010gw,Cvetic:2001tj,Cvetic:2001nr}
and therefore the K-theory constraint implies the existence of a discrete $\Z_2$ symmetry in any global D6-brane model.

\subsubsection{The orbifold $T^6/(\Z_6 \times \OR)$ with AAB lattice}\label{Sss:Z6}

The first toroidal orbifold under investigation with fractional D6-branes is $T^6/(\Z_6 \times \OR)$ with shift vector $\vec{v}=\frac{1}{6}(1,1,-2)$, allowing fractional three-cycles of the form~\cite{Honecker:2004kb}
\begin{equation}\label{Eq:Z6-3-cycle-expansion}
\Pi_a = \frac{1}{2} \left( X_a \, \rho_1 + Y_a \, \rho_2 + \sum_{\alpha=1}^5 \left[x_{a,\alpha} \,\varepsilon_{\alpha} + y_{a,\alpha}  \,\tilde{\varepsilon}_{\alpha}  \right] \right)  ,
\end{equation}
with bulk three-cycles $\rho_{i}$ and exceptional three-cycles $\varepsilon_{\alpha}, \tilde{\varepsilon}_{\alpha} $ at $\Z_2 \subset \Z_6$ orbifold singularities
and with bulk and exceptional wrapping numbers $X_a,Y_a, x_{a,\alpha}, y_{a,\alpha} \in \Z$. The basic non-vanishing intersection numbers~\cite{Honecker:2004kb}
with corrected sign of the bulk three-cycle intersection~\cite{Forste:2010gw} are given by  
\begin{equation}
\rho_1 \circ \rho_2 = 2,
\qquad 
\varepsilon_{\alpha} \circ \tilde{\varepsilon}_{\beta} = -2 \, \delta_{\alpha\beta}
.
\end{equation}
The orientifold projection of three-cycles depends on the choice of lattice orientation and is given in table~\ref{Tab:OR-on-Z6} for the {\bf AAB} lattice.
\begin{SCtable}
$\begin{array}{|c|c||c||c|}\hline
\muc{4}{|c|}{\text{\bf Orientifold projection on $T^6/(\Z_6 \times \OR)$ on {\bf AAB}}}
\\\hline\hline
\OR(\rho_1) & \OR(\rho_2) & \OR(\varepsilon_{\alpha}) & \OR(\tilde{\varepsilon}_{\alpha})
\\\hline\hline
\rho_2 & \rho_1 & - \tilde{\varepsilon}_{\beta} & - \varepsilon_{\beta} 
\\
& & \muc{2}{|c|}{\alpha=\beta=1,2,3 ; \;  \alpha=4,5  \leftrightarrow \beta=5,4 } 
\\\hline
\end{array}
$
\caption{Orientifold projection on the bulk three-cycles $\rho_{i, i \in \{1,2\}}$ and exceptional three-cycles $\varepsilon_{\alpha,\alpha \in \{1 \ldots 5\}}$, $\tilde{\varepsilon}_{\alpha,\alpha \in \{1 \ldots 5\}}$
on $T^6/(\Z_6 \times \OR)$ with {\bf AAB} background orientation.}
\label{Tab:OR-on-Z6}
\end{SCtable}
From 
\begin{equation}\label{Eq:Z6-3-cycle-expansion}
\begin{aligned}
\frac{\Pi_a +\Pi_{a'} }{2}&= \frac{1}{4} \Biggl(\left[X_a + Y_a\right] \left(\rho_1 + \rho_2\right) 
+ \sum_{\targ2{\alpha=\beta \in \{1,2,3\} \text{ and}}{(\alpha,\beta) \in \{(4,5),(5,4)\}}}  
\left[x_{a,\alpha} - y_{a,\beta} \right] \left(\varepsilon_{\alpha} -\tilde{\varepsilon}_{\beta}  \right)
\Biggr)
,
\\
\frac{\Pi_a - \Pi_{a'}}{2} &=\frac{1}{4} \Biggl(\left[X_a - Y_a\right] \left(\rho_1 - \rho_2\right) 
+  \sum_{\targ2{\alpha=\beta \in \{1,2,3\} \text{ and}}{(\alpha,\beta) \in \{(4,5),(5,4)\}}}  
\left[x_{a,\alpha} + y_{a,\beta} \right] \left(\varepsilon_{\alpha} +\tilde{\varepsilon}_{\beta}  \right)
\Biggr)
,
\end{aligned}
\end{equation}
we find the decomposition into $\OR$-even and $\OR$-odd three-cycles, which originate purely from the bulk or $\Z_2$ twisted sector,
\begin{equation}\label{Eq:Z6_even+odd_fractional}
\begin{array}{cc}
\Pi^{\text{even}}_0 =\rho_1 + \rho_2
,
&
\Pi^{\text{odd}}_0 = \rho_1 - \rho_2
,
\\
\Pi^{\text{even}}_{\alpha} = \left\{\begin{array}{c} 
\varepsilon_{\alpha} - \tilde{\varepsilon}_{\alpha}  
\\ \varepsilon_4 - \tilde{\varepsilon}_5 
\\ \varepsilon_5 - \tilde{\varepsilon}_4 
\end{array}\right. 
,
\qquad
&
\Pi^{\text{odd}}_{\alpha} = \left\{\begin{array}{cr}  
\varepsilon_{\alpha} + \tilde{\varepsilon}_{\alpha} & \alpha=1,2,3 \\ 
\varepsilon_5 + \tilde{\varepsilon}_4 & 4 \\ \varepsilon_4 + \tilde{\varepsilon}_5 & 5 \end{array}\right. 
,
\end{array}
\end{equation}
with intersection numbers 
\begin{equation}
\Pi^{\text{even}}_{\tilde{\alpha}} \circ \Pi^{\text{odd}}_{\tilde{\beta}} = - 4 \, \delta_{\tilde{\alpha}\tilde{\beta}}
\qquad \text{ with } \quad 
\tilde{\alpha}, \tilde{\beta} \in \{0 \ldots 5\}
.
\end{equation}
These three-cycles only form a sublattice of the full $\OR$-even plus $\OR$-odd lattice, and therefore
they provide the six {\it necessary} but not {\it sufficient} conditions on the existence of a discrete $\Z_n$ symmetry
on the first six lines in equation~(\ref{Eq:Z6_discrete_nec+suff}),
\begin{equation}\label{Eq:Z6_discrete_nec+suff}
{\footnotesize
\sum_a k_a N_a \left(\begin{array}{c} 
-(X_a-Y_a) \\ \hline-(x_{a,1} + y_{a,1}) \\ -(x_{a,2} + y_{a,2})\\ -(x_{a,3} + y_{a,3}) \\ -(x_{a,5} + y_{a,4})  \\ -(x_{a,4} + y_{a,5}) \\\hline\hline - \frac{(X_a - Y_a) + (x_{a,1} + y_{a,1}) +  (x_{a,2} + y_{a,2}) + (x_{a,3} + y_{a,3})}{2}
\\ - \frac{(X_a - Y_a) + (x_{a,1} + y_{a,1}) + (x_{a,5} + y_{a,4})}{2} \\ - \frac{(X_a - Y_a) + (x_{a,3} + y_{a,3}) + (x_{a,4} + y_{a,5})}{2}
\end{array}\right) \stackrel{!}{=}0 \text{ mod }n}
.
\end{equation}
The six {\it necessary} conditions obviously have integer entries since all bulk wrapping numbers $(X_a, Y_a)$ and exceptional wrapping numbers $(x_{a,\alpha}, y_{a,\alpha})$ are integers.

The three {\it sufficient} conditions on the last three lines of equation~(\ref{Eq:Z6_discrete_nec+suff}) also have integer entries within the parenthesis. This can be seen as follows.
The shortest possible $\OR$-even three-cycles are those fractional three-cycles, on which the gauge group is enhanced from a single $U(N_a)$ to a $USp(2N_a)$  or $SO(2N_a)$ factor.
These three-cycles are either parallel to one of the O6-plane orbits $\OR$ or $\OR\Z_2^{(3)}$ with $\Z_2^{(3)} \equiv \Z_2$ or perpendicular to them along a four-torus. Following the notation for $T^6/(\Z_2 \times \Z_{2M} \times \OR)$ 
orbifolds with discrete torsion, the latter planes are denoted $\OR\Z_2^{(1)}$ and $\OR\Z_2^{(2)}$ even though the $\Z_2^{(1)}$ and $\Z_2^{(2)}$ symmetries are not present in the $T^6/(\Z_6 \times \OR)$
orbifold. The conditions on the existence of a gauge enhancement can be obtained by truncating the conditions in table~\ref{Tab:Z2Z2M_Conditions-on_b+t+s-SOSp}
to only those on $\eta_{(3)}\equiv +1$ with $b_1=b_2=\frac{1}{2}$. 
\begin{table}[h!]
\renewcommand{\arraystretch}{1.3}
  \begin{center}
\begin{equation*}
\begin{array}{|c|c|}\hline
\multicolumn{2}{|c|}{\text{\bf Existence of $\OR$ invariant three-cycles on } T^6/(\Z_2 \times \Z_{2M} \times \OR) \text{ \bf  with } \eta=-1}
\\\hline\hline
\pp \text{ to O6-plane}
& (\eta_{(1)},\eta_{(2)},\eta_{(3)}) \stackrel{!}{=}
\\\hline\hline
\OR  & \left( -(-1)^{2(b_2\sigma^2\tau^2 + b_3\sigma^3 \tau^3)} , - (-1)^{2(b_1\sigma^1\tau^1 + b_3\sigma^3 \tau^3)} ,  -(-1)^{2(b_1\sigma^1\tau^1 + b_2\sigma^2 \tau^2)} \right)
\\
\OR\Z_2^{(1)}  & \left( -(-1)^{2(b_2\sigma^2\tau^2 + b_3\sigma^3 \tau^3)} ,  (-1)^{2(b_1\sigma^1\tau^1 + b_3\sigma^3 \tau^3)} ,  (-1)^{2(b_1\sigma^1\tau^1 + b_2\sigma^2 \tau^2)} 
\right)
\\
\OR\Z_2^{(2)}  &  \left( (-1)^{2(b_2\sigma^2\tau^2 + b_3\sigma^3 \tau^3)} , - (-1)^{2(b_1\sigma^1\tau^1 + b_3\sigma^3 \tau^3)} ,  (-1)^{2(b_1\sigma^1\tau^1 + b_2\sigma^2 \tau^2)}
\right)
\\
\OR\Z_2^{(3)}  &  \left( (-1)^{2(b_2\sigma^2\tau^2 + b_3\sigma^3 \tau^3)} ,  (-1)^{2(b_1\sigma^1\tau^1 + b_3\sigma^3 \tau^3)} ,  -(-1)^{2(b_1\sigma^1\tau^1 + b_2\sigma^2 \tau^2)}
\right)
\\\hline
\end{array}
\end{equation*}
\end{center}
\caption{Conditions on the existence of $\OR$ invariant fractional three-cycles on $T^6/(\Z_2 \times \Z_{2M} \times \OR)$ with
discrete torsion ($\eta=-1$) for $2M \in \{2,6,6'\}$ as derived in~\cite{Forste:2010gw}. The conditions on the existence of $\OR$-invariant three-cycles on $T^6/\Z_{2N}$ 
with $\Z_2^{(k)} \subset \Z_{2N} \subset \Z_2 \times \Z_{2M}$ are obtained by only considering the condition on $\eta_{(k)}\equiv +1$ and truncating
the other two $\eta_{(i \neq k)}$. $b_i =0,\frac{1}{2}$ parameterise the untilted or tilted shape per two-torus $T^2_{(i)}$. $\sigma^i \in \{0,1\}$ denote the displacements 
of a given three-cycle from the origin of $T^2_{(i)}$, and $\tau^i \in \{0,1\}$ parameterise the choice of discrete Wilson line.
}
\label{Tab:Z2Z2M_Conditions-on_b+t+s-SOSp}
\end{table}

The three-cycles parallel to some O6-plane orbit take the form on the {\bf AAB} lattice
\begin{eqnarray}
&\Pi^{\text{frac},\OR/\OR\Z_2^{(3)}}_{(\sigma^1,\sigma^2)=(1,0)}  \; \stackrel{\tau^1=1}{=} \; &  \frac{1|3}{2}\Pi^{\text{even}}_0 + (-1)^{\tau^{\Z_2}} \times \frac{
 -3 \, \Pi^{\text{even}}_1 -(-1)^{\tau^2} \times \left[\Pi^{\text{even}}_4  + 2 \; \Pi^{\text{even}}_5 \right]}{2}
 , \nonumber
\\
&\Pi^{\text{frac},\OR/\OR\Z_2^{(3)}}_{(\sigma^1,\sigma^2)=(0,1)} \; \stackrel{\tau^2=1}{=} \; & \frac{1|3}{2}\Pi^{\text{even}}_0 +(-1)^{\tau^{\Z_2}} \times \frac{
 -3 \, \Pi^{\text{even}}_2 + (-1)^{\tau^1} \times \left[\Pi^{\text{even}}_4  - \Pi^{\text{even}}_5 \right] }{2}
, \nonumber
\\
&\Pi^{\text{frac},\OR/\OR\Z_2^{(3)}}_{(\sigma^1,\sigma^2)=(1,1)} \; \stackrel{\tau^1 \neq \tau^2}{=} \; &  \frac{1|3}{2}\Pi^{\text{even}}_0 + (-1)^{\tau^{\Z_2}} \times \frac{
 -3 \, \Pi^{\text{even}}_3  + (-1)^{\tau^1} \times \left[2 \, \Pi^{\text{even}}_4  + \Pi^{\text{even}}_5 \right] }{2}
, \nonumber\\
\label{Eq:Z6_USp+SO_Part1}
\end{eqnarray}
where for the sake of shortness we introduced the notation $\frac{1|3}{2}\Pi^{\text{even}}_0$ as bulk contribution $\frac{\Pi^{\text{even}}_0}{2}$ for D6-branes parallel to the $\OR$-invariant orbit and 
$\frac{3 \; \Pi^{\text{even}}_0}{2}$ parallel to the $\OR\Z_2^{(3)}$-invariant orbit.
$\sigma^i, \tau^i \in \{0,1\}$ denote discrete displacements $(\sigma^i)$ from the origin and discrete Wilson lines ($\tau^i$) on $T^2_{(i)}$, and $\tau^{\Z_2}\in \{0,1\}$ parameterises the two allowed choices of $\Z_2$ 
eigenvalue per three-cycle.

The orientifold invariant three-cycles perpendicular to the O6-planes in the $T^6/(\Z_6 \times \OR)$ background can be concisely summarised as 
\begin{eqnarray}
&\Pi^{\text{frac},\OR\Z_2^{(1)}/\OR\Z_2^{(2)}}_{(\sigma^1,\sigma^2)=(0,0)} =& \frac{\Pi^{\text{even}}_0}{2} + (-1)^{\tau^{\Z_2}} \times 
\frac{(-1)^{\tau^1} \; \Pi^{\text{even}}_1 +(-1)^{\tau^2} \; \Pi^{\text{even}}_2 +(-1)^{\tau^1+\tau^2} \; \Pi^{\text{even}}_3  }{2}
, \nonumber
\\
&\Pi^{\text{frac},\OR\Z_2^{(1)}/\OR\Z_2^{(2)}}_{(\sigma^1,\sigma^2)=(1,0)} \; \stackrel{\tau^1=0}{=} \; &\frac{\Pi^{\text{even}}_0}{2}  +(-1)^{\tau^{\Z_2}} \times \frac{  - \Pi^{\text{even}}_1 -(-1)^{\tau^2}  \; \Pi^{\text{even}}_4 }{2}
,\label{Eq:Z6_USp+SO_Part2}
\\
&\Pi^{\text{frac},\OR\Z_2^{(1)}/\OR\Z_2^{(2)}}_{(\sigma^1,\sigma^2)=(0,1)}  \; \stackrel{\tau^2=0}{=} \; & \frac{\Pi^{\text{even}}_0}{2} + (-1)^{\tau^{\Z_2}} \times \frac{ - \Pi^{\text{even}}_2 +(-1)^{\tau^1} \times \left[\Pi^{\text{even}}_4  + \Pi^{\text{even}}_5 \right]
}{2}
, \nonumber
\\
&\Pi^{\text{frac},\OR\Z_2^{(1)}/\OR\Z_2^{(2)}}_{(\sigma^1,\sigma^2)=(1,1)} \; \stackrel{\tau^1=\tau^2=\tau}{=}\; &  \frac{\Pi^{\text{even}}_0}{2}  +(-1)^{\tau^{\Z_2}} \times \frac{ - \Pi^{\text{even}}_3 -(-1)^{\tau} \; \Pi^{\text{even}}_5 }{2}
. \nonumber 
\end{eqnarray}

A careful inspection shows that upon adding integer multiples of the basic $\Pi^{\text{even}}_{\tilde{\alpha}}$ defined in~(\ref{Eq:Z6_even+odd_fractional}), only three of the 64 cycles in equations~(\ref{Eq:Z6_USp+SO_Part1}) and~(\ref{Eq:Z6_USp+SO_Part2})
are linearly independent. These three cycles provide the {\it sufficient} conditions on the existence of a $\Z_n$ symmetry in the last three lines of equation~(\ref{Eq:Z6_discrete_nec+suff}).
For example, the entry in the parenthesis for the first {\it sufficient} condition is the integer valued intersection number with the shortest $\OR$-even fractional three-cycle 
$\Pi^{\text{frac},\OR\Z_2^{(1)}/\OR\Z_2^{(2)}}_{(\sigma^1,\sigma^2)=(0,0)}$ with the choice of discrete parameters $\tau^{\Z_2}=\tau^1=\tau^2=0$.

Since the lattice of $\OR$-even three-cycles on $T^6/(\Z_6 \times \OR)$ is only $h_{21}+1=6$ dimensional, not all combinations of {\it necessary} and {\it sufficient} conditions in equation~(\ref{Eq:Z6_discrete_nec+suff})
are independent. However, for practical purposes it is useful to first compute the simpler {\it necessary} conditions and verify the {\it sufficient} conditions afterwards.

In contrast to the $T^6/(\Z_2 \times \Z_2 \times \OR)$ orbifold without discrete torsion in section~\ref{Sss:Z2Z2}, the K-theory constraint on $T^6/(\Z_6 \times \OR)$ does not automatically imply a $\Z_2$ symmetry since not all orientifold 
invariant three-cycles support $USp(2N_a)$ gauge factors. The gauge group and matter content in the antisymmetric $(\Anti_a)$ and symmetric $(\Sym_a)$ representation depends on the choice 
of discrete Wilson lines $(\tau^1,\tau^2)$ and discrete displacements $(\sigma^1,\sigma^2)$ along \mbox{$T^2_{(1)} \times T^2_{(2)}$} as follows (but is independent of the choice of $\Z_2$ eigenvalues $(-1)^{\tau^{\Z_2}}$):
\begin{itemize}
\item
\framebox{All D6-branes $\pp$ $\OR$: $USp(2N_a)$} with one chiral multiplet in the $(\Sym_a)\equiv (\Adj_a)$ representation from the $aa'$ sector and no further matter states at the intersections $(\omega^k a)(\omega^k a)_{k=1,2}^{\prime}$ of orbifold images,
\item
\framebox{All D6-branes $\pp$ $\OR\Z_2^{(3)}$: $USp(2N_a)$} with one chiral multiplet in the $(\Sym_a) $ representation from the $aa'$ sector plus $12 \times (\Sym_a)$ from the $(\omega^k a)(\omega^k a)_{k=1,2}^{\prime}$
sector,
\item
\framebox{All  D6-branes $\pp$ $\OR\Z_2^{(1)}$, $\OR\Z_2^{(2)}$   with $(\sigma^i,\tau^i)_{i=1,2}\neq (1,1;1,1)$: $USp(2N_a)$} with one chiral multiplet in the $(\Anti_a)$ representation from the $aa'$ sector plus $2 \times (\Anti_a)$ from the  $(\omega^k a)(\omega^k a)_{k=1,2}^{\prime}$ sector,
\item
\framebox{D6-branes $\pp$ $\OR\Z_2^{(1)}$, $\OR\Z_2^{(2)}$ with $(\sigma^i,\tau^i)_{i=1,2}=(1,1;1,1)$: $SO(2N_a)$} with one chiral multiplet in the $(\Sym_a)$ representation from the $aa'$ sector plus $ 2 \times (\Anti_a)$  from the  $(\omega^k a)(\omega^k a)_{k=1,2}^{\prime}$ sector.
\end{itemize}
The last case deviates from the classification performed in table~15 of~\cite{Honecker:2011sm} due to the sublety of the sign $(-1)^{2 b_i \sigma^i\tau^i}$ in the M\"obius strip contribution to the beta function coefficient  for a D6-brane parallel to its own orientifold image along a tilted $T^2_{(i)}$ as argued in appendix~B.1 of~\cite{Honecker:2012qr}.

While the existence of some gauge enhancement to $SO(2N_a)$ annuls the argument that the K-theory constraint automatically implies a discrete $\Z_2$ symmetry, it was shown in~\cite{Gmeiner:2007we}
that the probe-brane condition is for global supersymmetric models on $T^6/(\Z_6 \times \OR)$ automatically fulfilled - including a probe brane of $SO(2N_a)$ type. Global supersymmetric models on $T^6/(\Z_6 \times \OR)$ thus possess at least one 
discrete $\Z_2$ symmetry.

\subsubsection{The $T^6/(\Z_6' \times \OR)$ with ABa lattice}\label{Sss:Z6p}

A second class of toroidal orientifolds providing consistent global models with fractional D6-branes consists of $T^6/(\Z_6' \times \OR)$
with $\vec{v}^{\; \prime}=\frac{1}{6}(1,2,-3)$~\cite{Gmeiner:2007zz,Gmeiner:2008xq}, for which a generic fractional three-cycle 
can be written as follows with bulk and exceptional wrapping numbers $P_a,Q_a,U_a,V_a,d_{a,\alpha}, e_{a,\alpha} \in \Z$,
\begin{equation}\label{Eq:Z6p-even+odd_expansion}
\Pi_a = \frac{1}{2} \left(P_a \, \rho_1 + Q_a \, \rho_2 + U_a \, \rho_3 + V_a \, \rho_4 
+ \sum_{\alpha=1}^4 \left\{d_{a,\alpha} \,\delta_{\alpha}  + e_{a,\alpha} \, \tilde{\delta}_{\alpha}  \right\} \right).
\end{equation}
The orientifold projection on bulk and exceptional three-cycles $\rho_i$ and $\delta_{\alpha}, \tilde{\delta}_{\alpha}$  displayed in table~\ref{Tab:OR-on-Z6p}
\begin{SCtable}
$\begin{array}{|c|c|c|c||c||c|}\hline
\muc{6}{|c|}{\text{\bf Orientifold projection on $T^6/(\Z_6' \times \OR)$ on {\bf ABa}}}
\\\hline\hline
\OR(\rho_1) & \OR(\rho_2) & \OR(\rho_3) & \OR(\rho_4) 
& \OR(\delta_{\alpha}) & \OR(\tilde{\delta}_{\alpha})
\\\hline\hline
\rho_2 & \rho_1 & -\rho_4 & -\rho_3 & - \tilde{\delta}_{\alpha} & -\delta_{\alpha} 
\\\hline
\end{array}
$
\caption{Orientifold projection on bulk three-cycles $\rho_{i, i \in \{1 \ldots 4\}}$ and exceptional three-cycles $\delta_{\alpha,\alpha \in \{1 \ldots 4\}}$, $\tilde{\delta}_{\alpha,\alpha \in \{1 \ldots 4\}}$ 
on the {\bf ABa} lattice of $T^6/(\Z_6' \times \OR)$.}
\label{Tab:OR-on-Z6p}
\end{SCtable}
leads to the following $\OR$-even and $\OR$-odd expansions on the {\bf ABa} lattice,
\begin{equation}\label{Eq:Z6p-even+odd_expansion}
\begin{aligned}
\frac{\Pi_a +\Pi_{a'}}{2} &=\frac{1}{4} \left( \left[P_a + Q_a\right] \left(\rho_1 + \rho_2 \right)  
+ \left[U_a - V_a\right] \left(\rho_3-\rho_4  \right) 
+ \sum_{\alpha=1}^4 \left[d_{a,\alpha} - e_{a,\alpha}\right] \left(\delta_{\alpha}  - \tilde{\delta}_{\alpha}\right)
\right)
,
\\
\frac{\Pi_a - \Pi_{a'}}{2} &=\frac{1}{4} \left( \left[P_a - Q_a\right] \left(\rho_1 - \rho_2 \right)  
+ \left[U_a + V_a\right] \left(\rho_3 + \rho_4  \right) 
+ \sum_{\alpha=1}^4\left[d_{a,\alpha} + e_{a,\alpha}\right] \left(\delta_{\alpha}  + \tilde{\delta}_{\alpha}\right)
\right)
,
\end{aligned}
\end{equation}
and thereby to the ansatz for an $\OR$-even and $\OR$-odd basis of three-cycles,
\begin{equation}\label{Eq:Z6p-even+odd_cycles}
\begin{array}{ll}
\Pi^{\text{even}}_1 = \rho_1 + \rho_2
, 
\qquad\qquad\quad
&
\Pi^{\text{odd}}_1 =\rho_3 + \rho_4
,
\\
\Pi^{\text{even}}_2 =\rho_3-\rho_4
,
&
\Pi^{\text{odd}}_2 =\rho_1 - \rho_2
,
\\
\Pi^{\text{even}}_{2+\alpha} =\delta_{\alpha}  - \tilde{\delta}_{\alpha}
,
&
\Pi^{\text{odd}}_{2+\alpha} =\delta_{\alpha}  + \tilde{\delta}_{\alpha},
\qquad \text{ with } \quad
\alpha=1 \ldots 4
.
\end{array}
\end{equation}
Using the basic non-vanishing intersection numbers~\cite{Gmeiner:2007zz}
\begin{equation}
\rho_1 \circ \rho_3 = \rho_2 \circ \rho_4 = 4
,
\qquad\quad
\rho_2 \circ \rho_3 = \rho_1 \circ \rho_4 = 2
,
\qquad\quad
\delta_{\alpha} \circ \tilde{\delta}_{\beta} = -2 \, \delta_{\alpha\beta}
,
\end{equation}
the ansatz~(\ref{Eq:Z6p-even+odd_cycles}) for a symplectic $\OR$-even and $\OR$-odd basis of three-cycles does not have an unimodular intersection form,
\begin{equation}
\Pi^{\text{even}}_{\tilde{\alpha}}\circ \Pi^{\text{odd}}_{\tilde{\beta}} = \delta_{\tilde{\alpha}\tilde{\beta}}\times \left\{\begin{array}{cl}
12 & \tilde{\alpha}=1 \\ -4 &  \tilde{\alpha}= 2 \ldots 6 \end{array}\right.
.
\end{equation}

In complete analogy to the $T^6/(\Z_6 \times \OR)$ background in section~\ref{Sss:Z6}, the conditions on the existence of a discrete $\Z_n$ symmetry can be decomposed into
2+4 {\it necessary} ones on the first lines,
\begin{equation}\label{Eq:Z6p_Condition_ZM}
{\footnotesize
\sum_a k_a \; N_a
\left(\begin{array}{c} 3 \, (U_a + V_a) \\-(P_a-Q_a)\\\hline  - (d_{a,1} + e_{a,1}) \\ - (d_{a,2} + e_{a,2}) 
\\  - (d_{a,3} + e_{a,3}) \\ - (d_{a,3} + e_{a,3})
\\\hline\hline \frac{3 \, (U_a + V_a) - (d_{a,1} + e_{a,1})-(d_{a,2} + e_{a,2})}{2}
\\ \frac{3 \, (U_a + V_a)-(d_{a,3} + e_{a,3})-(d_{a,4} + e_{a,4})}{2}
\\\frac{-(P_a-Q_a)-(d_{a,1} + e_{a,1})-(d_{a,4} + e_{a,4})}{2}
\\\frac{-(P_a-Q_a)-(d_{a,2} + e_{a,2})-(d_{a,3} + e_{a,3})}{2}
\end{array}\right) \stackrel{!}{=} 0 \text{ mod } n }
,
\end{equation}
and four {\it sufficient} ones from the shortest possible orientifold invariant three-cycles.
On $T^6/(\Z_6' \times \OR)$, table~\ref{Tab:Z2Z2M_Conditions-on_b+t+s-SOSp} can be truncated to the constraints on $\eta_{(2)} \equiv +1$ with $b_3=0$ for the untilted {\bf a}-type lattice and $b_1=\frac{1}{2}$ for 
the {\bf A}-type lattice. The fractional three-cycles that satisfy this topological condition and give rise to a gauge group enhancement can be written down schematically as 
\begin{equation}\label{Eq:Z6p_all_OR-in_cycles}
\Pi^{\text{frac},\OR}_{(\sigma^1, \tau^1)=(1,1)}
,
\qquad
\Pi^{\text{frac},\OR\Z_2^{(2)}}_{(\sigma^1,\tau^1)=(1,1)}
,
\qquad
\Pi^{\text{frac},\OR\Z_2^{(1)}}_{(\sigma^1,\tau^1)\in \{(0,0),(1,0),(0,1)\}} 
,
\qquad
\Pi^{\text{frac},\OR\Z_2^{(3)}}_{(\sigma^1,\tau^1)\in \{(0,0),(1,0),(0,1)\}} 
,
\end{equation}
with arbitrary values for $(\sigma^3,\tau^3)$ and arbitrary $\Z_2$ eigenvalues $(-1)^{\tau^{\Z_2}}$. Out of the initially 64 orientifold invariant three-cycles, only four are linearly independent if adding multiple integers of the basic $\Pi^{\text{even}}_{\tilde{\alpha}}$ defined in~(\ref{Eq:Z6p-even+odd_cycles}) are allowed.
These provide the four {\it sufficient} conditions in the lower lines of equation~(\ref{Eq:Z6p_Condition_ZM}). 
Again, since $h_{21}+1=6$, some of the {\it necessary} conditions are redundant, but for practical purposes it is useful to first compute these most simple constraints
and impose the {\it sufficient} conditions afterwards.

On the {\bf ABa} lattice, there are no orientifold invariant three-cycles supporting $SO(2N_a)$ gauge groups, suggesting that the K-theory constraint imply the automatic presence of a $\Z_2$ symmetry for this background.\footnote{Note that gauge group enhancement to an  $SO(2N_a)$ gauge factor on $T^6/(\Z_6' \times \OR)$ occurs only when the third two-torus is also tilted, i.e.~a {\bf b}-type lattice for $T^2_{(3)}$. 
In this configuration the factor $(-1)^{2b_3 \sigma_a^3 \tau_a^3}$ in the M\"obius-amplitude contribution to the beta-function coefficient~\cite{Honecker:2012qr} will no longer be idle, but depending on the values of the discrete Wilson lines $(\tau_a^1, \tau_a^3)$ and discrete displacements $(\sigma_a^1,\sigma_a^3)$ the gauge group can enhance to a $USp(2N_a)$ or a $SO(2N_a)$ factor, similar to the classification in section~\ref{Sss:Z6}.}~Indeed, a full classification of orientifold invariant three-cycles on the {\bf ABa} lattice shows that all of them give rise to a $U(N_a) \hookrightarrow USp(2N_a)$ enhancement in agreement with~\cite{Honecker:2011sm}, while the amount of matter in the antisymmetric ($\Anti_a$) or symmetric ($\Sym_a$) representation alters depending on which (pesudo) O6-plane the D6-brane $a$ is parallel to:
\begin{itemize}
\item
\framebox{All D6-branes  $\pp$ $\OR$: $USp(2N_a)$} with one chiral multiplet in the $(\Sym_a)\equiv (\Adj_a)$ representation from the $aa'$ sector and $3 \times (\Anti_a)$ at the intersections $(\omega^k a)(\omega^k a)_{k=1,2}^{\prime}$ of orbifold images, 
\item
\framebox{All D6-branes  $\pp$ $\OR\Z_2^{(3)}$: $USp(2N_a)$} with one chiral multiplet in the $(\Sym_a) $ representation from the $aa'$ sector plus $9 \times (\Anti_a)$ from the $(\omega^k a)(\omega^k a)_{k=1,2}^{\prime}$
sector,
\item
\framebox{All D6-branes  $\pp$ $\OR\Z_2^{(1)}$: $USp(2N_a)$} with one chiral multiplet in the $(\Anti_a)$ representation from the $aa'$ sector plus $2  \times (\Anti_a)$ and $1 \times (\Sym_a)$ from the  $(\omega^k a)(\omega^k a)_{k=1,2}^{\prime}$ sector.
\item
\framebox{All D6-branes  $\pp$ $\OR\Z_2^{(1)}$: $USp(2N_a)$} with one chiral multiplet in the $(\Anti_a)$ representation from the $aa'$ sector plus $1  \times (\Anti_a)$  from the  $(\omega^k a)(\omega^k a)_{k=1,2}^{\prime}$ sector.
\end{itemize}
In~\cite{Gmeiner:2007zz} it was shown, that for {\it all} choices of background lattices of $T^6(\Z_6' \times \OR)$ the combination of RR tadpole cancellation and supersymmetric fractional D6-branes 
renders the K-theory constraint~(\ref{Eq:K-theory}) for all `probe' D6-branes trivial - including those `probe' branes with $SO(2N_a)$ gauge symmetry. As a consequence, also the `diagonal' $\Z_2$
symmetry exists for any supersymmetric global D6-brane model on $T^6(\Z_6' \times \OR)$.

\subsubsection{The orbifold $T^6/(\Z_2 \times \Z_6' \times \OR)$ with discrete torsion and AAA lattice}\label{Sss:Z2Z6p}

The last toroidal orbifold discussed in this paper is $T^6/(\Z_2 \times \Z_6' \times \OR)$ with discrete torsion ($\eta=-1$) with shift vectors $\vec{v} = \frac{1}{2}(1,-1,0)$ and $\vec{w} = \frac{1}{6}(-2,1,1)$ generating the $\Z_2$ and $\Z_6'$ orbifold action. On this background, rigid three-cycles are constructed~\cite{Forste:2010gw} as the superposition of one-quarter of some bulk three-cycle with one-quarter of some exceptional three-cycles from each of the three $\Z_2^{(k)}$ twisted sectors,
\begin{equation}
\Pi_a = \frac{1}{4} \left( X_a \, \rho_1 + Y_a \, \rho_2 
+ \sum_{k=1}^3 \sum_{\alpha=1}^5 \left[x_{a,\alpha}^{(k)} \,\varepsilon_{\alpha}^{(k)} + y_{a,\alpha}^{(k)} \,\tilde{\varepsilon}_{\alpha}^{(k)}  \right]\right),
\end{equation}
where the $\rho_i$ correspond to the orbifold invariant bulk three-cycles, and $(\varepsilon_{\alpha}^{(k)}, \tilde\varepsilon_{\alpha}^{(k)})$ form the exceptional three-cycle basis at the $\Z_2^{(k)}$ orbifold singularities, with the non-vanishing intersection numbers:
\begin{equation}
\rho_1 \circ \rho_2 = 4. 
\hspace{0.4in} 
\varepsilon_{\alpha}^{(k)} \circ \tilde{\varepsilon}_{\beta}^{(l)} = -4 \, \delta_{\alpha\beta} \, \delta^{kl},
\end{equation}
\begin{SCtable}
$\begin{array}{|c|c||c||c|}\hline
\muc{4}{|c|}{\text{\bf Orientifold projection on $T^6/(\Z_2 \times \Z_6' \times \OR)$ on {\bf AAA}}}
\\\hline\hline
\OR(\rho_1) & \OR(\rho_2) & \OR(\varepsilon_{\alpha}^{(k)}) & \OR(\tilde{\varepsilon}_{\alpha}^{(k)}) 
\\\hline\hline
\rho_1 & \rho_1 - \rho_2 & -\eta_{(k)} \, \varepsilon_{\beta}^{(k)} &  \eta_{(k)} \, \left(\tilde{\varepsilon}_{\beta}^{(k)} - \varepsilon_{\beta}^{(k)}\right) 
\\
& & \muc{2}{|c|}{\alpha=\beta=1,2,3 ; \; \alpha=4,5 \leftrightarrow \beta=5,4 
}
\\\hline
\end{array}$
\caption{Orientifold projection of bulk three-cycles $\rho_{i, i \in \{1,2\}}$ and exceptional three-cycles $\varepsilon_{\alpha,\alpha \in \{1 \ldots 5\}}$, $\tilde{\varepsilon}_{\alpha,\alpha \in \{1 \ldots 5\}}$
on $T^6/(\Z_2 \times \Z_6' \times \OR)$ with $\eta=-1$ and  {\bf AAA}  orientation.}
\label{Tab:OR-on-Z2Z6p}
\end{SCtable}
Restricting to the {\bf AAA} lattice, the orientifold projection acts on the three-cycles as given in table~\ref{Tab:OR-on-Z2Z6p} and allows for the following decomposition in terms of $\OR$-even and 
$\OR$-odd three-cycles,
\begin{equation}\label{Eq:Z2Z6p_even+odd_expansion}\hspace{-10mm}
\begin{aligned}
\frac{\Pi_a +\Pi_{a'}}{2} &=\frac{1}{8} \left( \!\! \left[2 \, X_a + Y_a \right]  \rho_1  
+ \sum_{k=1}^3 \!\! \sum_{\targ2{\alpha=\beta \in \{1,2,3\} \text{ and}}{(\alpha,\beta) \in \{(4,5),(5,4)\}}} \!\!\!\!\!\!\!\!\!\!\!\!
\left\{\left( x_{a,\alpha}^{(k)} - \eta_{(k)} \left[ x_{a,\beta}^{(k)} +y_{a,\beta}^{(k)} \right]\right)\varepsilon_{\alpha}^{(k)} + \left( y_{a,\alpha}^{(k)} + \eta_{(k)} \, y_{a,\beta}^{(k)} \right)  \tilde{\varepsilon}_{\alpha}^{(k)} \right\} \!\!
\right)
,
\\
\frac{\Pi_a - \Pi_{a'}}{2} &=\frac{1}{8} \left( \!\! Y_a  \left(-\rho_1 + 2 \, \rho_2 \right)
+ \sum_{k=1}^3 \!\! \sum_{\targ2{\alpha=\beta \in \{1,2,3\} \text{ and}}{(\alpha,\beta) \in \{(4,5),(5,4)\}}} \!\!\!\!\!\!\!\!\!\!\!\!
\left\{\left( x_{a,\alpha}^{(k)} +\eta_{(k)} \left[ x_{a,\beta}^{(k)} +y_{a,\beta}^{(k)} \right]\right)\varepsilon_{\alpha}^{(k)}   + \left( y_{a,\alpha}^{(k)} - \eta_{(k)} \, y_{a,\beta}^{(k)} \right)  \tilde{\varepsilon}_{\alpha}^{(k)} \right\} \!\!
\right),
\end{aligned}
\end{equation}
where $\eta_{(k)} \equiv \eta_{\OR} \eta_{\OR\Z_2^{(k)}} = \pm 1$ with $\prod_{k=1}^3 \eta_{(k)} = \eta=-1$ depends on the choice of exotic O6-plane.
Focusing on configurations with the $\OR$-plane as the exotic O6-plane (i.e.~$\eta_{\OR}=\eta_{(k), k \in \{1,2,3\}}=-1$), the ansatz for the $\OR$-even and $\OR$-odd three-cycles reads
\begin{equation}\label{Eq:Z2Z6p_even+odd_basis_eta-1}
\begin{array}{l@{\hspace{0.4in}}l}
\Pi^{\text{even},\unity}_0 =\rho_1,
& \Pi^{\text{odd},\unity}_0 =-\rho_1 + 2 \, \rho_2 ,
\\
\Pi^{\text{even},\Z_2^{(k)}}_{\alpha \in \{1,2,3\}} = \varepsilon_{\alpha}^{(k)},
&\Pi^{\text{odd},\Z_2^{(k)}}_{\alpha \in \{1,2,3\}} = - \varepsilon_{\alpha}^{(k)}   + 2 \, \tilde{\varepsilon}_{\alpha}^{(k)},
\\
\Pi^{\text{even},\Z_2^{(k)}}_{4} = \varepsilon_4^{(k)} +  \varepsilon_5^{(k)},
&\Pi^{\text{odd},\Z_2^{(k)}}_{4} = 2 \, (\tilde{\varepsilon}_4^{(k)} + \tilde{\varepsilon}_5^{(k)})  - ( \varepsilon_4^{(k)} +  \varepsilon_5^{(k)}),
\\
\Pi^{\text{even},\Z_2^{(k)}}_{5} = 2 \, (\tilde{\varepsilon}_4^{(k)} - \tilde{\varepsilon}_5^{(k)})  - (\varepsilon_4^{(k)} -  \varepsilon_5^{(k)}),
&\Pi^{\text{odd},\Z_2^{(k)}}_{5} = \varepsilon_4^{(k)} -  \varepsilon_5^{(k)},
\end{array}
\end{equation}
for which the intersection numbers clearly do not form a unimodular basis,
\begin{equation}
\Pi^{\text{even},\Z_2^{(k)}}_{\tilde{\alpha}} \circ \Pi^{\text{odd},\Z_2^{(l)}}_{\tilde{\beta}} = \delta^{kl} \delta_{\tilde{\alpha}\tilde{\beta}} \times 
\left\{\begin{array}{cr}
8 & \tilde{\alpha}=0
\\ - 8 & 1 \ldots 3
\\ - 16 & 4
\\ 16 & 5
\end{array} \right. \, 
\qquad \text{ with } \quad 
\Z_2^{(0)} \equiv \unity
.
\end{equation}
Analogously to the two other orbifolds, the {\it necessary} conditions for the existence of discrete $\Z_n$ symmetries follow from inserting the sixteen basic $\OR$-even three-cycles of 
equation~(\ref{Eq:Z2Z6p_even+odd_basis_eta-1})  into equation~(\ref{Eq:Zn-condition}),
\begin{equation}\label{Eq:Z2Z6p-necessary_discrete}
{\footnotesize
\sum_a k_a N_a \; \left(\begin{array}{c}  Y_a \\\hline -y_{a,1}^{(1)}  \\ -y_{a,2}^{(1)}  \\ -y_{a,3}^{(1)} \\  - (y_{a,4}^{(1)} + y_{a,5}^{(1)})\\ 2 \, (x_{a,4}^{(1)} - x_{a,5}^{(1)}) + (y_{a,4}^{(1)} - y_{a,5}^{(1)})
\\\hline -y_{a,1}^{(2)}  \\ -y_{a,2}^{(2)}  \\ -y_{a,3}^{(2)} \\  - (y_{a,4}^{(2)} + y_{a,5}^{(2)})\\ 2 \, (x_{a,4}^{(2)} - x_{a,5}^{(2)}) + (y_{a,4}^{(2)} - y_{a,5}^{(2)})
\\\hline -y_{a,1}^{(3)}  \\ -y_{a,2}^{(3)}  \\ -y_{a,3}^{(3)} \\  - (y_{a,4}^{(3)} + y_{a,5}^{(3)})\\ 2 \, (x_{a,4}^{(3)} - x_{a,5}^{(3)}) + (y_{a,4}^{(3)} - y_{a,5}^{(3)})
\end{array}\right)
\stackrel{!}{=} 0 \text{ mod } n
\qquad \text{for} \qquad
\eta_{\OR}=-1}.
\end{equation}
These sixteen {\it necessary} conditions have to be supplemented by sixteen {\it sufficient} conditions arising from the shortest $\OR$-even rigid three-cycles with an enhanced $USp(2N_a)$ or $SO(2N_a)$ gauge group. These
shortest three-cycles are parallel to one of the four $\OR\Z_2^{(i)}$-invariant planes and satisfy the topological conditions  in table~\ref{Tab:Z2Z2M_Conditions-on_b+t+s-SOSp} 
with $b_k=\frac{1}{2}$  and $\eta_{(k)}=-1$ for $k \in \{1,2,3\}$. In section~3.1 of~\cite{Honecker:2012qr}, we found that a total of 256 fractional three-cycles are orientifold invariant.
These are split into 108+4 three-cycles parallel to the $\OR$-plane with gauge enhancement to $USp(2N_a)$ and $SO(2N_a)$, respectively, and $3 \times (36+12)$ three-cycles parallel to the 
$\OR\Z_2^{(k), k\in \{1,2,3\}}$-planes with gauge enhancements to $USp(2N_a)$ and $SO(2N_a)$.\footnote{In reviewing the three-cycles with enhanced gauge group discussed in~\cite{Honecker:2012qr} a typo was discovered, which interchanges the number of three-cycles carrying a $USp(2N_a)$ gauge factor with the number of three-cycles carrying a $SO(2N_a)$ gauge factor in equation~(36) of~\cite{Honecker:2012qr}. Nevertheless, the typo neither affects the total number of three-cycles with enhanced gauge group, nor does it influence the discussion on the existence of three quark generations for $SU(2)_L \simeq USp(2)$ in section~3.4 of~\cite{Honecker:2012qr}.}~However, only $h_{21}+1=16$ $\OR$-even fractional three-cycles are linearly independent.
As an example, the fractional three-cycle parallel to the $\OR$-invariant plane but displaced by $(\vec{\sigma})=(1,1,1)$ has the following expansion in terms of the basic 
$\OR$-even cycles in equation~(\ref{Eq:Z2Z6p_even+odd_basis_eta-1}),
\begin{equation}
\Pi^{\text{frac},\OR}_{(\sigma^1,\sigma^2,\sigma^3)=(1,1,1)} 
\stackrel{\tau^1=\tau^2=\tau^3=\tau}{=} \;\frac{\Pi^{\text{even}}_0}{4} + \sum_{i=1}^3  \frac{(-1)^{\tau^{\Z_2^{(i)}}}}{4}\; \left( -\Pi^{\text{even},\Z_2^{(i)}}_3 + 
(-1)^{\tau} \, \frac{-\Pi^{\text{even},\Z_2^{(i)}}_4 + \Pi^{\text{even},\Z_2^{(i)}}_5}{2} \right)
.
\end{equation}
For $(\vec{\tau})=(\vec{0})$, the gauge group is of $USp(2N_a)$ type for any choice of $\Z_2$ eigenvalues $(-1)^{\tau^{\Z_2^{(i)}}}$. The fourth line from the bottom in equation~(\ref{Eq:Z2Z6p-6-sufficient_discrete})
contains the intersection number of the D6-branes $a$ with such a three-cycle with $\Z_2$ eigenvalues $(+,+,+)$. The third and second line from the bottom are obtained by adding the analogous intersection number with
$\OR$-invariant three-cycles with
$\Z_2$ eigenvalues $(+,-,-)$ and $(-,+,-)$, respectively. For $(\vec{\tau})=(1,1,1)$ the gauge group is enhanced to $SO(2N_a)$, and the last entry in the parenthesis of equation~(\ref{Eq:Z2Z6p-6-sufficient_discrete}) contains the sum of intersection numbers with the two $\OR$-invariant cycles with $(\vec{\tau})=(\vec{0})$ and $(\vec{\tau})=(1,1,1)$, both with $\Z_2$ eigenvalues $(+,+,+)$.

Regarding the rigid three-cycles parallel to $\OR\Z_2^{(k), k\in \{1,2,3\}}$-planes with enhanced gauge groups, their bulk part is three times as long as the bulk part of the three-cycles parallel to the $\OR$-plane. Upon closer inspection the rigid three-cycles parallel to $\OR\Z_2^{(k), k\in \{1,2,3\}}$-planes with enhanced gauge groups can be written as linear combinations of $\OR$-even rigid three-cycles parallel to the $\OR$-plane and $\OR$-even three-cycles from the basis introduced in equation~(\ref{Eq:Z2Z6p_even+odd_basis_eta-1}).

\newpage
\begin{equation}\label{Eq:Z2Z6p-6-sufficient_discrete}
{\footnotesize
\sum_a k_a N_a \; \left(\begin{array}{c}  \frac{Y_a - \sum_{i=1}^3 [ y^{(i)}_{a,1} +  y^{(i)}_{a,2} + y^{(i)}_{a,3} ]}{4} \\ \frac{Y_a - [ y^{(1)}_{a,1} +  y^{(1)}_{a,2} + y^{(1)}_{a,3} ]}{2} \\ 
\frac{Y_a - [ y^{(2)}_{a,1} +  y^{(2)}_{a,2} + y^{(2)}_{a,3} ]}{2} \\ - \frac{   y^{(2)}_{a,1} +y^{(2)}_{a,3}+ y^{(3)}_{a,1} +y^{(3)}_{a,3}  }{2} \\ - \frac{y^{(1)}_{a,1} + y^{(1)}_{a,3} + y^{(3)}_{a,2} +y^{(3)}_{a,3} }{2} \\ - \frac{y^{(1)}_{a,2} + y^{(1)}_{a,3} + y^{(2)}_{a,2} +y^{(2)}_{a,3}}{2} 
\\\hline
\frac{Y_a + [y^{(1)}_{a,3} + x^{(1)}_{a,4} + y^{(1)}_{a,4} - x^{(1)}_{a,5}] + \sum_{j=2}^3 [y^{(j)}_{a,2} - (y^{(j)}_{a,4} + y^{(j)}_{a,5})]}{4}
\\
\frac{Y_a + \sum_{j=1,2} [y^{(j)}_{a,1} - x^{(j)}_{a,4}  + x^{(j)}_{a,5} + y^{(j)}_{a,5}] + [y^{(3)}_{a,3} + x^{(3)}_{a,4} + y^{(3)}_{a,4} - x^{(3)}_{a,5}] }{4}
\\
\frac{Y_a + [y^{(2)}_{a,2} - (y^{(2)}_{a,4} + y^{(2)}_{a,5})]}{2}
\\
\frac{Y_a + [y^{(1)}_{a,1} - x^{(1)}_{a,4}  + x^{(1)}_{a,5} + y^{(1)}_{a,5}]}{2}
\\
- \frac{y^{(2)}_{a,4} + y^{(2)}_{a,5} +  y^{(3)}_{a,4} + y^{(3)}_{a,5} }{2}
\\
-\frac{ x_{a,4}^{(1)} - x^{(1)}_{a,5}  + y^{(1)}_{a,5}+ x_{a,4}^{(2)} - x^{(2)}_{a,5} + y^{(2)}_{a,5} }{2}
 \\\hline
 \frac{Y_a + \sum_{i=1}^3 [y^{(i)}_{a,3} + x^{(i)}_{a,4} + y^{(i)}_{a,4} - x^{(i)}_{a,5}] }{4} \\ \frac{Y_a + y^{(1)}_{a,3}  + x^{(1)}_{a,4} + y^{(1)}_{a,4} - x^{(1)}_{a,5}}{2} \\
\frac{Y_a + y^{(2)}_{a,3}  + x^{(2)}_{a,4} + y^{(2)}_{a,4} - x^{(2)}_{a,5} }{2} \\ \frac{Y_a + \sum_{i=1}^3 y^{(i)}_{a,3}}{2}
\end{array}\right)
\stackrel{!}{=} 0 \text{ mod } n}
.
\end{equation}
As in the previous examples, each entry in the parenthesis in the {\it sufficient} conditions~(\ref{Eq:Z2Z6p-6-sufficient_discrete}) on the existence of some $\Z_n$ symmetry is an intersection number or sum of two intersection numbers with $\OR$-even three-cycles. Each number in the parenthesis is therefore integer valued.

\subsection{Discrete $\Z_n$ symmetries in supersymmetric field theories}\label{Ss:DGSSUSYFT}

In supersymmetric field theories - and more specifically the MSSM -  discrete $\Z_n$ symmetries are invoked  \cite{Ibanez:1991pr,Dreiner:2005rd}  to argue that lepton- and/or baryon-number violating operators
are absent, and to guarantee proton stability in agreement with experimental bounds. More explicitly, three-point and four-point couplings in the superpotential leading to dimension-four- and dimension-five-operators in the scalar potential violating baryon ($B$)- and/or lepton ($L$)-number, as listed schematically in table~\ref{Tab:LBViolatingOp}, have to be forbidden while the usual Yukawa-couplings in the superpotential are allowed.
Even though the dimension-five-operators are suppressed by some mass scale, $ {\cal O}_5 \sim 1/M$, they are  capable of inducing proton decay at an unacceptable rate, if they appear with a coefficient which is of order one.
\begin{table}[h]
\begin{center}
\begin{tabular}{|c|c||c||c||c|}
\hline \multicolumn{5}{|c|}{\bf Lepton-and/or baryon-number violating operators in MSSM}\\
\hline \hline \multicolumn{2}{|c||}{} & $\Delta L \neq 0$ & $\Delta B \neq 0$ & $\Delta L, \Delta B \neq 0$\\
\hline \hline 
3-point & $F$ & $LL \ov E, L Q \ov D, $ &  $\ov U\, \ov D\, \ov D, $ & \\
& $D$ & $\ov U\, \ov D^\dagger \, \ov E, H_u^\dagger H_d \ov E, Q \ov U L^\dagger$ & $ Q Q \ov D^\dagger$ & \\
4-point & $F$ & $Q \ov U\, \ov E H_d, L H_u L H_u,$ & $Q Q Q H_d$ & $QQQL, \ov U\, \ov U\, \ov D\, \ov E$ \\
&$F$& $L H_u H_d H_u$ &&\\
\hline
\end{tabular}
\caption{Overview of three-and four-point operators violating lepton ($\Delta L \neq 0$)- and/or baryon ($\Delta B \neq 0$)-number for the MSSM. The subscripts denoting the handedness (L or R) are left out for simplicity. The second column indicates whether the operator appears from the $F$- or $D$-term.\label{Tab:LBViolatingOp}
}
\end{center}
\end{table}
 Discrete $\Z_n$ symmetries are less constrained by anomalies than their continuous counter parts, and they  arise naturally in string theory, as discussed in the previous section. Discrete symmetries
 thus offer excellent extensions to the MSSM in order to prohibit the lepton-and/or baryon-number violating operators. 

Generically, a discrete $\Z_n$ symmetry acts on the chiral superfields by a global phase rotation,
\begin{equation}
\Phi_j \rightarrow e^{i\, \alpha_j 2 \pi/n} \Phi_j,
\end{equation}
where the integer charge $\alpha_j$ is defined modulo an integer shift $n$,
\begin{equation}
\alpha_j \sim \alpha_j + m_j n, \hspace{0.4in} m_j \in \Z.
\end{equation}
Hence, charges under the $\Z_n$ symmetry should be seen as equivalence classes, as it is customary for representations of cyclic groups. A generation-blind discrete $\Z_n$ symmetry acting on the MSSM states can be represented by a six-tuple of integer charges\footnote{In the original papers \cite{Ibanez:1991pr,Dreiner:2005rd} right-handed neutrinos were not included in the analysis. In a recent paper \cite{Anastasopoulos:2012zu} it was pointed out that the presence of right-handed neutrinos eliminates all $\Z_9$ and $\Z_{18}$ symmetries, which would otherwise be allowed.},
\begin{eqnarray}
\vec{\alpha} = (\alpha_Q, \alpha_u, \alpha_d, \alpha_L, \alpha_e, \alpha_{\nu}),
\end{eqnarray}
as the charges of the two Higgs-doublets are fixed by the Yukawa-couplings to quarks and leptons. Given that every particle can have $n$ different phases, we might be lead to the conclusion that there are $n^6 -1$
 possible non-trivial $\Z_n$ symmetries, where the $-1$ eliminates the case where the charges for the MSSM particles are all zero. However, not all of these symmetries are independent. As the down-type  Higgs-doublet $H_d$ couples both to $(Q_L, \ov D_R)$ and to $(L,\ov E_R)$ one of the charges of the four chiral states is fixed by the three other. A similar argument can be used to fix one of the charges of the chiral states $(Q_L, \ov U_R)$, $(L, \ov N_R)$, using their coupling to the up-type Higgs-doublet $H_u$. In addition, one can set one of the remaining four independent charges to zero by a discrete hypercharge rotation. These considerations reduce the number of independent phases to $n^3$, indicating a possible $\Z_n^3$ freedom. Hence, each of the $\Z_n$ symmetries $g_n$ can be decomposed in terms of three independent discrete generators  $ {\cal R}_n=e^{i 2 \pi {\cal R}/n}$, $ \mathfrak{L}_n = e^{i 2 \pi \mathfrak{L}/n}$ and ${\cal A}_n = e^{i 2 \pi {\cal A}/n} $ as follows~\cite{Ibanez:1991pr},
\begin{equation}\label{Eq:GenRALMSSM}
g_n = {\cal R}_n^m \cdot {\cal A}_n^k \cdot {\mathfrak{L}_n^p} , \hspace{0.4in}  m, k, p = 0, 1, \ldots, n-1 
\end{equation}
with the charge assignments of MSSM states given in table \ref{Tab:ZnsymmChargesMSSM}. 
\begin{SCtable}
$\begin{array}{|c||c|c|c|c|c|c|c|c|c|}\hline
\muc{9}{|c|}{\text{\bf Charges of generation-independent $\Z_n$ symmetries}}
\\\hline\hline
\text{Generator} & Q_L & \ov U_R & \ov D_R & L & \ov E_R & \ov N_R &  H_u & H_d \\
\hline \hline \cal R & 0 & n-1 & 1 & 0& 1 & n-1 & 1 &n-1\\
 \mathfrak{L} & 0 & 0 & 0 &n-1 & 1 & 1 & 0 & 0 \\
\cal A & 0 & 0 & n-1 &  n-1 & 0 &  1  & 0 &  1 \\
\hline
\end{array}$
\caption{Generation-independent charge generators ${\cal R}$, ${\cal L}$, ${\cal A}$
of the generic discrete $\Z_n$ symmetry in equation~(\ref{Eq:GenRALMSSM}) with the corresponding charges of the chiral MSSM states. 
\label{Tab:ZnsymmChargesMSSM}}
\end{SCtable}

For a generic discrete $\Z_n$ symmetry, the MSSM particles transform under the corresponding $g_n$ as follows:
\begin{eqnarray}
\begin{array}{l@{\hspace{0.4in}}l@{\hspace{0.4in}}l}
\alpha_Q = 0, & \alpha_u = -m, & \alpha_d = m-k,\\
\alpha_L = -k-p, & \alpha_e = m + p & \alpha_\nu = -m + k + p, \\
\alpha_{H_u} = m, & \alpha_{H_d}= - m + k. &
\end{array} \label{Eq:MSSMcharges}
\end{eqnarray}
Provided that the discrete symmetry arises from the breaking of a continuous gauge symmetry $U(1)_X$, the discrete symmetry will also have to obey anomaly cancelation conditions.\footnote{If the anomaly of the discrete symmetry  does not vanish, it appears as a non-perturbative anomaly, i.e. a mixed $\Z_n \times SU(3)^2$ anomaly will induce terms generated by QCD instantons violating the discrete symmetry.}~There exist two types of anomaly cancelation conditions: linear anomaly constraints and cubic anomaly constraints. The linear anomaly constraints can be seen as the necessary conditions for the existence of a discrete symmetry. They boil down to the vanishing of the anomaly coefficients for the mixed non-Abelian $SU(3)-SU(3)-\Z_n$ and $SU(2)-SU(2) - \Z_n$ anomalies, as well as for the gravitational $G-G-\Z_n$ anomaly (modulo $n$ times the dimension of the fundamental representation of the gauge group). The cubic $\Z_n^3$ anomaly is the only non-linear anomaly constraint that plays a non-trivial r\^ole, such that this constraint can be interpreted as the sufficient condition for the existence of a discrete $\Z_n$ gauge symmetry. Writing down the four anomaly constraints for $N_g$ particle generations and $N_h$  Higgs doublet two-tuples $(H_u, H_d)$ with discrete $\Z_n$ charge assignments given in  table~\ref{Tab:ZnsymmChargesMSSM}, the following relations define a discrete $\Z_n$ symmetry:\footnote{ Additional chiral fermions (such as Dirac fermions, Majorana fermions, etc.) charged under $\Z_n$ can influence the vanishing of anomaly constraints drastically, and thereby also which discrete $\Z_n$ gauge symmetries are conserved in a specific model~\cite{Dreiner:2005rd}. For now the anomaly constraints will be written down disregarding the possible extra chiral fermions, but in the global models discussed in section \ref{S:GICBM} all additional and exotically charged chiral states are taken into account.}

\begin{enumerate}
\item[$(i)$] \underline{$SU(3)-SU(3) - \Z_n$ anomaly constraint:} 
\begin{equation}\label{Eq:Anomaly_SU3-U1}
N_g \left( 6 \alpha_Q + 3 \alpha_u + 3 \alpha_d \right)  \stackrel{!}{\in} 3 n \Z 
\end{equation}
\item[$(ii)$] \underline{ $SU(2)-SU(2) - \Z_n$ anomaly constraint:}
\begin{equation}\label{Eq:Anomaly_SU2-U1}
N_g \left(6 \alpha_Q + 2 \alpha_L\right) + N_h \left( 2 \alpha_{H_u} + 2 \alpha_{H_d}  \right)   \stackrel{!}{\in} 2 n \Z 
\end{equation}
\item[$(iii)$] \underline{$G-G - \Z_n$ anomaly constraint:}
\begin{equation}\label{Eq:Anomaly_G-U1}
N_g \left(6 \alpha_Q + 3 \alpha_u + 3 \alpha_d + 2 \alpha_L + \alpha_e + \alpha_\nu  \right) + N_h \left( 2 \alpha_{H_u} + 2 \alpha_{H_d}  \right)   \stackrel{!}{\in} n \Z
\end{equation}
\item[$(iv)$] \underline{$\Z_n - \Z_n - \Z_n $ anomaly constraint}\footnote{The discussion presented here follows the line of thoughts developed in~\cite{Ibanez:1991pr,Dreiner:2005rd}. In~\cite{Banks:1991xj} it was pointed out that the cubic anomaly constraints are unnecessary for the consistency of the low energy theory. Violation of the cubic anomaly constraints indicates that the discrete $\Z_n$ symmetry acts as an anomalous subgroup of an anomaly-free enlarged discrete group at higher energies. Moreover, a direct computation of the anomalies via the path integral formalism~\cite{Araki:2008ek} shows the absence of cubic anomalies for discrete gauge symmetries in field theory. We would like to thank Michael Ratz and Maximilian Fallbacher for pointing out this aspect of discrete symmetries to us. The discrete symmetries discussed in the context of global intersecting D6-brane models indeed automatically satisfy the cubic anomaly constraints, as the generalized Green-Schwarz-mechanism ensures the vanishing of all cubic $U(1)$ anomalies.}
\begin{eqnarray}\label{Eq:Anomaly_U1-U1}
N_g \left(6 \alpha_Q^3 + 3 \alpha_u^3 + 3 \alpha_d^3 + 2 \alpha_L^3 + \alpha_e^3 +  \alpha_\nu^3  \right) + N_h \left( 2 \alpha_{H_u}^3 + 2 \alpha_{H_d}^3  \right)  \stackrel{!}{\in} n \Z
\end{eqnarray}
\end{enumerate}
The expression for the first anomaly constraint $(i)$ given in the form here clearly stipulates that the integer $n$ has to be divisible by the number of generations $N_g$.  In the absence of additional heavy chiral fermions the cubic anomaly constraint $(iv)$ eliminates~\cite{Dreiner:2005rd} most of the candidate $\Z_n$ discrete gauge symmetries. Only three discrete symmetries satisfy all anomaly constraints for the MSSM: matter parity ${\cal R}_2$ ($\Z_2$), baryon-triality ${\cal B}_3 \equiv {\cal R}_3 \mathfrak{L}_3$ ($\Z_3$) and proton-hexality  ${\cal P}_6 \equiv {\cal R}_6^5 \mathfrak{L}_6^2$ ($\Z_6$). The corresponding charges are listed in table~\ref{Tab:DGS27fund}.
\begin{table}[h]
\begin{center}
\begin{tabular}{|c|c|c||c||c|c|c||c|c||c|c|}\hline
\muc{11}{|c|}{\text{\bf Charges of phenomenologically relevant $\Z_n$ symmetries}} 
\\\hline\hline
\text{Name} & & $\Z_n$& $Q_L$ & $\ov U_R$ & $\ov D_R$ & $L$ & $\ov E_R$ & $\ov N_R$& $H_u$ &$H_d$\\
\hline \hline
\text{R-parity}& ${\cal R}_2$ & $\Z_2$ &  0 &$1$ & $1$& 0& $1$ & $1$&$1$&$1$\\
\hline  
\text{baryon triality} & ${\cal L}_3 {\cal R}_3$ & $\Z_3$&  0 &$2$ & $1$& $2$& $2$ & $0$&$1$&$2$\\
\hline 
\text{proton hexality} & ${\cal L}_6^2 {\cal R}_6^5$& $\Z_6$&  0 &$5$ & $1$& $1$& $0$ & $2$&$5$&$1$\\
\hline
\end{tabular}
\caption{Charge assignments for R-parity, baryon triality and proton hexality with charges chosen in the range $0 \leqslant \alpha <n$ and $\gcd(\alpha_Q,\alpha_u, \dots,n)=1$. 
The entries in the second column denote the expression in terms of generators according to equation~(\protect\ref{Eq:GenRALMSSM}).
\label{Tab:DGS27fund}}
\end{center}
\end{table}

The `stringy' constraints on discrete $\Z_n$ symmetries turn out much stronger than the purely field theoretical constraints as discussed for some examples in the following
section. However, we find several new $\Z_2$, $\Z_4$ and $\Z_6$ symmetries in extensions of the MSSM that respect all `stringy' constraints. 
 A first kind of extensions consists of left-right symmetric models with gauge group $SU(3)\times SU(2)_L \times SU(2)_R \times U(1)_{B-L}$. The recombination of the right-handed fermions 
$(u_R,d_R)$ and $(e_R,\nu_R)$ and the Higgs-doublets $(H_u,H_d)$ into fundamental representations under the $SU(2)_R$ gauge group imposes a non-trivial relation on the discrete charges in equation~(\ref{Eq:MSSMcharges}),
\begin{equation}\label{Eq:ConstraintLRS}
k = 2m \hspace{0.2in} \text{ mod } n,
\end{equation}
which relates the two discrete generators $\cal R$ and $\cal A$ such that only two independent discrete generators remain. The second type of extensions studied here are Pati-Salam models, for which the gauge group $SU(3) \times U(1)_{B-L}$ of the left-right symmetric models is enhanced to a $SU(4)$ gauge group. Under this enhancement, a second constraint for the discrete charges in equation~(\ref{Eq:MSSMcharges}) arises,
\begin{equation}\label{Eq:ConstraintPS}
p = -k \hspace{0.2in} \text{ mod } n,
\end{equation}
such that the three discrete generators reduce to a single independent discrete generator, when the MSSM is embedded in a Pati-Salam GUT.\footnote{One can use the same reasoning for a Pati-Salam model as the one applied above for the MSSM, in order to derive the independent discrete generators and the corresponding discrete charges from scratch for a given chiral spectrum and Yukawa-couplings. For a Pati-Salam model, there are a priori three independent charges. The requirement of having a suitable Yukawa-coupling between these three different multiplets yields one relation between the three different discrete charges. Hence, we can have at most two independent generators. In this sense, the discrete symmetries of a MSSM arising from a Pati-Salam model only form a subgroup of the possible discrete symmetries in a Pati-Salam model. The discrepancy can be traced back to the fact that the massless hypercharge, which emerges from the spontaneous breaking of $SU(4)\times SU(2)_R \rightarrow SU(3)\times U(1)_Y$, is not yet at hand to set the charges of the left-handed fermions to zero. Meanwhile, in the case of a left-right symmetric extension of the MSSM, the massless $B-L$ symmetry can be used to set the charges of the left-handed fermions to zero.

}

\section{Global Intersecting D-Brane Models}\label{S:GICBM}

In this section, the presence of discrete $\Z_n$ gauge symmetries in global D6-brane models on various orbifold backgrounds
is investigated for explicit examples constructed in~\cite{Honecker:2004kb,Gmeiner:2008xq,Honecker:2012qr}. Our method to identify the discrete gauge symmetries consists of three steps:
first we identify all linear combinations of $U(1)$ symmetries that remain massless according to the first line of equation~(\ref{Eq:Zn-condition}). In the second step, we perform a complete scan for all possible discrete $\Z_n$  symmetries
satisfying the second line of equation~(\ref{Eq:Zn-condition}), or in other words the necessary and sufficient conditions derived in section~\ref{Ss:DiscreteOrbifolds}.
In the third step, we shift the discrete charges by those of some massless $U(1)$, e.g.  a gauged $B-L$ symmetry, such that the left-handed quarks are neutral, $\alpha_Q=0$. 
The shifted charges are then interpreted in terms of the field theory description of section~\ref{Ss:DGSSUSYFT}. 

The known global models on fractional D6-branes fall into two classes: left-right symmetric models with gauge groups containing $SU(3)_a\times SU(2)_b\times USp(2)_c$, and Pati-Salam models with gauge groups  
containing $SU(4)_a\times SU(2)_b \times SU(2)_c$.

\subsection{Left-right symmetric models}
\subsubsection{A global model on $T^6/(\Z_6 \times \OR)$}\label{Sss:T6-model}

The first global model under investigation was constructed~\cite{Honecker:2004kb,Gmeiner:2009fb} on five stacks of D6-branes in the orbifold background $T^6/(\Z_6 \times \OR)$ with  {\bf AAB} lattice orientation.
The gauge group is a priori given by $U(3)_a\times U(2)_b \times USp(2)_c\times U(1)_d \times USp(2)_e$. Applying the generalized Green-Schwarz-mechanism, the Abelian gauge factors $U(1)_a$, $U(1)_b$ and $U(1)_d$ 
provide two massive gauge bosons and one massless gauge boson, which generates a gauged $B-L$ symmetry, 
\begin{equation} 
 Q_{B-L} = \frac{Q_a}{3} + Q_d.
\end{equation}
The gauge groups $USp(2)_c$ and $USp(2)_e$ result from fractional D6-branes wrapped along three-cycles parallel to the $\OR$-plane, with displacements and Wilson lines along $T^2_{(1)} \times T^2_{(2)}$ providing the 
enhancement of the gauge groups from $U(1)_c$ and $U(1)_e$, as discussed in subsection~\ref{Sss:Z6}. Upon continuous displacement of the three-cycles or when turning on a Wilson-line along $T^2_{(3)}$, each gauge group $USp(2)_{c,e}$ 
is broken to the diagonal massless and anomaly-free subgroup $U(1)_{c,e}$,  
\begin{table}[h]
\begin{center}
{\footnotesize
\begin{tabular}{|c|c||c|c|c|c|}
\hline \multicolumn{6}{|c|}{\bf  Chiral states and Higgses for a five-stack model on the AAB lattice of $T^6(\Z_6 \times \OR)$} \\
\hline \hline
\bf Matter & \bf Sector & $U(3)_a\times U(2)_b$& $(Q_a, Q_b, Q_c, Q_d, Q_e)$ & $Q_Y$&$Q_{B-L}$\\
\hline
$Q_L$ & $ab'$ & 3 $\times ({\bf  3}, {\bf  2})$ & $(1,1,0,0,0)$& $\frac 1 6$& $\frac 1 3$ \\
$\ov U_R$ & $ac$ & 3 $({\bf  \ov 3}, \1)$ & $(-1,0,1,0,0)$& $-\frac 2 3$& $-\frac 1 3$ \\
$\ov D_R$ & $ac'$ & 3 $({\bf  \ov 3}, \1)$ & $(-1,0,-1,0,0)$& $\frac 1 3$& $-\frac 1 3$ \\
$L$ & $bd'$ & 3$(1, {\bf  \ov 2})$ & $(0,-1,0,-1,0)$& $-\frac 1 2$& $-1$ \\
$\ov E_R$ & $cd$ & 3 $\times (\1, \1)$ & $(0,0,-1,1,0)$& $1$& $1$ \\
$\ov N_R$ & $cd'$ & 3 $\times (\1, \1)$ & $(0,0,1,1,0)$& $0$& $1$ \\
$X_{be}$ & $be$ & 3$\times (\1, {\bf  \ov 2})$ & $(0,-1,0,0,1)$& $0$& $0$ \\
$X_{be'}$ & $be'$ & 3$\times (\1, {\bf  \ov 2})$ & $(0,-1,0,0,-1)$& $0$& $0$ \\
\hline \hline
$H_d^{(2)}$ &$bc'$ & $(\1,{\bf 2}) $ & $(0,1,1,0,0)$& $-\frac 1 2$& $0$ \\
$H_d^{(1)}$ & $bc$ &$(\1,{\bf \ov 2}) $ & $(0,-1,1,0,0)$ & $-\frac 1 2$& $0$ \\
$H_u^{(1)}$ &$bc$ & $(\1,{\bf 2}) $ & $(0,1,-1,0,0)$& $\frac 1 2$& $0$ \\
$H_u^{(2)}$ &$bc'$ & $(\1,{\bf \ov 2})  $ & $(0,-1,-1,0,0)$& $\frac 1 2$& $0$ \\
\hline
\end{tabular}}
\caption{Chiral states and non-chiral Higgs doublets arising from a five-stack left-right symmetric model on the {\bf AAB} lattice of $T^6/(\Z_6 \times \OR)$ after spontaneously breaking the $USp(2)_{c,e}$ factors to their Abelian subgroups
$U(1)_{c,e}$.\label{Tab:ChirSpecModelT6Z6}}
\end{center}
\end{table}
and  the hypercharge $Y$ or $Y'$ is identified as 
\begin{equation}
Q_Y  =\frac{Q_{B-L}}{2} + \frac{Q_c}{2} \qquad \text{ or } \qquad Q_Y' = \frac{Q_{B-L}}{2} + \frac{Q_c}{2} + \frac{Q_e}{2}.
\end{equation}  
It is noteworthy that  the massless $U(1)_c$ symmetry acts as a gauged R-symmetry as can be seen from the fact that all charges of 
the MSSM states are (up to a global sign) identical to those given in table~\ref{Tab:ZnsymmChargesMSSM}.

Discrete $\Z_n$ symmetries arise from massive gauge symmetries of the form $U(1)_{\text{massive}} = k_a \, U(1)_a + k_b \, U(1)_b + k_d \, U(1)_d$, which can then be shifted by any linear combination of the 
$U(1)_{B-L}$, $U(1)_c$ and $U(1)_e$ symmetries to bring the charge assignments of MSSM states into the form in table~\ref{Tab:DGS27fund}.

The full chiral spectrum for this model is listed in table~\ref{Tab:ChirSpecModelT6Z6}. The Higgs states in this model stem from the  non-chiral  $bc$ plus $bc'$ sector and are also given in table~\ref{Tab:ChirSpecModelT6Z6}. The non-chiral origin of the Higgses implies that the charges under the discrete $\Z_n$ symmetry are opposite,
\begin{equation}
\alpha_{H_d^{(1)}} = -\alpha_{H_u^{(1)}}, \hspace{0.4in} \alpha_{H_d^{(2)}} = -\alpha_{H_u^{(2)}}.
\end{equation}
Applying this constraint in terms of the charges given in (\ref{Eq:MSSMcharges}), one can immediately deduce that $k=0$ (mod $n$), or equivalently that a $\Z_n$ discrete gauge symmetry in this model will only be generated by ${\cal R}$ and/or $\mathfrak{L}$. Comparing the charges of the states under the various $U(1)$'s in table \ref{Tab:ChirSpecModelT6Z6} with the field theory charges in table \ref{Tab:ZnsymmChargesMSSM}, one would naively identify $\mathfrak{L} = Q_d$. 
However, such a discrete ${\cal L}_n$ symmetry does not exists due to the constraint~(\ref{Eq:Z6_LRSModelNECSUFF}) as discussed in detail below.

The necessary and sufficient conditions~(\ref{Eq:Z6_discrete_nec+suff}) on the existence of discrete $\Z_n$ symmetries read for this model,
\begin{equation}\label{Eq:Z6_LRSModelNECSUFF}
{\footnotesize
\left\{ k_a  \left(\begin{array}{c} 0 \\\hline 3 \\ 3 \\ 0 \\ -3  \\ 6  \\\hline\hline  3 \\ 0 \\ 3  \end{array}\right)
+ k_b  \left(\begin{array}{c}  0 \\\hline -2 \\ 2 \\  0 \\ 2  \\ -4  \\\hline\hline 0 \\  0 \\ -2 \end{array}\right)
+ k_d  \left(\begin{array}{c} 0 \\\hline -1 \\ -1 \\ 0 \\ 1  \\ -2 \\\hline\hline -1 \\ 0 \\ -1 \end{array}\right)
\right\} }
\stackrel{!}{=}0 \text{ mod }n\,
,
\end{equation}
with $(k_a,k_b,k_d) \neq (1,0,3)$, since this particular linear combination provides the massless gauged $B-L$ symmetry. 

As expected, not all conditions are independent: the second, fifth and  ninth line are identical, and the sixth line is twice the fifth line.  The third line is linearly dependent on the second and seventh line. 
The two remaining independent conditions lead to the shape $(k_a,k_b,k_d)=(k_a,0,k_d)\neq (1,0,3)$ or $(0,k_b,0)$ of solutions. The corresponding charge assignments after shift by the gauged $B-L$ symmetry, which is used to 
 set the charge $\alpha_Q$ of the left-handed quark doublets to zero, are  given in table~\ref{Tab:DiscreteSymmT6Z6EndI}.

\begin{table}[h]
\begin{center}
\hspace*{-0.3in}\begin{tabular}{|c|c||c|c|c|c|c|c|c|c|c|c|}
\hline \multicolumn{12}{|c|}{\bf Discrete charges for the L-R symmetric model on $T^6/(\Z_6\times \OR)$}\\
\hline
\hline  \multicolumn{2}{|c||}{\bf Discrete symmetries}&\multicolumn{10}{|c|}{\bf Charge assignment for the MSSM states}\\
\hline
\hline $\Z_n$  & $U(1) = \sum_{x \in \{a,b,d \}} k_x\, U(1)_x$& $Q_L$ & $\ov U_R$ & $\ov D_R$ & $L$ & $\ov E_R$ & $\ov N_R$& $H_u^{(1)}$ &$H_u^{(2)}$ & $H_d^{(1)}$ &$H_d^{(2)}$ \\
\hline \hline
$\Z_2$ & $U(1)_a + U(1)_d$&  0 &0 &0 & 0 & 0 & 0 & 0 & 0 & 0  & 0 \\
$\Z_2$& $U(1)_b$ &$0$ &$1$ &$1$ &$0$ &$1 $&$1$&$1$ &$1$ &$1$  &$1$ \\
$\Z_3$ &$U(1)_a$&  0 &0 & 0 &0&0&0&0&0&0&0 \\
\hline
\end{tabular}
\caption{The three $\Z_n$ discrete symmetries for a left-right symmetric five-stack model on $T^6/(\Z_6\times \OR)$. The charges of the MSSM states are listed after a $B-L$ rotation. Upon breaking of the $USp(2)_{c,e}$ gauge group the charges can be further rotated by the massless $U(1)_{c,e}$. A $Q_e$ rotation will not have any impact on the charges of the MSSM states since none of them carries a $Q_e$ charge, cf. table~\protect\ref{Tab:ChirSpecModelT6Z6}.
\label{Tab:DiscreteSymmT6Z6EndI}}
\end{center}
\end{table}

The full set of solutions to the constraints~(\ref{Eq:Z6_LRSModelNECSUFF}) on discrete $\Z_n$ symmetries has the following form.
\begin{itemize}
\item The combination $(k_a,k_b,k_d)=(1,0,0)$ points towards a discrete `baryon' $\Z_3$ symmetry arising from $U(1)_a$. However, as all states
in the last line of table~\ref{Tab:DiscreteSymmT6Z6EndI} (even all chiral and non-chiral ones) have charge $0$ mod $3$ under this discrete $\Z_3$ symmetry after a discrete $B-L$ rotation. The discrete $\Z_3$ symmetry is thus trivial in field theory. 
\item The combination $(k_a,k_b,k_d)=(1,0,1)$ in the first line of table~\ref{Tab:DiscreteSymmT6Z6EndI}
 allows for a discrete $\Z_2$ gauge symmetry. The charge assignment in the first row of table~\ref{Tab:DiscreteSymmT6Z6EndI} indicates that the discrete symmetry acts trivially after a $B-L$ shift to set the charge of left-handed quarks to zero. 
 \item
 A second discrete $\Z_2$ symmetry arises from the combination  $(k_a,k_b,k_d)=(0,1,0)$. The second row in table~\ref{Tab:DiscreteSymmT6Z6EndI} lists the charges of the MSSM fields after a $B-L$ shift. Comparing the charges to the ones in table~\ref{Tab:DGS27fund} suggests that this $\Z_2$ symmetry could be a matter parity ${\cal R}_2$. 
Without spontaneous breaking of the $USp(2)$ gauge factors, the effective gauge symmetry of the model is thus given by $SU(3)_a \times SU(2)_b \times USp(2)_c \times U(1)_{B-L} \times USp(2)_e$, taking into account that the discrete $\Z_2 \subset U(2)_b$ symmetry does not impose any new constraints on cubic matter couplings beyond the ones provided by the $SU(2)$ charges. As pointed out above the ${\cal R}$ symmetry is a gauged symmetry generated by $U(1)_c$ after breaking $USp(2)_c$, and applying an additional rotation over $Q_c$ sets all MSSM charges to zero. A third shift over the massless $Q_e$ direction sets the remaining charges of the exotic states charged under $U(1)_e$ to zero, implying that this ${\Z_2}$ symmetry  acts in fact trivially
 on the massless open string sector after continuous symmetry breaking of the effective gauge symmetry by displacements or Wilson lines to $SU(3)_a \times SU(2)_b \times U(1)_Y \times U(1)_{B-L} \times U(1)_e$. 
 \\
 The superposition of this $\Z_2$  with $(k_a,k_b,k_d)=(0,1,0)$ with the first one  with $(k_a,k_b,k_d)=(1,0,1)$ corresponds exactly to the $\Z_2$ discrete symmetry guaranteed by the K-theory constraints.\\
\end{itemize}

Finally, we should also have a look at combinations $(k_a,k_b,k_d)=(0,0,m)$ with $m\in \Z$ to investigate if discrete symmetries generated by $\mathfrak{L}$ occur in this model. From the necessary and sufficient conditions (\ref{Eq:Z6_LRSModelNECSUFF}) one can clearly see that they can never be satisfied for $0\leq k_d<n$ and $n\geq 2$. This consideration confirms our analysis above, where we did not find any discrete $\Z_n$ gauge symmetry generated by $\mathfrak{L}$. Purely based upon the spectrum and the four anomaly constraints one would actually expect discrete symmetries generated by $\mathfrak{L}$. As they do not arise from string theory, this example clearly shows that the necessary and sufficient conditions in equation (\ref{Eq:Z6_discrete_nec+suff}) are stronger than the field theoretic conditions of anomaly cancellations\footnote{It is easy to check that the purely field theoretical linear and non-linear 
anomaly constraints in section~\ref{Ss:DGSSUSYFT} are satisfied for e.g.~a discrete $\Z_2$ symmetry generated by $\mathfrak{L}$.}. Given that the ${\cal R}$ symmetry corresponds to a gauged symmetry, the absence of discrete symmetries generated by $\mathfrak{L}$ also implies that baryon triality ${\cal B}_3$ and proton hexality ${\cal P}_6$ do not appear as discrete symmetries in this model.

To complete the study of discrete $\Z_n$ gauge symmetries in this example, the field theory anomaly constraints for the chiral spectrum in table~\ref{Tab:ChirSpecModelT6Z6} ought to be investigated. However, given that the discrete symmetries found above all act trivially (i.e.~the charges of all chiral states are zero under the symmetry), the anomaly constraints are trivially satisfied.

\subsubsection{A global model on $T^6/(\Z_6' \times \OR)$ without a hidden sector}\label{Ss:T6Z6pNoHidden}

Considering the {\bf ABa} lattice configuration for the orbifold $T^6/(\Z_6' \times \OR)$, a first type of models with a phenomenologically appealing chiral spectrum has been constructed \cite{Gmeiner:2008xq} on four stacks of D6-branes, assuming that the complex structure modulus takes the value $\varrho=\frac{1}{4}$. 
The four stacks of D6-branes give a priori rise to the gauge group  $U(3)_a\times U(2)_b\times USp(2)_c \times U(1)_d$, and the RR tadpoles vanish without invoking hidden D6-branes. Turning on continuous Wilson lines or displacing  the D6$_c$-brane along $T^2_{(2)}$ breaks the enhanced gauge group $USp(2)_c$ to a massless Abelian gauge factor $U(1)_c$. The generalized Green-Schwarz mechanism provides the cancellation of the remaining non-vanishing anomalies, and in that process $U(1)_a\times U(1)_b\times U(1)_d$ recombine into two massive linear combinations and one massless linear combination, which is a gauged $B-L$ symmetry,
\begin{equation}\label{Eq:T6Z6pB-L}
Q_{B-L} = \frac{1}{3} Q_a + Q_d.
\end{equation}
Together with the massless $U(1)_c$ factor after $USp(2)_c$ breaking along some flat direction, we can identify a massless hypercharge,
\begin{equation}\label{Eq:T6Z6pHypercharge}
Q_Y = \frac{1}{2} Q_{B-L} + \frac{1}{2} Q_c = \frac{1}{6} Q_a + \frac{1}{2} Q_c + \frac{1}{2}Q_d.
\end{equation}
As in the model on $T^6/(\Z_6 \times \OR)$, the massless $U(1)_c$ group acts as a gauged R-symmetry.
The chiral spectrum of the model is presented in table~\ref{Tab:T6Z6pModel2Spectrum}.
\begin{table}[h]
\begin{center}
{\footnotesize
\hspace*{-0.3in}\begin{tabular}{|c|c||c|c|c|c|}
\hline \multicolumn{6}{|c|}{\bf  Chiral states for the global model without hidden matter on the ABa lattice of $T^6/(\Z_6' \times \OR)$} \\
\hline \hline
\bf Matter & \bf Sector & $U(3)_a\times U(2)_b$& $(Q_a, Q_b, Q_c, Q_d)$ & $Q_Y$&$Q_{B-L}$\\
\hline
$Q_L$ & $ab'$ & 3 $\times ({\bf  3}, {\bf  2})$ & $(1,1,0,0)$& $\frac 1 6$& $\frac 1 3$ \\
$\ov D_R$ & $ac$ & 3 $\times ({\bf  \ov 3}, \1)$ & $(-1,0,1,0)$& $\frac 1 3$& $-\frac 1 3$ \\
$\ov U_R$ & $ac'$ & 3 $\times ({\bf  \ov 3}, \1)$ & $(-1,0,-1,0)$&$-\frac 2 3$& $-\frac 1 3 $ \\
$L$ & $bd$ & 6 $\times (\1, {\bf  2})$ & $(0,1,0,-1)$& $-\frac 1 2$& $-1$ \\
$\ov L$ & $bd'$ & 3 $\times (\1, {\bf  2})$ & $(0,1,0,1)$& $\frac 1 2$& $1$ \\
$\ov N_R$ & $cd$ & 3 $\times (\1, \1)$ & $(0,0,-1,1)$& $0$& $1$ \\
$\ov E_R$ & $cd'$ & 3 $\times (\1, \1)$ & $(0,0,1,1)$& $1$& $1$ \\
$H_u$ & $bc$ & 18 $\times (\1, {\bf \ov 2})$ & $(0,-1,1,0)$& $\frac 1 2$& $0$ \\
$H_d$ & $bc'$ & 18 $\times (\1, {\bf \ov 2})$ & $(0,-1,-1,0)$& $-\frac 1 2$& $0$ \\
$\Sigma_b$ & $bb'$ & 9 $\times (\1,\1_{\ov\Anti}) $ & $(0,-2,0,0) $ & $ 0$ & $0 $\\
\hline
\end{tabular}}
\caption{Chiral states arising from a left-right symmetric model without hidden sector on the $T^6/(\Z_6' \times \OR)$ orbifold after spontaneously breaking the $USp(2)_c$ gauge factor to an massless Abelian $U(1)_c$ subgroup. \label{Tab:T6Z6pModel2Spectrum}}
\end{center}
\end{table}

In the search for discrete $\Z_n$ gauge symmetries arising from  $U(1)_{\text{massive}}^2$, the generation-independent generators $\mathfrak{L}$ and ${\cal A}$ are naively identified respectively with $Q_d$ and $\frac{1}{2}\left( Q_a - Q_b - Q_c + Q_d \right)$. As a consequence the additional chiral states $\ov L$ and $\Sigma_b$ in table \ref{Tab:T6Z6pModel2Spectrum} transform under a generic discrete symmetry $g_n$ with charges 
\begin{equation} 
\alpha_{\ov L} = p, \hspace{0.4in} \alpha_{\Sigma_b} = k.
\end{equation}

Discrete $\Z_n$ gauge symmetries are present provided that the necessary and sufficient conditions from equation~(\ref{Eq:Z6p_Condition_ZM}) are satisfied,
\begin{equation}
{\footnotesize
\left\{ k_a  \left(\begin{array}{c} 0 \\ 0 \\\hline 0 \\ 6 \\ 6 \\ 0 \\\hline\hline 3 \\ 3 \\ 0 \\ 6 \end{array}\right)
+ k_b  \left(\begin{array}{c} 36 \\ -6 \\\hline 6 \\ 6 \\ 0 \\ 0 \\\hline\hline 24 \\ 18 \\ 0 \\ 0 \end{array}\right)
+ k_d  \left(\begin{array}{c} 0 \\ 0 \\\hline 0 \\ -2 \\ -2 \\ 0 \\\hline\hline -1 \\ -1 \\ 0 \\ -2 \end{array}\right)
\right\}}
 \stackrel{!}{=}0 \text{ mod }n \, .
\end{equation}
The second and third line are identical  with the first line being the (trivial) multiple of six of the latter.
The fourth line is the sum of third and fifth line, which in turn is identical to the tenth line.
In consequence, within the first six lines only the conditions on the second and fifth line are independent,
and any $\Z_2$ symmetry is trivially respected by these lines. The seventh and eighth line are related by adding the third line, and they
 do {\it not} automatically respect $\Z_2$ symmetries for arbitrary $(k_a,k_d)$. This example thus clearly shows that the sufficient conditions have to be taken into account when determining the appearance of discrete $\Z_n$ gauge symmetries from string theory. 
\begin{table}[h]
\begin{center}
\begin{tabular}{|c|c||c|c|c|c|c|c|c|c|c|c|}
\hline \multicolumn{12}{|c|}{\bf Discrete charges of L-R symmetric models on $T^6/(\Z_6' \times \OR)$}\\
\hline
\hline  \multicolumn{2}{|c||}{\bf Discrete symmetries}&\multicolumn{10}{|c|}{\bf Charge assignment for the chiral states}\\
\hline
\hline $\Z_n$  & $U(1) = \sum_{x \in \{a,b,d \}} k_x\, U(1)_x$& $Q_L$ & $\ov U_R$ & $\ov D_R$ & $L$ & $\ov L$ & $\ov E_R$ & $\ov N_R$& $H_u$ & $H_d$ & $\Sigma_b$  \\
\hline
\hline 
$\Z_2$ &$U(1)_a + U(1)_d$& 0&0&0&0&0&0&0&0&0&0\\
$\Z_3$&$U(1)_a$ & $0$ &$0$ &$0$ & 0 &0 &0 &0 &0 &0  &0 \\
$\Z_6$ &   $ U(1)_b$ & 0 & $1$& $1$& $4$&$4$ &$3$ &$3$ &$5$ &$5$& $4$ \\
&& 0 & 0 &2 &4&4&4&2&0&4&4\\
\hline
\end{tabular}
\caption{Overview of discrete $\Z_n$  symmetries for a four-stack model on $T^6/(\Z_6'\times \OR)$ without hidden sector (and for a five-stack model with hidden sector if the singlet $\Sigma_b$ is taken out). For each linear combination $(k_a,k_b,k_d)$ the corresponding charges for the chiral states are given after having performed a $B-L$ rotation to set the charge of $Q_L$ to zero. For the $\Z_6$ symmetry, an additional rotation over $Q_c$ also sets the charge of $\ov U_R$ to zero, as shown in the last row. The $\Z_2$ and $\Z_3$ symmetry are trivial, while the $\Z_6$ symmetry corresponds to $({\cal R}_6^5)\mathfrak{L}_6^4 {\cal A}_6^4$. 
\label{Tab:DiscreteSymmT6Z6primeModel2EndI}}
\end{center}
\end{table}

Solving the four independent conditions for various $(k_a, k_b, k_d) \neq (1,0,3)$ for a fixed $n$ gives the following list of discrete symmetries, with the corresponding charges of all chiral fields given in table~\ref{Tab:DiscreteSymmT6Z6primeModel2EndI}:

\begin{itemize}
\item The combination $(k_a, k_b, k_d) = (1,0,0)$ corresponds to a discrete $\Z_3$ symmetry embedded in $U(1)_a$, representing the baryon number. After a $B-L$ rotation setting the charge $\alpha_Q$ for the left-handed quarks to zero, the symmetry acts trivially on the chiral and non-chiral spectrum (i.e.~the charges for all states are 0 mod 3).
\item The combination $(k_a, k_b, k_d) = (1,0,1)$ gives rise to a $\Z_2$ symmetry, which after a $B-L$ rotation acts trivially on the chiral states of table \ref{Tab:T6Z6pModel2Spectrum} as displayed in table
\ref{Tab:DiscreteSymmT6Z6primeModel2EndI}. 
\item The last possibility giving rise to discrete symmetries is $(k_a, k_b, k_d) = (0,1,0)$, for which one can identify a discrete $\Z_6$ symmetry with charges given in table~\ref{Tab:DiscreteSymmT6Z6primeModel2EndI}. The $\Z_6$ symmetry corresponds in field theory terms to the symmetry ${\cal R}_6^5 {\cal A}_6^4 {\cal L}_6^4$, after a $B-L$ rotation on the charges. Before spontaneously breaking $USp(2)_c$, this $\Z_6$ symmetry is a discrete gauge symmetry within the full gauge group $SU(3)_a \times SU(2)_b \times USp(2)_c \times U(1)_{B-L} \times \Z_6$.~\footnote{\label{Ft:discreteZ6} 
Given the unwritten lore to mod out the center $\Z_2$ of the non-Abelian $SU(2)_b$ gauge group to extract the non-trivial discrete symmetry, we verified manually that the discrete $\Z_6$ charges of the massless open string states can be fitted into a $\Z_3$ group. Prior to any shifts over massless gauge symmetries, a consistent $\Z_3$ charge assignment for the MSSM states can be summarized by the following table (where the state $\Sigma_b$ only appears in the chiral spectrum of the model without hidden sector): 
  {\footnotesize
$$\begin{array}{|c||c|c|c|c|c|c|c|c|c|c|}\hline & Q_L & \ov{D}_R & \ov{U}_R & L & \ov{L} & \ov{N}_R & \ov{E}_R & H_u & H_d & \Sigma_b
\\\hline\hline
\Z_3&1 & 0 & 0 & 1 & 1 & 0 & 0 & 2 & 2 & 1
\\\hline
\end{array}$$
}This table follows naturally from the investigation of perturbative three-point couplings under the $SU(3)_a \times SU(2)_b \times USp(2)_c \times U(1)_{B-L} \times \Z_6$ selection rules. Nonetheless, it remains to be verified whether the reduction from $\Z_6$ to a $\Z_3$ symmetry by imposing `mod 3' on the discrete charges remains valid for higher order and non-perturbative couplings.
 }

 A spontaneous breaking of the gauge group $USp(2)_c$ to a massless $U(1)_c$ gauge symmetry, however, permits a shift of the charges over $Q_c$ which sets the charge of $\ov U_R$ to zero. Upon comparison with the MSSM charges in table~\ref{Tab:ZnsymmChargesMSSM}, the resulting symmetry acting on the massless open string states can be identified as ${\cal A}_6^4 {\cal L}_6^4$. Applying  the latter discrete  $\Z_6$ symmetry twice yields a discrete $\Z_3$ symmetry, corresponding to the symmetry $ {\cal A}_3 \mathfrak{L}_3$. The $\Z_2$ subgroup of this discrete $\Z_6$ symmetry acts trivially on the massless open string sector, and added to the previous combination $(k_a, k_b, k_d) = (1,0,1)$ the discrete $\Z_2$ symmetry guaranteed by the K-theory constraints emerges. 
Hence, the effective gauge group acting on the MSSM states in table~\ref{Tab:DiscreteSymmT6Z6primeModel2EndI} corresponds to $SU(3)_a \times SU(2)_b\times U(1)_c \times U(1)_{B-L} \times \Z_3$. 
\end{itemize}

In summary, in this global model none of the field theoretically known discrete $\Z_n$ symmetries of the MSSM, matter parity ${\cal R}_2$,  baryon triality ${\cal B}_3$ or proton hexality ${\cal P}_6$ occur as discrete symmetries. Nevertheless, a new and somewhat exotic discrete $\Z_3(\Z_6)$ symmetry has been identified as $({\cal R}_6^5) {\cal A}_6^4 {\cal L}_6^4$ in this model.     The discussion ends by showing that the field theory anomaly constraints are satisfied for the discrete symmetries listed above. For a generic $\Z_n$ symmetry the field theory anomalies for the chiral spectrum in table~\ref{Tab:T6Z6pModel2Spectrum} read: 
\begin{enumerate}
\item[$(i)$] \underline{$SU(3)-SU(3) - \Z_n$ anomaly constraint:} 
\begin{equation}
N_g \left( 6 \alpha_Q + 3 \alpha_u + 3 \alpha_d \right)  \in 3 n \Z \hspace{0.2in} \Rightarrow \hspace{0.2in} k N_g \in n \Z\, .
\end{equation}
\item[$(ii)$] \underline{ $SU(2)-SU(2) - \Z_n$ anomaly constraint:}
\begin{equation}\label{Eq:T6Z6SU2SU2ZnAnomaly}
\begin{aligned}
&N_g \left(6 \alpha_Q + 2 \alpha_L\right) + N_h \left( 2 \alpha_{H_u} + 2 \alpha_{H_d}  \right) + 3 (2 \alpha_{L} + 2 \alpha_{\ov L})  \in 2n \Z \\
&\hspace{0.2in}\Rightarrow \hspace{0.2in} k (N_g + 3 - N_h) + p N_g \in n \Z\, .
\end{aligned}
\end{equation}
\item[$(iii)$] \underline{$G-G - \Z_n$ anomaly constraint:}
\begin{equation}
\begin{aligned}
&N_g \left(6 \alpha_Q + 3 \alpha_u + 3 \alpha_d + 2 \alpha_L + \alpha_e + \alpha_\nu  \right) \\
& + N_h \left( 2 \alpha_{H_u} + 2 \alpha_{H_d}  \right) + 3 (2 \alpha_{L} + 2 \alpha_{\ov L}) + N_{\Sigma_b} \alpha_{\Sigma_b} \in n \Z \\
&\hspace{0.2in}\Rightarrow \hspace{0.2in}   k \left( 4 N_g - 2 N_h - N_{\Sigma_b} \right) + 6 p \in n \Z\, .
\end{aligned}
\end{equation}
\item[$(iv)$] \underline{$\Z_n - \Z_n - \Z_n $ anomaly constraint}
\begin{equation}
\begin{aligned}
&N_g \left(6 \alpha_Q^3 + 3 \alpha_u^3 + 3 \alpha_d^3 + 2 \alpha_L^3 + \alpha_e^3 + \alpha_\nu^3  \right) \\ 
 &+ N_h \left( 2 \alpha_{H_u}^3 + 2 \alpha_{H_d}^3  \right)
 + 3 (2 \alpha_{L}^3 + 2 \alpha_{\ov L}^3) + N_{\Sigma_b} \alpha^3_{\Sigma_b} \in n \Z\\
 \hspace{0.2in}\Rightarrow \hspace{0.2in}&\left[ -k \left\{k(k+3p) + 3(k-m)^2 + 3(p+m)^2  \right\} +   6 m^2 p \right] N_g  \\
  &+ 2k [ k^2 - 3 km + 3m^2 ] N_h  - 6 k [ k^2 + 3 kp + 3p^2 ] + N_{\Sigma_b} k^3 \in n \Z\, .
\end{aligned}
\end{equation}
\end{enumerate}
Evaluating the anomaly constraints for the $({\cal R}_6^5) {\cal A}_6^4 {\cal L}_6^4$
symmetry explicitly is cumbersome, but by inserting the details of the chiral spectrum $(N_g = 3, N_h = 18, N_{\Sigma_b} = 9)$ one can clearly see that all constraints are satisfied.

\subsubsection{A global model on $T^6/(\Z_6' \times \OR)$ with a hidden sector}\label{Ss:T6Z6pWithHidden}

The {\bf ABa} lattice configuration for the $T^6/(\Z_6' \times \OR)$ orbifold also allows to construct \cite{Gmeiner:2008xq} global phenomenologically appealing models with a hidden sector. For a complex structure modulus $\varrho = \frac{1}{2}$ five-stack models give a priori rise to left-right symmetric models with gauge groups $U(3)_a\times U(2)_b \times USp(2)_c \times U(1)_d\times G_{\text{hidden}}$, with the hidden gauge group $G_{\text{hidden}} = USp(6)$, $USp(4)\times USp(2) $ or $USp(2)$. 
Similarly to the model without hidden sector, the enhanced gauge group $USp(2)_c$ can be broken to a massless Abelian gauge factor $U(1)_c$ by turning on continuous Wilson lines or displacing the D6-branes $c$ along $T^2_{(2)}$. The other three $U(1)$ factors give rise to a gauged $B-L$ symmetry as in equation (\ref{Eq:T6Z6pB-L}) and two massive linear combinations. The massless hypercharge can be constructed in exactly the same way as in equation (\ref{Eq:T6Z6pHypercharge}).  
The  details for the hidden gauge groups are not given here, as the hidden stacks do not give rise to chiral matter at the intersections with the SM stacks.  The chiral spectrum for all models with hidden sector is identical, and we can discuss the discrete $\Z_n$ symmetries independently of the hidden sector. Note that the only differences between the chiral spectrum of these models and the model without hidden matter given in table~\ref{Tab:T6Z6pModel2Spectrum} are the number of Higgs doublet two-tuples $(H_u, H_d)$ ($N_h = 9$ instead of $N_h = 18$) and the absence of the singlet states $\Sigma_b$ in the models with hidden gauge groups. This implies that the naive identification of the gauged R-symmetry as $U(1)_c$ and 
generators of discrete $\Z_n$ symmetries  $\mathfrak{L}$ and ${\cal A}$ in terms of $Q_d$ and $\frac{1}{2}\left( Q_a - Q_b - Q_c + Q_d \right)$, respectively, still holds.
 
The necessary and sufficient conditions for this type of models follow from equation~(\ref{Eq:Z6p_Condition_ZM}):
\begin{equation}
{\footnotesize
\left\{ k_a  \left(\begin{array}{c} 0 \\ 0 \\\hline 0 \\ -6 \\ 6 \\ 0 \\\hline\hline -3 \\ 3 \\ 0 \\ 0 \end{array}\right)
+ k_b  \left(\begin{array}{c} 18 \\ -6 \\\hline 6 \\ 0 \\ 6 \\ 0\\\hline\hline 12 \\ 12 \\ 0 \\ 0 \end{array}\right)
+ k_d  \left(\begin{array}{c} 0 \\ 0 \\\hline 0 \\ 2 \\ -2 \\ 0\\\hline\hline 1 \\ -1 \\ 0 \\ 0 \end{array}\right)
\right\} }
 \stackrel{!}{=}0 \text{ mod }n\, .
\end{equation}
The first three lines are identical (up to an overall factor of 3 in the first line), and the fifth line is the sum 
(with possible minus signs that are irrelevant 
due to $0 \text{ mod }n$) of the conditions from the second and fourth line. As a consequence, within the first six lines 
only the conditions on the second and fourth line are independent. These lines automatically respect $n=2$ for any choice of 
$(k_a,k_b,k_d)$. The two non-trivial sufficient conditions are related by adding multiples of the second line. As in the model without hidden sector on $T^6/(\Z_6' \times \OR)$, they only respect a $\Z_2$ symmetry for particular values of $(k_a, k_d)$, emphasizing once more the importance of fulfilling the sufficient conditions as well. Solving the three independent conditions for various combinations of $(k_a, k_b, k_d)$ yields the same list of discrete $\Z_n$ gauge symmetries as for the model without hidden sectors in  table~\ref{Tab:DiscreteSymmT6Z6primeModel2EndI}. Also the expressions for the field theoretical anomaly constraints for the discrete symmetries are very similar to the ones in subsection \ref{Ss:T6Z6pNoHidden}. The only differences are the number of Higgs doublet two-tuples ($N_h = 9$ instead of $N_h=18$) and the absence of the chiral singlets $\Sigma_b$ (which boils down to setting $N_{\Sigma_b} = 0$), but one can easily show that the anomaly constraints are also satisfied for the discrete symmetries in this model, independently of the hidden sector. Although the three generators ${\cal R}$, $\mathfrak{L}$ and ${\cal A}$ can be identified in the models with hidden sectors, there are only three discrete symmetries satisfying the string theory conditions, which is less than expected from field theoretic considerations. More importantly, baryon triality and proton hexality are not explicitly realized.

\subsection{Pati-Salam models}\label{Ss:PS-models}

By unifying $SU(3) \times U(1)_{B-L} \subset SU(4)$, a left-right symmetric model turns into a Pati-Salam type GUT. These arise very naturally in intersecting D6-brane models. The apparent unification of quarks and leptons in Pati-Salam models will be reflected in the structure of the superpotential, which one naively expects to be constrained by discrete $\Z_n$ symmetries not present in left-right symmetric models or the MSSM. In this section this naive expectation will be confirmed  by investigating the discrete $\Z_n$ symmetries for global Pati-Salam models constructed on $T^6/( \Z_6' \times \OR)$ and $T^6/(\Z_2 \times \Z_6' \times \OR)$ with discrete torsion.

\subsubsection{A global model on $T^6/(\Z_6' \times \OR)$}\label{Sss:PS-model_Z6p}

A first global Pati-Salam model considered here arises~\cite{Gmeiner:2007zz} from five stacks of D6-branes wrapping fractional three-cycles on the {\bf ABa} lattice of the $T^6/(\Z_6' \times \OR)$ orbifold with complex structure modulus $\varrho=\frac{3}{2}$. The five stacks of D6-branes give rise to the gauge groups $U(4)_a \times U(2)_b \times U(2)_c \times U(1)_d \times U(1)_e$. The five $U(1)$ gauge factors combine into one massless gauged $U(1)_X = U(1)_b - 2 U(1)_e$ symmetry with conserved charge,
\begin{equation}\label{Eq:QXchargePS}
Q_X = Q_b -2 Q_e,
\end{equation}
while the four orthogonal directions give rise to four massive linear combinations by virtue of the generalized Green-Schwarz mechanism. The left-handed quarks and leptons arise from the $ab$ sector, whereas the right-handed quarks and leptons are found at the $ac'$ intersections. The full chiral spectrum together with the non-chiral Higgs-sector is given in table~\ref{Tab:T6Z6pPSModelChiralHiggs}. 
\begin{table}[h]
\begin{center}
{\footnotesize
\hspace*{-0.4in}\begin{tabular}{|c|c||c|c|c|}
\hline \multicolumn{5}{|c|}{\bf  Chiral spectrum and Higgs states for the Pati-Salam model on the ABa lattice of $T^6/(\Z_6' \times \OR)$} \\
\hline \hline
\bf Matter & \bf Sector & $U(4)_a\times U(2)_b \times U(2)_c$& $(Q_a, Q_b, Q_c, Q_d, Q_e)$ & $Q_X$\\
\hline
$(Q_L,L)$ & $ab$ & 3 $({\bf  4}, {\bf  \ov 2}, 1)$ & $(1,-1,0,0,0)$& $-1$ \\
$(Q_R,R)$ & $ac'$ & 3 $({\bf  \ov 4}, 1, {\bf  \ov 2})$ & $(-1,0,-1,0,0)$& $0$ \\
$X_{ae}$& $ae$& 3 $({\bf 4},1,1)$ &$(1,0,0,0,-1)$& $2$ \\
$X_{ad'}$& $ad'$& 3 $({\bf \ov 4},1,1)$ &$(-1,0,0,-1,0)$& 0 \\
\hline
$X_{bd}$& $bd$& 6 $(1, {\bf 2},1)$ &$(0,1,0,-1,0)$& 1 \\
$X_{be'}$& $be'$& 6 $(1,{\bf \ov 2},1)$ &$(0,-1,0,0,-1)$& $1$ \\
$X_{de}$&$de$& 6 $(1,1,1)$ &$(0,0,0,-1,1)$ &$-2$\\
$\Sigma_b$&$bb'$& 6 $(1,1_{\ov \Anti},1)$ &$(0,-2,0,0,0)$&$-2$\\
$\Sigma_c$&$cc'$& 6 $(1,1,1_{\ov \Anti})$ &$(0,0,-2,0,0)$&0\\
$S_d$&$dd'$& 6 $(1,1,1)$&$(0,0,0,2_{\Sym},0)$&0\\ 
\hline \hline
$(H_u,H_d)$ & $bc$ & 5 $[(1,{\bf 2}, {\bf \ov 2})_{(0,0)} + h.c.  ]$ &  $(0,\pm 1,\mp 1,0,0)$ &$ \pm 1$ \\
$(H_u,H_d)$ & $bc'$ & 6 $[(1,{\bf 2}, {\bf 2})_{(0,0)} + h.c.  ]$ &  $(0,\pm 1,\pm 1,0,0)$ &$ \pm 1$ \\
\hline
\end{tabular}}
\caption{Chiral states and non-chiral Higgs doublets arising from a five-stack global Pati-Salam model on the {\bf ABa} lattice of $T^6/(\Z_6'\times \OR)$.\label{Tab:T6Z6pPSModelChiralHiggs}}
\end{center}
\end{table}

For this model with five $U(1)$ gauge groups, discrete gauge symmetries survive as $\Z_n$ subgroups of linear combinations of the U(1)'s with coefficients $(k_a, k_b, k_c, k_d, k_e) \neq (0,1,0,0,n-2)$ satisfying the necessary and sufficient conditions from equation~(\ref{Eq:Z6p_Condition_ZM}),
\begin{equation}
{\footnotesize
\left\{ k_a  \left(\begin{array}{c} 0 \\ 0 \\\hline 0 \\ -8 \\ -8 \\ 0 \\\hline\hline -4 \\ -4 \\ 0 \\ -8\end{array}\right)
+ k_b  \left(\begin{array}{c} 6 \\ -2 \\\hline 2 \\ 0 \\ 2 \\ 0 \\\hline\hline 4 \\ 4 \\ 0 \\ 0 \end{array}\right)
+ k_c  \left(\begin{array}{c} 6 \\ -2 \\\hline  -2 \\ 0 \\ -2 \\ 0 \\\hline\hline 2 \\ 2 \\ -2 \\ -2 \end{array}\right)
+ k_d  \left(\begin{array}{c} 3 \\ -3 \\\hline 1 \\ 0 \\ 1 \\ 0 \\\hline\hline 2 \\ 2 \\ -1 \\ -1 \end{array}\right)
+ k_e  \left(\begin{array}{c} 3 \\ -1 \\\hline  1 \\ 0 \\ 1 \\ 0 \\\hline\hline 2 \\ 2 \\ 0 \\ 0   \end{array}\right)
\right\} }\stackrel{!}{=} 0  \text{ mod }n\, .
\end{equation}
The sixth line is trivially satisfied, whereas the fifth line is the sum of the third and fourth line. Line seven and eight are identical, and the last line is the sum of the fourth line and the second-to-last line. This brings the total number of independent conditions down to at most six. Solving them for various combinations $(k_a, k_b, k_c, k_d, k_e)$ provides the following list of inequivalent discrete gauge symmetries:  
\begin{itemize}
\item One notices immediately that the conditions are solved for $(k_a, k_b, k_c, k_d, k_e) = (1,0,0,0,0)$, $(k_a, k_b, k_c, k_d, k_e) = (0,1,0,0,0)$ and $(k_a, k_b, k_c, k_d, k_e) = (0,0,1,0,0)$, corresponding respectively to a discrete $\Z_4 \subset U(1)_a$, and two discrete $\Z_2$ symmetries, one as a subgroup of $U(1)_b$ and one as a subgroup of $U(1)_c$. The discrete $\Z_2 \subset U(1)_c$ symmetry can be interpreted as the matter parity ${\cal R}_2$, cf. the charges in
table~\ref{Tab:T6Z6pPSModelChiral}. Regarding the other $\Z_2$ symmetry, in order to set the $Q_L$ charge to zero the charges are shifted by $Q_X$, making the $\Z_2$ symmetry a trivial one, as can be seen from the second row of table~\ref{Tab:T6Z6pPSModelChiral}. 
 The discrete $\Z_4$ symmetry emerging from $U(1)_a$ is in first instance equivalent to the center of the non-Abelian $SU(4)_a$ subgroup within the $U(4)_a$ gauge group. However, the presence of a massless gauged $U(1)_X$ symmetry allows to shift the charges over $Q_X$, setting the $Q_L$ charge to zero as shown in the third row of table~\ref{Tab:T6Z6pPSModelChiral}. Upon comparison with the charges in table~\ref{Tab:ZnsymmChargesMSSM}, the $\Z_4$ symmetry can be interpreted as the discrete symmetry generated by ${\cal R}_4 \mathfrak{L}_4^2 {\cal A}_4^2$, satisfying the conditions~(\ref{Eq:ConstraintLRS}) and (\ref{Eq:ConstraintPS}). The perturbative and non-perturbative couplings in this Pati-Salam model are fully constrained by the $SU(4)_a$ gauge symmetry. The non-trivial character of the discrete $\Z_4$ symmetry manifests itself when the gauge group $SU(4)_a$ is spontaneously broken to $SU(3) \times U(1)$, provided that the $SU(3)\times U(1)$ invariant vacuum preserves the $\Z_4$ symmetry. This situation occurs for instance when the spontaneous symmetry breaking is driven by a matter multiplet in the adjoint representation of $U(4)_a$.  
\begin{table}[h]
\begin{center}
{\scriptsize
\hspace*{-0.3in}\begin{tabular}{|c|c||c|c|c@{\hspace{0.12in}}|c|c|c|c|c|c|c|c|c|}
\hline \multicolumn{14}{|c|}{\bf Discrete charges of the Pati-Salam model on $T^6/(\Z_6' \times \OR)$}\\
\hline
\hline  \multicolumn{2}{|c||}{\bf Discrete Symmetries}&\multicolumn{12}{|c|}{\bf Charge assignment for the chiral states and Higgses}\\
\hline
\hline $\Z_n$  & $U(1) = \sum_{x } k_x\, U(1)_x$& $(Q_L,L)$ & $(Q_R,R)$ & \multicolumn{2}{|c|}{$\begin{array}{c}  (H_u,H_d)  \\ bc \hspace{0.28in} bc'\end{array}$} & $X_{ae}$ & $X_{ad'}$ & $X_{bd}$ & $X_{be'}$ & $X_{de}$& $\Sigma_{b}$ & $\Sigma_{c}$ & $S_{d}$  \\
\hline
\hline 
$\Z_2$ &  $U(1)_c$ & 0 &$1$& \hspace{0.1in}$1$&$1$&0&0&0&0&0&0&0&0\\ 
 &  $U(1)_b$ & $0$ &0& \hspace{0.1in}$0$&$0$&0&0&$0$&$0$&0&$0$&0&0\\ 
$\Z_4$&  $U(1)_a $ & $0$ &$3$ & \hspace{0.1in}$1$&$1$&$3$&$3$&$1$&$1$&$2$&$2$&0&0\\ 
$\Z_6$&$ U(1)_c + 4 U(1)_d + 4 U(1)_e$ &0 &$5$& \hspace{0.1in} $5$ & $1$ &$2$& $2$& $2$ &$2$ & 0 & 0& $4$ & $2$\\ 
 \hline
\end{tabular}}
\caption{Overview of the discrete $\Z_n$ symmetries for a five-stack Pati-Salam model on $T^6/(\Z_6'\times \OR)$. For each $\Z_n$ symmetry the corresponding charges for the chiral states and the Higgs-sector are given, where for the latter the charges of the states from the $bc$ sector can be different from those of the states from the $bc'$ sector. For the second row and third row a rotation over $Q_X$ defined in equation~(\ref{Eq:QXchargePS}) was used to set the charge of $Q_L$ to zero. \label{Tab:T6Z6pPSModelChiral}}
\end{center}
\end{table}
\item The combination $(k_a, k_b, k_c, k_d, k_e) = (0,0,1,4,4)$ yields a genuine discrete $\Z_6$ symmetry, with the charges of the chiral states listed in table~\ref{Tab:T6Z6pPSModelChiral} after a $Q_X$-shift. This $\Z_6$ symmetry can be interpreted as ${\cal R}_6 \mathfrak{L}_6^4 {\cal A}_6^2$ upon comparison with the charges in table~\ref{Tab:ZnsymmChargesMSSM} and satisfies the constraints in equations~(\ref{Eq:ConstraintLRS}) and (\ref{Eq:ConstraintPS}). Applying the symmetry twice corresponds to a discrete $\Z_3$ symmetry embedded in $U(1)_c + U(1)_d + U(1)_e$ and interpretable as ${\cal R}_3 \mathfrak{L}_3 {\cal A}_3^2$.
The $\Z_6$ symmetry only survives as a discrete MSSM symmetry if the $U(2)_R$ gauge group is spontaneously broken by e.g.~a multiplet in the adjoint representation. 
\end{itemize}

In total there are three discrete $\Z_n$ symmetries: one $\Z_2$, one  $\Z_4$ and one $\Z_6$, 
 where the first two symmetries do not yield additional selection rules for the perturbative couplings beyond those of the non-Abelian $SU(4)_a \times SU(2)_c$ charges. Hence, below the string scale the effective gauge symmetry group of this model corresponds to $SU(4)_a\times SU(2)_b \times SU(2)_c \times U(1)_X \times \Z_6$.\footnote{ \label{Ft:discreteZ6PS} Analogously to the considerations in footnote~\ref{Ft:discreteZ6}, one can reduce this $\Z_6$ symmetry acting on the massless open string states manually to a $\Z_3$ symmetry. In this case the discrete $\Z_3$ charges for the chiral states in the massless open string sector read:
$${\footnotesize \begin{array}{|c||c|c|c@{\hspace{0.12in}}|c|c|c|c|c|c|c|c|c|}
\hline
&(Q_L,L) & (Q_R,R) &   
\multicolumn{2}{|c|}{ \begin{array}{c}  (H_u,H_d)  \\ bc \hspace{0.28in} bc' \end{array}} 
& X_{ae} & X_{ad'} & X_{bd} & X_{be'} & X_{de}& \Sigma_{b} & \Sigma_{c} & S_{d}\\
\hline\hline 
\Z_3&0 &2& \hspace{0.1in} 2 & 1 &2& 2& 2 &2 & 0 & 0& 1 & 2\\
\hline
\end{array}}$$
The charge assignment in this table has been checked to agree with the $\Z_6$ selection rules for the perturbative cubic couplings between the massless states.
 }
From a low-energy point of view, the discrete $\Z_n$ symmetries only survive as discrete MSSM symmetries if they are left unbroken by the vacuum configuration spontaneously breaking $SU(4)\times SU(2)_R$ to $SU(3)\times U(1)_Y$.

\subsubsection{Five-stack global Pati-Salam models on $T^6/(\Z_2 \times \Z_6' \times \OR)$ with $\eta = -1$}\label{Ss:PST6Z2Z6pmodel1}
Next, we consider global Pati-Salam models constructed on D6-branes wrapping rigid three-cycles on the {\bf AAA} lattice of the $T^6/(\Z_2 \times \Z_6' \times \OR)$ orbifold with discrete torsion.  The first type of Pati-Salam models consists~\cite{Honecker:2012qr} of five stacks of D6-brane with gauge groups $U(4)_a\times U(2)_b \times U(2)_c \times U(2)_d \times U(2)_e$, and with one generation of left-handed (right-handed) quarks and leptons coming from the $ab$ $(ac)$ sector and two generations from the $ab'$ $(ac')$ sector. As the three generations of leptons and quarks do not come from a single sector, discrete $\Z_n$ symmetries can be generation-dependent. The Higgs-sector arising from the $bc$ sector is minimal, and additional chiral supermultiplets have to be added to cancel the non-Abelian gauge anomalies, as can be seen from table~\ref{Tab:CSdecentPatiSalam1}.
\begin{table}[h]
\begin{center}
{\footnotesize
\hspace*{-0.25in}\begin{tabular}{|c||c|c|c|}
\hline \multicolumn{4}{|c|}{\bf Chiral spectrum of a five-stack Pati-Salam model on the AAA lattice of $T^6/(\Z_2 \times \Z_6' \times \OR)$}\\
\hline \hline
Matter & Sector & $U(4)\times U(2)_L \times U(2)_R \times U(2)_d \times U(2)_e$& $(Q_a, Q_b, Q_c, Q_d, Q_e)$ \\
\hline $\left(Q_L , L\right)  $&$ab$&$({\bf 4}, {\bf\bar 2}, \1,\1,\1)$ &(1,-1,0,0,0)\\
$\left( Q_L, L \right)$&$ab'$&  $2 \times ({\bf4}, {\bf 2}, \1,\1,\1)$&(1,1,0,0,0)\\
$\left( \bar u_R, \bar d_R, \bar\nu_R, \bar e_R \right)  $ & $ac$ & $({\bf\bar 4}, \1,{\bf 2},\1,\1)$ &(-1,0,1,0,0) \\
$\left( \bar u_R, \bar d_R, \bar\nu_R, \bar e_R \right)  $& $ac'$ &  $2 \times ({\bf\bar 4}, \1,{\bf\bar 2},\1,\1)$&(-1,0,-1,0,0) \\
$\left(H_d, H_u\right)$&$bc$&$(\1, {\bf2}, {\bf\bar 2},\1,\1)$&(0,1,-1,0,0) \\
$X_{bd}$& $bd$ & $(\1, {\bf 2}, \1, {\bf \bar 2},\1 ) $&(0,1,0,-1,0)\\
$X_{bd'}$&$bd'$&  $3 \times (\1, {\bf \bar  2}, \1, {\bf \bar 2},\1 ) $ &(0,-1,0,-1,0)\\
$X_{be'}$&$be'$&$(\1, {\bf \bar  2}, \1, \1, {\bf \bar 2}) $&(0,-1,0,0,-1)\\
$X_{cd}$&$cd$& $(\1, \1, {\bf \bar 2}, {\bf 2},\1 ) $   &(0,0,-1,1,0) \\
$X_{cd'}$&$cd'$&  $3 \times (\1, \1, {\bf  2}, {\bf 2},\1 ) $ & (0,0,1,1,0)\\
$X_{ce'}$&$ce'$&  $(\1, \1, {\bf 2}, \1, {\bf 2}) $ & (0,0,1,0,1)\\
\hline
\end{tabular}}
\caption{Chiral spectrum of a five-stack Pati-Salam model on $T^6/(\Z_2 \times \Z_6' \times \OR)$ with discrete torsion. 
All exotic states denoted by $X_{xy}$ are chiral with respect to the anomalous $U(1) \subset U(2)$ factors,
but non-chiral with respect to the Pati-Salam group.\label{Tab:CSdecentPatiSalam1}}
\end{center}
\end{table}
In this particular global D-brane model no massless linear combinations of the $U(1)$'s survive, and all $U(1)$'s acquire a St\"uckelberg mass via the generalized Green-Schwarz mechanism.

Due to the presence of three distinct $\Z_2$ twisted sectors for the rigid three-cycles, the number of necessary and sufficient conditions is drastically increased, as discussed in section \ref{Sss:Z2Z6p}. The sixteen a priori necessary conditions from equation (\ref{Eq:Z2Z6p-necessary_discrete}) read for this particular model:
\begin{equation}\label{Eq:Z2Z6p-Ex1-necessary_discrete}
{\tiny
k_a \; \left(\begin{array}{c} 0 \\\hline 0 \\ 0 \\ 0 \\ 0 \\ 16 \\\hline 0 \\ 0 \\ 0 \\ 0 \\ -16 \\\hline 0 \\ 0 \\ 0 \\ 8 \\ 8  \end{array}\right)
+ k_b \; \left(\begin{array}{c} 0 \\\hline 0 \\ 0 \\ 0 \\ 0 \\ 0 \\\hline 0 \\ 0 \\ 0 \\ 0 \\ 8 \\\hline 0 \\ 0 \\ 0 \\  0 \\ 8 \end{array}\right)
+ k_c \; \left(\begin{array}{c} 0 \\\hline 0 \\  0 \\ 0 \\ 0 \\ -8 \\\hline 0 \\ 0 \\ 0 \\ 0 \\ 0 \\\hline 0 \\ 0 \\ 0 \\  4 \\ -4 \end{array}\right)
+ k_d \; \left(\begin{array}{c} 0 \\\hline 0 \\  0 \\ 4 \\ -4 \\ -12 \\\hline 0 \\ 0 \\ -4 \\  4 \\ 12 \\\hline 0 \\ 0 \\ 0 \\ -4 \\ -4 \end{array}\right)
+ k_e \; \left(\begin{array}{c} 0 \\\hline -4 \\  0 \\ 0 \\ 4 \\ -4 \\\hline 4 \\ 0 \\ 0 \\ -4 \\ 4 \\\hline 0 \\ 0 \\ 0 \\  0 \\ 0 \end{array}\right)}
\stackrel{!}{=} 0 \text{ mod } n\, .
\end{equation}
One immediately observes that the first, third, eight, twelfth, thirteenth and fourteenth line are trivially satisfied. The fifth and tenth line are linear combinations of the second and fourth line, which are equivalent respectively to the seventh and ninth line. Hence, there are only six independent necessary conditions left. Adding the sufficient conditions from equation (\ref{Eq:Z2Z6p-6-sufficient_discrete}) constrains the viable discrete gauge symmetries even more:

\begin{equation}\label{Eq:Z2Z6p-Ex1-sufficient_discrete}
{\tiny k_a \; \left(\begin{array}{c} 0 \\ 0 \\ 0 \\ 0 \\ 0 \\ 0 \\\hline 4 \\ 0 \\ 0 \\ - 4 \\ 4 \\ 0
\\\hline 0\\ 4 \\ -4 \\ 0 \end{array}\right)
+ k_b \; \left(\begin{array}{c} 0 \\ 0 \\ 0 \\ 0 \\ 0 \\ 0 \\\hline 0 \\ 0 \\ 0  \\ 0 \\ 0 \\ -2 
\\\hline 2 \\ 0 \\2 \\ 0 \end{array}\right)
+ k_c \; \left(\begin{array}{c} 0 \\ 0 \\ 0 \\ 0\\ 0 \\ 0 \\\hline 0 \\ 0 \\ 0  \\ 2 \\2 \\2
\\\hline -2 \\ -2 \\ 0 \\ 0 \end{array}\right)
+ k_d \; \left(\begin{array}{c} 0 \\ 2 \\ -2 \\ -2 \\ 2 \\ 0 \\\hline -2 \\ 0 \\ 2  \\ 4 \\ 0 \\ 0
\\\hline 0 \\ -4 \\ 4 \\ 0 \end{array}\right)
+ k_e \; \left(\begin{array}{c} 0 \\ -2 \\ 2 \\ 2 \\ -2 \\ 0 \\\hline -2 \\ 0 \\ -2  \\ 2 \\ -2 \\ 0
\\\hline 0 \\ -2 \\ 2 \\ 0 \end{array}\right)}
\stackrel{!}{=} 0 \text{ mod } n\, .
\end{equation}
Of the sixteen sufficient conditions, four are trivially satisfied (the first, sixth, eigth and sixteenth line).  The second, third, fourth, fifth and ninth are equivalent, as well as the tenth with the fourteenth and the twelfth with the thirteenth, such that there are only six independent sufficient conditions. 
A careful analysis of the necessary and sufficient conditions shows that there exist seven relations among them, implying that only five conditions are truly independent.  Solving these five independent conditions for all combinations of $(k_a, k_b, k_c, k_d, k_e)$ leads to the following discrete symmetries:
\begin{itemize}
\item The most simple solutions occur when only one gauge factor is involved. The combination $(k_a = 1, k_{x\neq a} = 0)$ corresponds to a discrete $\Z_4 \subset U(1)_a$  symmetry. If one of the $k_x = 1$ with $x \in \{b,c,d,e\}$ while all the other coefficients $k_x = k_y = 0$, a discrete $\Z_2$ symmetry emerges from $U(1)_x$. All of these discrete symmetries are generation-independent, as can be seen from the charges in table~\ref{Tab:T6Z2Z6pPSModel1Chiral}. The $\Z_2$ symmetries arising from $U(1)_d$ and $U(1)_e$ act trivially on the MSSM states, but non-trivially on some exotic matter. The $\Z_2\subset U(1)_c$ symmetry corresponds to the R-parity ${\cal R}_2$. The $\Z_2$ symmetry emerging from $U(1)_b$ and the $\Z_4$ symmetry emerging from $U(1)_a$ cannot be interpreted as discrete symmetries generated by ${\cal R}$, $\mathfrak{L}$ or $\cal A$, as the $Q_L$ charge is non-zero and cannot be set to zero by shifting by a massless gauged $U(1)$.  
\begin{table}[h]
\begin{center}
{\scriptsize
\hspace*{-0.3in}\begin{tabular}{|c|c||c@{\hspace{0.12in}}|c|c@{\hspace{0.16in}}|c|c|c|c|c|c|c|c|}
\hline \multicolumn{13}{|c|}{\bf Discrete charges for the five-stack Pati-Salam model on $T^6/(\Z_2 \times \Z_6' \times \OR)$}\\
\hline
\hline    \multicolumn{2}{|c||}{\bf Discrete Symmetries}&\multicolumn{11}{|c|}{\bf Charge assignment for the chiral states}\\
\hline
\hline $\Z_n$  & $U(1) = \sum_{x } k_x\, U(1)_x$& \multicolumn{2}{|c|}{$\begin{array}{c}(Q_L,L)\\ ab \hspace{0.24in} ab' \end{array}$} &  \multicolumn{2}{|c|}{ $\begin{array}{c}(Q_R,R)\\ ac \hspace{0.24in} ac' \end{array}$} & $(H_d,H_u)$  &$X_{bd}$ & $X_{bd'}$ & $X_{be'}$ & $X_{cd}$ & $X_{ cd'}$& $X_{ce'}$  \\
\hline
\hline 
$\Z_2$&   $U(1)_e$ &\hspace{0.08in}0&0&\hspace{0.08in}0&0&0&0&0&$1$&0&0&$1$\\
&$U(1)_d$ &\hspace{0.08in}0&0&\hspace{0.08in}0&0&0&$1$& $1$&0&$1$&$1$&0\\
&$ U(1)_c$ &\hspace{0.08in}$0$ &$0$&\hspace{0.08in}$1$&$1$&$1$&$0$&$0$&$0$&$1$&$1$&$1$\\
&$U(1)_b $ &\hspace{0.08in}$1$ &$1$&\hspace{0.08in}$0$&$0$&1&$1$&$1$&$1$&$0$&$0$&$0$\\
$\Z_4$&$U(1)_a$ &\hspace{0.08in} $1$ &$1$&\hspace{0.08in}$3$&$3$&0&0&0&0&0&0&0\\
& $U(1)_b+U(1)_c+U(1)_d+U(1)_e$ &\hspace{0.08in} $3$&$1$&\hspace{0.08in}1&$3$&0&0&$2$&$2$&0&2&2\\
\hline
\end{tabular}}
\caption{Overview of the discrete symmetries  for the five-stack Pati-Salam model on $T^6/(\Z_2 \times \Z_6' \times \OR)$ with discrete torsion. For each $\Z_n$ symmetry the corresponding charges of the chiral states are listed. For the left-handed and right-handed quarks and leptons the charges depend on the sector ($ab$, $ab'$, $ac$, $ac'$) the state originates from. In this model there is no massless gauge symmetry available to shift the charge of $Q_L$ to zero.\label{Tab:T6Z2Z6pPSModel1Chiral}}
\end{center}
\end{table}
\item In this model there are no discrete $\Z_3$ symmetries, as the rank of the QCD gauge group for a Pati-Salam model is $N_a=4$ instead of $N_a=3$. 
\item Surprisingly, an additional discrete $\Z_4$ symmetry arises from the massive, linear combination  $U(1)_b+U(1)_c + U(1)_d + U(1)_e$. This $\Z_4$ symmetry acts on the quarks and leptons in a  {\bf generation-dependent} way, and leaves the Higgs-sector untouched, see table~\ref{Tab:T6Z2Z6pPSModel1Chiral}. The symmetry also transforms some of the exotic chiral matter. In this sense, the discrete $\Z_4$ symmetry can be used to exclude certain couplings between the exotic and visible matter. 
For example, the three-point coupling $(\4,\ov{\2},\1,\1,\1) .  (\ov{\4},\1,\1,\2,\1). (\1,\ov{\2},\1,\ov{\2},\1)$ is perturbatively forbidden by $U(1)_b$ charge conservation, or from the low-energy perspective by the $\Z_4$ charge 
\mbox{3 +1 + 2 = 2\text{ mod }4}. D2-brane instantons in the antisymmetric representation of some $U(2)$ gauge factor carry exactly the $\Z_4$ charge 2 needed for compensation. A closer inspection of D2-brane instanton contributions to the couplings, however, goes beyond the scope of the present article.
\end{itemize}
In summary, there are five generation-independent $\Z_n$ symmetries (four $\Z_2$'s and one $\Z_4$ with one linear dependence, which do not provide additional selection rules for the perturbative couplings beyond the ones of the non-Abelian charges), and one generation-dependent $\Z_4$ symmetry.\footnote{ Given the unwritten folklore to mod out the center of the non-Abelian gauge factors from the set of independent discrete symmetries, as also pointed out in footnotes \ref{Ft:discreteZ6} and \ref{Ft:discreteZ6PS}, one is led to a remaining $((\Z_4)^2 \times (\Z_2)^3)/(\Z_4 \times (\Z_2)^4) \simeq \Z_2$ symmetry as the only non-trivial discrete symmetry. We verified manually that a consistent $\Z_2$ charge assignment for the massless open string spectrum gives the following table:
{\footnotesize
$$\begin{array}{|c||c|c|c|c|c|c|c|c|c|c|c|c|c|}\hline  &\multicolumn{2}{|c|}{ \begin{array}{c}(Q_L,L)\\ ab \hspace{0.24in} ab' \end{array}} 
 & \multicolumn{2}{|c|}{ \begin{array}{c}(Q_R,R)\\ ac \hspace{0.24in} ac' \end{array}}  & (H_d,H_u) & X_{bd} & X_{bd'} &  X_{be'} & X_{cd} & X_{cd'} & X_{ce'}\\\hline\hline
\Z_2 & \hspace{0.09in} 0  \hspace{0.09in} & 1 &  \hspace{0.09in} 0  \hspace{0.09in} & 1 & 0 & 0 & 1 & 1 & 0 & 1 & 1
\\\hline
\end{array}$$}
By investigating perturbative three-point couplings, we have ensured that this $\Z_2$ symmetry imposes the same selection rules as the $\Z_4$ symmetry. From the perturbatively allowed coupling $(\4,\ov{\2},\1,\1,\1) . (\ov{\4},\1,\2,\1,\1)  . (\1,\2,\ov{\2},\1,\1)$ and the perturbatively forbidden coupling $(\4,\ov{\2},\1,\1,\1) .  (\ov{\4},\1,\ov{\2},\1,\1) . (\1,\2,\ov{\2},\1,\1)$ one notices immediately that this $\Z_2$ remains a generation-dependent discrete symmetry and cannot be obtained by imposing `mod 2' on the $\Z_4$ charges.
  }
Below the string scale, the effective gauge group for this model thus corresponds to $SU(4)_a\times SU(2)_b\times SU(2)_c \times SU(2)_d \times SU(2)_e \times \Z_4$.

\subsubsection{Six-stack global Pati-Salam models on $T^6/(\Z_2 \times \Z_6' \times \OR)$ with $\eta=-1$}
On the {\bf AAA} lattice a second type of global Pati-Salam models can be constructed~\cite{Honecker:2012qr,Honecker:2013kda} on six stacks of D6-branes wrapping fractional three-cycles. They give rise to the gauge group $U(4)_a\times U(2)_b \times U(2)_c \times U(4)_d \times U(2)_e  \times U(2)_f$. Just like in the previous Pati-Salam model, the three generations of quarks and leptons arise from different sectors: two generations of left-handed (right-handed) quarks arise from the $ab$ ($ac'$) sector, while the third generations arises from the $ab'$ $(ac)$ sector. Hence, also in this model $\Z_n$ symmetries can be generation-dependent. The Higgs-sector is not minimal in this model: two chiral Higgs-doublets $(H_d, H_u)$ arise from the $bc'$ sector, and one non-chiral pair arises at the $bc$ intersections. Similarly to the five-stack model, exotic chiral matter has to be added in order for the non-Abelian gauge anomalies to vanish, see table~\ref{Tab:OtherDecentPatiSalamC}.
\begin{table}[h]
\begin{center}
{\footnotesize
\hspace*{-0.4in}\begin{tabular}{|c||c|c|c|c|}
\hline \multicolumn{5}{|c|}{\bf Chiral spectrum of a six-stack Pati-Salam model on the AAA lattice of $T^6/(\Z_2 \times \Z_6' \times \OR)$}\\
\hline \hline
Matter & Sector & $U(4)\times U(2)_L \times U(2)_R \times U(4)_d \times U(2)_e\times U(2)_f$& $(Q_a, Q_b, Q_c, Q_d, Q_e, Q_f)$ & $Q_Z$ \\
\hline $\left(Q_L , L\right)  $&$ab$&$2 \times ({\bf 4}, {\bf\bar 2}, \1,\1,\1,\1)$ &(1,-1,0,0,0,0)&$-1$\\
$\left( Q_L, L \right)$&$ab'$&  $({\bf4}, {\bf 2}, \1,\1,\1,\1)$&(1,1,0,0,0,0)&$1$\\
$\left( \bar u_R, \bar d_R, \bar\nu_R, \bar e_R \right)  $ & $ac$ & $({\bf\bar 4}, \1,{\bf 2},\1,\1,\1)$ &(-1,0,1,0,0,0)&$-1$ \\
$\left( \bar u_R, \bar d_R, \bar\nu_R, \bar e_R \right)  $& $ac'$ &  $2 \times ({\bf\bar 4}, \1,{\bf\bar 2},\1,\1,\1)$&(-1,0,-1,0,0,0)&$1$ \\
$\left(H_d, H_u\right)$&$bc'$&$2 \times (\1, {\bf2}, {\bf 2},\1,\1,\1)$&(0,1,1,0,0,0)& 0 \\
$X_{bd}$& $bd$ & $(\1, {\bf 2}, \1, {\bf \bar 4},\1,\1) $&(0,1,0,-1,0,0)& 1\\
$X_{bd'}$&$bd'$&  $(\1, {\bf \bar  2}, \1, {\bf \bar 4},\1,\1) $ &(0,-1,0,-1,0,0)&$-1$\\
$X_{bf}$&$bf$&$(\1, {\bf 2}, \1, \1,\1, {\bf \bar 2}) $&(0,1,0,0,0,-1)&0\\
$X_{bf'}$&$bf'$&$(\1, {\bf \bar  2}, \1, \1, \1, {\bf \bar 2}) $&(0,-1,0,0,0,-1)&$-2$\\
$X_{cd}$&$cd$& $(\1, \1, {\bf \bar 2}, {\bf 4},\1,\1) $   &(0,0,-1,1,0,0)&1 \\
$X_{cd'}$&$cd'$&  $(\1, \1, {\bf  2}, {\bf 4},\1,\1) $ & (0,0,1,1,0,0)&$-1$\\
$X_{cf}$&$cf$&  $(\1, \1, {\bf 2}, \1, \1, {\bf \bar 2}) $ & (0,0,1,0,0,-1)&$-2$\\
$X_{cf'}$&$cf'$&  $(\1, \1, {\bf \bar 2}, \1, \1, {\bf \bar 2}) $ & (0,0,-1,0,0,-1)&0\\
$\left(H_d, H_u\right)$&$bc$&$(\1, {\bf2}, {\bf \ov 2},\1,\1,\1) + h.c.$&(0,$\pm1$,$\mp1$,0,0,0)& $\pm 2$ \\
\hline
\end{tabular}}
\caption{Chiral spectrum and non-chiral Higgs pair arising from a global six-stack Pati-Salam model on $T^6/(\Z_2 \times \Z_6' \times \OR)$ with discrete torsion. 
All exotic states denoted by $X_{xy}$ are chiral with respect to the anomalous $U(1) \subset U(2)$ factors,
but non-chiral with respect to the Pati-Salam group.\label{Tab:OtherDecentPatiSalamC}}
\end{center}
\end{table}

The six stacks of D6-branes in this model provide for six Abelian gauge factors. Four of them combine into a massless gauged $U(1)_Z$ with conserved charge:
\begin{equation}
Q_Z = Q_b - Q_c + Q_e + Q_f,
\end{equation}
while the five perpendicular directions acquire a St\"uckelberg mass via the generalized Green-Schwarz mechanism. The massless linear combination can be found by investigating the {\it necessary} conditions (\ref{Eq:Zn-condition}) for this particular model, which seem -- inferring from equation~(\ref{Eq:Z2Z6p-necessary_discrete}) -- at first sight quite involved:  
\begin{equation}\label{Eq:Z2Z6p-Ex2-necessary_discrete}
{\tiny
k_a \; \left(\begin{array}{c} 0 \\\hline  -8 \\ 0 \\ 0 \\ -8 \\ 8 \\\hline 0 \\ 0 \\ 0 \\ 0 \\ 0  \\\hline 0 \\ 0 \\ -8 \\ 8 \\ 8 \end{array}\right)
+ k_b \; \left(\begin{array}{c} 0 \\\hline  2 \\ 0 \\ 2 \\ 0 \\  4 \\\hline -2 \\ 0 \\ -2 \\ 0 \\ -4 \\\hline 0 \\ 0 \\ 0 \\  0 \\ 0 \end{array}\right)
+ k_c \; \left(\begin{array}{c} 0 \\\hline -2 \\ 0 \\ -2 \\  0 \\ -4 \\\hline -2 \\ 0 \\ -2 \\  0 \\ -4 \\\hline 0 \\ 0 \\ 0 \\ 0 \\ 0  \end{array}\right)
+ k_d \; \left(\begin{array}{c} 0 \\\hline  0 \\ 0 \\ 8 \\  8 \\ 8 \\\hline 0 \\ 0 \\ 0 \\  0 \\ 0 \\\hline 0 \\ 0 \\ 8 \\ -8 \\ -8 \end{array}\right)
+ k_e \; \left(\begin{array}{c} 0 \\\hline  0 \\ 0 \\ -4 \\ -4 \\ -4 \\\hline 0 \\ 0 \\ 0 \\  0 \\ 0 \\\hline 0 \\ -4 \\ 0 \\ 0 \\ 8  \end{array}\right)
+ k_f \; \left(\begin{array}{c} 0 \\\hline -4 \\ 0 \\ 0 \\  4 \\ -4 \\\hline 0 \\ 0 \\ 0 \\ 0 \\ 0 \\\hline 0 \\ 4 \\ 0 \\ 0 \\ -8  \end{array}\right)}
\stackrel{!}{=} 0 \text{ mod } n\, .
\end{equation}
A closer look reveals that the first, third, eight, tenth and twelfth line are trivially satisfied, while the sum of the thirteenth and fourteenth line reproduce the fifth and sixteenth line. Furthermore, the seventh, ninth and eleventh line are equivalent, such that  only six lines are independent.  It is then not difficult to verify that the combination $(k_a,k_b,k_c,k_d,k_e,k_f) = (0,1,-1,0,1,1)$ solves the constraints with $n=0$ and implies the existence of a massless $U(1)$ gauge symmetry.

In order to find the discrete symmetries, the {\it necessary} conditions have to be supplemented with the {\it sufficient} conditions from equation~(\ref{Eq:Z2Z6p-6-sufficient_discrete}) :
\begin{equation}\label{Eq:Z2Z6p-Ex2-sufficient_discrete}
{\tiny k_a \; \left(\begin{array}{c} -4 \\ -4 \\ 0 \\ -4 \\ -8 \\ 0 \\\hline 4 \\ 4 \\ 0  \\ 4 \\4 \\ 0
\\\hline 4 \\ 4 \\ 0 \\ 4 \end{array}\right)
+ k_b \; \left(\begin{array}{c} 0 \\ 2 \\ -2 \\ -2 \\ 2 \\ 0 \\\hline 0 \\ 0 \\ 0  \\ -2 \\ 0 \\ 0 
\\\hline 0 \\ 0 \\0 \\ 0 \end{array}\right)
+ k_c \; \left(\begin{array}{c} -2 \\ -2 \\- 2 \\ -2 \\ -2 \\ -2 \\\hline 0 \\ 2 \\ 0  \\ 2 \\0 \\2
\\\hline 0 \\ 0 \\ 0 \\ 2 \end{array}\right)
+ k_d \; \left(\begin{array}{c} 4 \\ 4 \\ 0 \\ 4 \\ 8 \\ 4 \\\hline -4 \\ -4 \\ 0  \\ -4 \\ -4 \\ -4
\\\hline -4 \\ -4 \\ 0 \\ -8 \end{array}\right)
+ k_e \; \left(\begin{array}{c} -2 \\ -2 \\ 0 \\ 0 \\ -4 \\ -2 \\\hline 2 \\ 2 \\ 0  \\ 2 \\ 0\\ 2
\\\hline 2 \\ 2 \\ 0 \\ 2 \end{array}\right)
+ k_f \; \left(\begin{array}{c} 0 \\ -2 \\ 0 \\ 0 \\ 0 \\ 0 \\\hline -2 \\ 0 \\ 0  \\ 2 \\ 0 \\ 0
\\\hline -2 \\ -2 \\ 0 \\ 0 \end{array}\right)}
\stackrel{!}{=} 0 \text{ mod } n\, .
\end{equation}
The intricate form or the {\it sufficient} conditions demands for simplification: the first and eighth line are equivalent, just like the second line is equivalent to the tenth line, and the sixth line to the twelfth line. Furthermore, the seventh, thirteenth and fourteenth line are equivalent, and the ninth and fifteenth line are trivially satisfied. The third and eleventh line already appeared in the {\it necessary} conditions, such that only seven lines form independent conditions. A further manipulation of the six necessary and seven sufficient conditions reveals seven additional relations among them, implying that only six conditions are truly independent. Solving these six conditions for all integer coefficients $k_x$ with $(k_a,k_b,k_c,k_d,k_e,k_f)\neq(0,1,n-1,0,1,1)$ leads to the discrete $\Z_n$ symmetries displayed in table~\ref{Tab:T6Z2Z6pPSModel2Chiral}. The only solutions are the discrete symmetries emerging from a single gauge factor $U(1)_{x \in \{a,b,c,d,e,f\}}$, i.e.~the solutions with one of the $k_x \neq 0$ while all other integer coefficients are set to zero. These solutions are identified as one discrete $\Z_4$ symmetry emerging $U(1)_a$ and a second one from  $U(1)_d$. The latter acts trivially on the quarks, leptons and Higgses, but non-trivially on the exotic chiral matter, as can be seen from the fifth row in table~\ref{Tab:T6Z2Z6pPSModel2Chiral}. The $\Z_4\subset U(1)_a$ symmetry acts only on the quarks and leptons in a generation-independent way, see the sixth row in table~\ref{Tab:T6Z2Z6pPSModel2Chiral}. Performing a rotation over $Q_Z$ to set the $Q_L$ charges for two generations  to zero, the discrete symmetry turns into a generation-dependent symmetry, inherited from the generation-dependent $U(1)_Z$ symmetry. For each of the $U(2)$ gauge factors, a discrete $\Z_2$ symmetry is found. None of these $\Z_2$ discrete symmetries are generation-dependent. The $\Z_2$ symmetries arising from $U(1)_c$ and $U(1)_b$ can both be interpreted as the matter parity ${\cal R}_2$. In case of the latter a $Q_Z$-shift was used to set the charges of $Q_L$ to zero. The $\Z_2$ symmetry emerging from $U(1)_e$ acts completely trivially on the chiral spectrum, but non-trivially on the non-chiral states charged under $U(1)_e$. Finally, the $\Z_2$ symmetry from $U(1)_f$ acts trivially on the quarks, leptons and Higgses, but might serve as a discrete symmetry forbidding interactions between the SM particles and chiral exotics charged under $U(1)_f$. Below the string scale, the effective gauge group acting on the massless open string states is thus given by $SU(4)_a\times SU(2)_b\times SU(2)_c \times SU(4)_d \times SU(2)_e \times SU(2)_f \times U(1)_Z $, since the discrete $\Z_4^2 \times \Z_2^3$ symmetries do not constrain the perturbative couplings any further beyond the selection rules of the non-Abelian gauge factors.
\begin{table}[h]
\hspace{-0.3in}
{\scriptsize
\hspace*{-0.2in}\begin{tabular}{|c|c||c@{\hspace{0.2in}}|c|c@{\hspace{0.12in}}|c|c|c|c|c|c|c|c|c|c|}
\hline \multicolumn{15}{|c|}{\bf Discrete charges for the six-stack Pati-Salam model on $T^6/(\Z_2 \times \Z_6' \times \OR)$}\\
\hline
\hline  \multicolumn{2}{|c||}{\bf Discrete Symmetries}&\multicolumn{13}{|c|}{\bf Charge assignment for the chiral states}\\
\hline
\hline $\Z_n$  & $U(1) = \sum_{x } k_x\, U(1)_x$& \multicolumn{2}{|c|}{$\begin{array}{c}(Q_L,L)\\ ab \hspace{0.3in} ab' \end{array}$} &  \multicolumn{2}{|c|}{ $\begin{array}{c}(Q_R,R)\\ ac \hspace{0.3in} ac' \end{array}$}& $(H_d,H_u)$  &$X_{bd}$ & $X_{bd'}$ & $X_{bf}$ & $X_{bf'}$ & $X_{ cd}$& $X_{cd'}$ & $X_{ cf}$& $X_{cf'}$  \\
\hline
\hline 
$\Z_2$  &  $U(1)_f$ &\hspace{0.02in} 0 & 0&\hspace{0.02in} 0&0&0&0&0&$1$& $1$&0&0&$1$&$1$\\
 & $U(1)_e$  &\hspace{0.06in}0&0&\hspace{0.06in}0 &0&0&0&0&0&0 &0&0&0&0\\
 &  $U(1)_c$  &\hspace{0.06in}0&0&\hspace{0.06in}$1$ &$1$&$1$&0&0&0& 0&$1$&$1$&$1$&$1$\\
 & $U(1)_b$ &\hspace{0.06in}$0$&0&\hspace{0.06in}1 &1&$1$&$0$&$0$&1&$1$ &1&1&0&0\\
 $\Z_4$  &  $U(1)_d$  &\hspace{0.06in}0&0&\hspace{0.06in}0 &0&0&$3$&$3$&0& 0&1&1&0&0\\
  & $U(1)_a$ & \hspace{0.06in}$1$&$1$&\hspace{0.06in}$3$ &$3$&0&0&0&0&0 &0&0&0&0\\
  &&\hspace{0.06in}$0$&$2$&\hspace{0.06in}$2$ &$0$&0&1&3&0&2 &1&3&2&0\\
  \hline
\end{tabular}}
\caption{Overview of the discrete symmetries  for the six-stack Pati-Salam model on $T^6/(\Z_2 \times \Z_6' \times \OR)$. For each $\Z_n$ symmetry the corresponding charges of the chiral states are listed. For the left-handed and right-handed quarks and leptons the charges depend on the sector ($ab$, $ab'$, $ac$, $ac'$) the state originates from. In the last row a $Q_Z$ shift sets the charge for two generations of $Q_L$ to zero and makes the symmetry generation dependent. \label{Tab:T6Z2Z6pPSModel2Chiral}}
\end{table}

\section{Discussion and Conclusions}\label{S:Conclusions}

In this paper, a thorough study of discrete $\Z_n$ gauge symmetries in global models (i.e. models satisfying the RR tadpole cancellation conditions and K-theory constraint) of intersecting D6-branes has been presented on orientifolds of toroidal orbifolds, where the D6-branes wrap fractional or rigid three-cycles along the internal directions. These discrete symmetries emerge from (the linear combinations of) the Abelian parts of the $U(N) = SU(N)\times U(1)$ gauge groups living on the D6-branes, once the $U(1)$'s acquired a mass via the St\"uckelberg mechanism and do no longer operate as local symmetries. The conditions for the existence of discrete $\Z_n$ symmetries are briefly reviewed for the six-torus and generalized for factorisable toroidal orbifolds with at least one titled two-torus. For such backgrounds it is rather non-trivial to find the complete basis of $\OR$-even and $\OR$-odd three-cycles, as they no longer form a unimodular lattice. The approach presented here consists of two steps: first  construct the basic $\OR$-even three-cycles from the orbifold-invariant bulk and exceptional three-cycles. With these $\OR$-even three-cycles the {\it necessary} conditions on the existence of discrete $\Z_n$ symmetries can be worked out. Secondly, the three-cycles for which the gauge group enhances to a $USp(2N)$ or $SO(2N)$ factor are intrinsically $\OR$-even, and the linearly independent ones give rise to additional constraints, which ought to be interpreted as ${\it sufficient}$ conditions. The necessary and sufficient conditions found in this process cannot be completely independent, as the total number of constraints clearly exceeds the dimension of the space of $\OR$-even three-cycles, given by the number $h_{21} +1$. In practice many of the constraints will turn out to be trivially satisfied or will be linearly dependent.      

The toroidal orbifolds in this paper either allow for the construction of fractional three-cycles, such as $T^6/(\Z_6\times \OR)$ and $T^6/(\Z_6'\times \OR)$, or of rigid three-cycles, such as $T^6/(\Z_2\times\Z_6'\times \OR)$ with discrete torsion. As a warm-up example, $T^6/(\Z_2\times\Z_2\times \OR)$ without discrete torsion with tilted tori iluminates how the presence of tilted two-tori takes away the unimodular character of the lattice of $\OR$-even and $\OR$-odd three-cycles. From a model building perspective however, the other three orbifolds are more appealing, as they allow for the construction of global intersecting D6-brane models. Each of the orbifolds has of at least two tilted two-tori whose complex structure moduli are fixed by a $\Z_6$ action, and consequently the $\OR$-even and $\OR$-odd three-cycles do not form a unimodular lattice for these orbifolds either. Following the strategy discussed above, the necessary and sufficient conditions on the existence of discrete $\Z_n$ symmetries are derived for all three orbifolds. As a byproduct to arrive at the sufficient conditions, a full classification of $\OR$-even three-cycles giving rise to a $USp(2N)$ or $SO(2N)$ gauge group is included for the {\bf AAB} lattice on $T^6/(\Z_6\times \OR)$ and the {\bf ABa} lattice on $T^6/(\Z_6'\times \OR)$.  

For the three lattice choices, global left-right symmetric models and global Pati-Salam models can be constructed on the respective toroidal orbifolds. The left-right symmetric character automatically implies the presence of the R-parity ${\cal R}_2$. If the gauge group $SU(2)_R$ is realized by an enhanced gauge factor $USp(2)$, the R-parity remains a gauged symmetry, whereas if the gauge group $SU(2)_R$ comes from a genuine $U(2)$, the R-parity is an actual discrete symmetry. For left-right symmetric models, the other two generators $\mathfrak{L}$ and ${\cal A}$ can also be explicitly identified in terms of $U(1)$ charges: $\mathfrak{L}$ corresponds to $Q_d$, the $U(1)$ charge of the Abelian gauge group in a four-stack model; ${\cal A}$ can be realized by the linear combination $\frac{1}{2}(Q_a - Q_b - Q_c + Q_d)$, if the Higgses $H_u$ and $H_d$ are not realized as each others' hermitian conjugates in the D6-brane construction. Although discrete symmetries generated purely by $\mathfrak{L}$ or ${\cal A}$ are not found for global D6-brane models, $\Z_6$ symmetries generated by a combination of them do occur explicitly in certain models. 

In the left-right symmetric models, additional gauged $U(1)$ symmetries (like the gauged $B-L$ symmetry) can occur that allow to set the charges under the $\Z_n$ symmetry to zero, making various discrete symmetries trivial from the field theory perspective. For the global left-right symmetric model on $T^6/(\Z_6\times \OR)$, all $\Z_2$ and $\Z_3$ discrete symmetries can be rotated to trivial symmetries in this way. The non-trivial $\Z_6$ and $\Z_3$ discrete symmetries arising from global left-right symmetric models and a global Pati-Salam model on $T^6/(\Z_6'\times \OR)$ are listed in table~\ref{Tab:DiscreteSymmLBViolatingOperators}, together with their effect on lepton- and/or baryon-number violating operators. The only operator allowed by all gauge and discrete $\Z_n$ symmetries is the four-point coupling $\ov U \, \ov U\, \ov D\, \ov E $, which violates lepton- and baryon-number but preserves the difference $B-L$. All other operators violate the discrete symmetries and are therefore forbidden in perturbation theory. In general, D-brane instantons are expected to produce non-perturbative corrections to the superpotential, provided that the three-cycle wrapped by the instanton is a rigid three-cycle. However, the D-brane instanton corrections are constrained by the existence of remnant discrete $\Z_n$ symmetries. As there are no rigid three-cycles on the $T^6/(\Z_6'\times \OR)$ orbifold, non-perturbative contributions to the superpotential from D-brane instantons are not expected in this particular model.
 
Besides the R-parity ${\cal R}_2$ additional discrete $\Z_2$ symmetries appear from other $U(2)$ gauge factor in the global Pati-Salam models. Remarkably, $\Z_3$ symmetries are not found, instead a discrete $\Z_4$ symmetry emerges from the $U(4)$ `QCD stack' in all models investigated here.  Depending on the model, $\Z_6$ symmetries are viable, as is the case for the Pati-Salam model on $T^6/(\Z_6'\times \OR)$ (see right block in table~\ref{Tab:DiscreteSymmLBViolatingOperators}). In the two global models on $T^6/(\Z_2\times\Z_6'\times \OR)$ with discrete torsion,
the three generations of quarks and leptons are realized by adding up two different sectors: $ab$ and $ab'$ for the left-handed states, $ac$ and $ac'$ for the right-handed states. As the three generations do not arise from a single sector, continuous and discrete symmetries can become generation-dependent. These generation-dependent symmetries exclude couplings between different generations of quark and leptons, without having to determine the Yukawa-couplings explicitly.  
       
The main objective of the paper was the study of discrete gauge symmetries as they arise purely from the string theoretic framework of intersecting D6-branes. Various global left-right symmetric models come with additional massless gauged $U(1)$ symmetries, such as the gauged $B-L$ symmetry, which if surviving as local symmetries at energies close to the electro-weak scale might be visible as $Z'$ bosons~\cite{Anchordoqui:2012wt,Ghilencea:2002da,Anchordoqui:2011eg}, dark photons~\cite{Abel:2008ai,Andreas:2011in} or cosmological fluctuations~\cite{Dudas:2012pb}. The $U(1)$ symmetries thus have to be broken spontaneously by field theoretic effects, such that they act at most as global symmetries  in agreement with the phenomenology of the Standard Model. Upon spontaneous symmetry-breaking, new discrete $\Z_n$ symmetries may emerge.
Similarly, when the $SU(4)_{QCD}$ and the $SU(2)_R$ gauge groups in the Pati-Salam models are broken to a $SU(3)_{QCD}$ and $U(1)_Y$ by field theoretic effects,  additional $\Z_n$ symmetries might arise.

\begin{table}[h]
\begin{center}
\hspace*{-0.2in}\begin{tabular}{|c||c|c|c||c|c|c|c||c|c|}
\hline \multicolumn{10}{|c|}{\bf Discrete Symmetries and their physical effect for models in $T^6/(\Z_6'\times \OR)$} \\
\hline
\hline &\multicolumn{3}{|c||}{\bf Field Theory} &\multicolumn{4}{|c||}{\bf left-right symmetric} & \multicolumn{2}{|c|}{\bf Pati-Salam}\\
\hline
\hline Operator & ${\cal R}_2$ & ${\cal B}_3$ & ${\cal P}_6$ & ${\cal R}_6^5{\cal A}_6^4 \mathfrak{L}_6^4$ & ${\cal R}_3^2{\cal A}_3 \mathfrak{L}_3$ & $\mathfrak{L}_3 {\cal A}_3$  & $B-L$ & ${\cal R}_6 \mathfrak{L}_6^4 {\cal A}_6^2$ & ${\cal R}_3 \mathfrak{L}_3 {\cal A}_3^2$  \\
\hline \hline
$H_d H_u$ & \checkmark& \checkmark& \checkmark& &  & &\checkmark&&\\
\hline
$L H_u$&& \checkmark& &&\checkmark&&&&\\
 $LL \ov E$&& \checkmark& &&&\checkmark&&&\\
$LQ \ov D$ && \checkmark& &&&\checkmark&&&\\
$Q \ov U\, \ov E H_d$ && \checkmark& &&\checkmark&&&&\\
$L H_u L H_u$& \checkmark& \checkmark& \checkmark&\checkmark&\checkmark&&&&\\
$L H_u H_d H_u$& & \checkmark&&&&&&&\checkmark\\
\hline
$\ov U \, \ov D\, \ov D$ &&& &&\checkmark&&&&\checkmark\\
$QQQ H_d$ & &&&&&&&&\\
\hline
$QQQL$ &  \checkmark&&&&&&\checkmark&\checkmark&\checkmark\\
$\ov U\, \ov U\, \ov D\, \ov E$& \checkmark&&&\checkmark&\checkmark&\checkmark &\checkmark&&\\
\hline \hline
$\ov U \, \ov D^\dagger \ov E$&& \checkmark& &&\checkmark&&&&\\
$H_u^\dagger H_d \ov E$ && \checkmark& &&\checkmark&&&&\\
$Q \ov U L^\dagger$&& \checkmark& &&\checkmark&&&&\\
\hline
$Q Q\, \ov D^\dagger$ &&& &&&&&&\\
\hline
\end{tabular}
\caption{Overview of discrete $\Z_n$ symmetries arising from the global left-right symmetric models and global Pati-Salam model on $T^6/(\Z_6'\times \OR)$: the left column lists the lepton- and/or baryon-number violating operators (see table~\ref{Tab:LBViolatingOp}) ordered in the following way: the first four blocks list the $\mu$-term, the lepton-number violating operators, the baryon-number violating operators and the lepton-and baryon-number violating operators from the superpotential; the last two blocks list the $D$-terms yielding the lepton- and baryon-number violating operators. The second block of columns marks (\checkmark) which of them are allowed by the usual field theoric discrete $\Z_n$ symmetries: R-parity ${\cal R}_2$, baryon triality ${\cal B}_3$ and proton hexality ${\cal P}_6$. The third block of columns lists the non-trivial discrete symmetries and the gauged $B-L$ symmetry of the left-right symmetric model
and marks (\checkmark) which operators are allowed for a particular symmetry. The last block of columns gives the non-trivial discrete $\Z_n$ symmetries for the Pati-Salam model with the allowed operators.   Operators might be forbidden by massless Abelian symmetries such as the ${\cal R}_6$-symmetry of the left-right symmetric model.\label{Tab:DiscreteSymmLBViolatingOperators}}
\end{center}
\end{table}
The classification of discrete symmetries in the global Pati-Salam models on $T^6/(\Z_2 \times \Z_6' \times \OR)$ with discrete torsion~\cite{Honecker:2012qr} facilitates the computation of D2-brane instanton corrections, which might generate hierarchically suppressed missing couplings. Besides this phenomenological search, the conditions on the existence of discrete $\Z_n$ symmetries will be derived also for  $T^6/(\Z_2 \times \Z_6 \times \OR)$ with discrete torsion, which we expect to be a fertile background for MSSM and GUT vacua. It will also be interesting to search for non-Abelian discrete symmetries along the lines in~\cite{BerasaluceGonzalez:2012vb}.

Abelian symmetries on D6-branes in generic Calabi-Yau manifolds have only been investigated to leading order~\cite{Grimm:2011dx,Kerstan:2011dy}. For future model building, it will be of great interest to extend the conditions on the existence of discrete $\Z_n$ symmetries to such backgrounds.

\acknowledgments

The authors would like to thank Michael Blaszczyk for helpful discussions.
This work is partially supported by the {\it Cluster of Excellence `Precision Physics, Fundamental Interactions and Structure of Matter' (PRISMA)} DGF no. EXC 1098,
the DFG research grant HO 4166/2-1 and the Research Center {\it `Elementary Forces and Mathematical Foundations' (EMG)} at JGU Mainz.

\clearpage


\addcontentsline{toc}{section}{References}
\bibliographystyle{ieeetr}
\bibliography{refsDiscrete}

\end{document}